\documentclass[11pt]{article}
\usepackage[margin=1in]{geometry}
\usepackage[T1]{fontenc}
\usepackage{lmodern}
\usepackage{microtype}
\usepackage{amsmath,amssymb,amsthm,mathtools,bm}
\usepackage{booktabs,enumitem}
\usepackage[hidelinks]{hyperref}
\usepackage[capitalize]{cleveref}
\hypersetup{pdftitle={CLS-Completeness of Exact Equilibrium Computation in Linear Fisher Markets with Chores},pdfsubject={CLS-completeness with two earning levels and explicit rational precision bounds}}
\setlist[itemize]{topsep=4pt,itemsep=2pt}
\setlist[enumerate]{topsep=4pt,itemsep=3pt}
\allowdisplaybreaks[2]

\usepackage{algorithm}
\usepackage[noend]{algpseudocode}

\usepackage{tikz}

\newtheorem{theorem}{Theorem}
\newtheorem{lemma}{Lemma}
\newtheorem{proposition}{Proposition}
\newtheorem{corollary}{Corollary}
\theoremstyle{definition}
\newtheorem{definition}{Definition}
\newtheorem{remark}{Remark}

\newcommand{\R}{\mathbb R}
\newcommand{\Q}{\mathbb Q}
\newcommand{\Z}{\mathbb Z}
\newcommand{\CLS}{\mathsf{CLS}}
\newcommand{\one}{\mathbf 1}
\newcommand{\bA}{\mathbf A}
\newcommand{\bQ}{\mathbf Q}
\newcommand{\e}{\mathbf e}
\newcommand{\bx}{\mathbf x}
\newcommand{\by}{\mathbf y}
\newcommand{\bz}{\mathbf z}
\newcommand{\bw}{\mathbf w}
\newcommand{\bv}{\mathbf v}
\newcommand{\ba}{\mathbf a}
\newcommand{\bc}{\mathbf c}
\newcommand{\bp}{\mathbf p}

\newcommand{\bV}{\mathbf V}
\newcommand{\bW}{\mathbf W}
\newcommand{\bbeta}{\bm \beta}
\newcommand{\bg}{\boldsymbol\gamma}
\newcommand{\bu}{\mathbf u}
\newcommand{\bt}{\mathbf t}
\newcommand{\bs}{\mathbf s}
\newcommand{\bd}{\mathbf d}
\newcommand{\bC}{\mathbf C}
\newcommand{\bP}{\mathbf P}
\newcommand{\bgg}{\mathbf g}
\newcommand{\bG}{\mathbf G}
\newcommand{\bcalG}{\boldsymbol{\mathcal G}}
\newcommand{\calG}{\mathcal{G}}
\newcommand{\calV}{\mathcal{V}}
\newcommand{\calE}{\mathcal{E}}
\newcommand{\diag}{\operatorname{diag}}
\newcommand{\supp}{\operatorname{supp}}
\newcommand{\poly}{\operatorname{poly}}
\newcommand{\htn}{\operatorname{ht}}
\newcommand{\pos}[1]{\left(#1\right)_{+}}
\newcommand{\gap}{\mathfrak g}

\DeclareMathOperator*{\argmin}{arg\,min}
\DeclareMathOperator*{\argmax}{arg\,max}

\usepackage{xcolor} 
\usepackage{hyperref}

\hypersetup{
    colorlinks=true,       
    linkcolor=blue,        
    citecolor=teal,       
    filecolor=magenta,     
    urlcolor=cyan          
}
\usepackage{authblk}

\title{The Complexity of Computing a Competitive Equilibrium \\
in Fisher Markets with Chores}
\author[$\dagger$]{Christian Kroer}
\author[$\ddagger$]{Ruta Mehta}
\author[$\dagger$]{Tianlong Nan}

\affil[$\dagger$]{Columbia University}
\affil[$\ddagger$]{University of Illinois Urbana-Champaign}

\date{}

\begin{document}

\maketitle

\begin{abstract}
    We study the complexity of computing competitive equilibria in Fisher markets with divisible chores and positive rational disutility coefficients and earning requirements.
    For linear disutilities, 
    we prove that exact equilibrium computation is $\CLS$-complete.
    In particular, the problem remains $\CLS$-hard even when the earning requirements are binary in $1$ and $2^k$, where $k$ is an integer.
    We prove hardness by encoding the first-order conditions of a box-constrained quadratic program in equilibrium prices and earning flows, using a compact minimum-cost flow construction.

    For convex constant-elasticity-of-substitution (CES) disutilities, 
    exact equilibrium prices may be irrational.
    We formulate the problem of finite-precision equilibrium price computation, in which
    we seek rational prices with a certified market clearing error at most a given rational tolerance encoded in binary.
    We prove that this problem is $\CLS$-complete for every fixed rational exponent $\rho>1$.
    To prove this hardness, 
    we encode a quartic polynomial through the fourth-order expansion of the CES price potential function corresponding to the Eisenberg-Gale (EG) dual program, and show that every sufficiently accurate approximate equilibrium of the constructed market lies near uniform prices.

    Additionally, we also identify a tractable special case of chores Fisher markets.
    In particular, we provide a strongly polynomial algorithm for exact equilibrium computation in linear markets whose disutility matrix has rank at most two and earning requirements are arbitrary positive rationals.
    For $n$ agents and $m$ chores, 
    the algorithm uses $O(nm(n+m))$ rational arithmetic operations and comparisons, 
    with polynomial intermediate encoding length.
\end{abstract}

\setcounter{tocdepth}{2}
\tableofcontents

\section{Introduction}
\label{sec:introduction}

Competitive equilibrium (CE)~\cite{walras1874elements} is a fundamental solution concept in game theory and economics.
At CE, agents make individually optimal choices at common prices, and the resulting allocation clears the market.
This connection between individual optimality and market clearing is central to the use of competitive equilibrium in predicting market outcomes and designing allocation rules.
A core question in algorithmic game theory is whether agents or a market designer can find the equilibrium prices and allocations efficiently.
Therefore, understanding the computational complexity of finding a competitive equilibrium is central to algorithmic game theory.

We study this question for Fisher markets with divisible chores, where agents are paid to perform undesirable chores.
In chores Fisher markets, 
each agent has a fixed earning requirement, and each chore has a fixed supply.
At given prices, an agent chooses a bundle of chores that minimizes her disutility subject to satisfying her earning requirement.
A competitive equilibrium consists of prices and an allocation such that each agent receives an optimal bundle and every chore is fully allocated.
We consider linear disutilities and non-linear convex constant-elasticity-of-substitution (CES) disutilities, which are most common in economics.

For linear Fisher markets with goods, \cite{devanur2008market} provides a polynomial-time combinatorial algorithm, 
and \cite{orlin2010improved} later obtains a strongly polynomial algorithm.
In contrast, the computation of chores CE is considered to be hard due to its nonconvex nature, and the fact that the equilibrium set can be disconnected~\cite{bogomolnaia2019dividing}.
For linear chores Fisher markets, $\CLS$-membership is known in the literature~\cite{chaudhury2022competitive,fearnley2022complexity}.
However, the hardness of computing a CE in chores Fisher market has remained open, even in the linear chores Fisher market.
We close this gap by proving $\CLS$-completeness for positive rational linear chores Fisher markets, and $\CLS$-completeness of finite-precision equilibrium price computation for each fixed rational exponent $\rho>1$ in convex CES markets.
On the other hand, we identify a special case of tractable chores Fisher markets---there is a strongly polynomial algorithm to computing an exact CE in linear chores Fisher markets whose disutility matrix has rank at most two.

\subsection{Our results}
\label{sec:results}

\paragraph{Linear markets.}
We first classify exact equilibrium computation in linear chores Fisher markets.
\begin{theorem}
    \label{thm:main}
    Computing an exact competitive equilibrium in a linear chores Fisher market with positive rational disutility coefficients and earning requirements is $\CLS$-complete.
\end{theorem}

Our reduction starts from a rational quadratic program on a unit box.
From every equilibrium price vector of the constructed market, we can recover in polynomial time a point satisfying the first-order conditions of that quadratic program.
Consequently, a polynomial-time algorithm for exact linear chores equilibrium computation would give polynomial-time algorithms for every problem in $\CLS$.

\paragraph{Convex CES markets.}
We next consider convex CES markets with $m$ chores, where agent $i$'s disutility for a bundle $\bx_i=(x_{ij})_j$ is $D_i(\bx_i):=(\sum_{j=1}^m a_{ij}x_{ij}^{\rho})^{1/\rho}$.
Here, $a_{ij}>0$ and agents share an exponent $\rho>1$.
Since exact equilibrium prices can be irrational even for rational market data, we formulate finite-precision equilibrium price computation as a search problem in which we seek rational prices with a certified market clearing error of at most a given tolerance $\epsilon$.

For every fixed rational exponent $\rho>1$, we show that finite-precision equilibrium price computation is $\CLS$-complete.
Because the tolerance is also encoded in binary, the result concerns computation at binary-specified accuracy.
We prove these statements in~\cref{sec:ces}, using a reduction from quadratic programming to the EG dual objective for each fixed exponent.
We provide the irrational-price example and a representation for exact CES equilibrium computation in~\cref{app:ces-exact}.

\paragraph{Rank-two linear markets.}
For positive rational linear disutility matrices of ordinary rank at most two, 
we propose an algorithm that computes an exact equilibrium using \mbox{$O(nm(n+m))$} rational arithmetic operations and comparisons, where $n$ is the number of agents and $m$ is the number of chores.
The rational numbers computed by the algorithm have encoding length polynomial in the total binary input length.
The key idea of the algorithm is finding a common order of agents and chores, then updates prices and agent indices until the indices stabilize.
We give the algorithm and its analysis in~\cref{sec:tractable}.

\subsection{Our techniques}
\label{sec:techniques}

For both hardness results, we start from the problem of finding a KKT point of a rational quadratic polynomial on the unit box, 
which is proved to be $\CLS$-complete by~\cite{fearnley2025complexity}. 
Our goal is to encode the equilibrium conditions in a chores market such that we can recover an exact KKT point of the original quadratic program from every valid output of the corresponding market problem.
The linear and CES reductions use different representations of the same source problem.

\paragraph{Linear markets: difference constraints and earning flows.}
We use the compact counting network of~\cite{disser2018simplex} to approximate quadratic terms by minimum-cost flow value functions.
Through the dual flow formulation, we transfer box-constrained quadratic stationarity to a quadratic program whose constraints bound differences between variables.
After a small perturbation and a translation, the quadratic objective becomes a negative weighted sum of squares with strictly positive weights.
This form is well suited to a market interpretation: the weights become earning requirements, and we encode the difference constraints in disutility ratios.

For a given source problem, we construct an $n$-agent-$n$-chore Fisher market instance. 
The main market argument is to prove that, at every equilibrium, each agent performs a positive amount of the chore paired with that agent.
We first show that prices are close to earning requirements and that positive off-diagonal earning flows are acyclic.
Then, we utilize a Directed Acyclic Graph (DAG), where each node is a pair of agent and chore with the same index, i.e., an agent and the chore pre-assigned to her by the market construction. 
By decomposing the earning flows in this DAG into paths, 
we bound earnings from other chores and obtain the required positive earnings on each paired chore.
The conditions of competitive equilibrium then yield stationarity of the QPs that are reduced to the chores marker problem.

To recover an exact KKT point, we use uniform bounds for rational LPs.
In particular, from each equilibrium in the chores Fisher market, we make the stationarity and complementary-slackness errors are sufficiently small, then identify appropriate constraints and solve a rational LP.
By the classic theorem on the rational LP, the solution of such an LP has to correspond an exact KKT point.
The required precision and the market coefficients have polynomial binary encoding length.
Thus, even we introduce some approximations in the reduction, from an approximate KKT points of the QPs we can still extract an exact KKT point from every CE in the linear chores Fisher market.
We develop this reduction in~\cref{sec:linear-completeness}.

\paragraph{CES markets: encoding a quartic model near uniform prices.}
For each fixed exponent, 
we first encode the box-constrained QP into a quartic polynomial whose sufficiently accurate stationary points yield exact KKT points.
From the unconstrained optimization problem that minimizes that quartic polynomial, by a change of variables, we obtain another quartic program with an optimizaiton domain $H := \{ \by: \one^\top \by = 0 \}$.
Such an constrained optimization problem then can be connected to the Eisenberg-Gale (EG) dual program introduced by~\cite{chaudhury2024competitive}, which was extended to the CES disutilities by~\cite{tao2026fisher,chaudhury2026tatonnement}.

We achieve this connection by leveraging Taylor expansion.
In particular, we construct a CES chores Fisher market whose price potential function has the required fourth-order Taylor expansion near uniform prices.
Since the aggregate demand in the market is the gradient of this potential function, 
small market clearing error corresponds to approximate stationarity along directions that preserve the sum of prices.
A key step is to prove that every sufficiently accurate approximate CE in the constructed market lies in the neighborhood where the expansion is valid.
We then consider rounding the disutility coefficients to rationals, and show that the correspondence between the CE and the KKT points of source problems survives the effect of rounding the market coefficients to rationals.
These steps make it possible to decode every certified approximate equilibrium price, as explained in~\cref{sec:ces-proof-overview}.

\paragraph{Rank two: a common order and monotone price updates.}
After normalizing each row of a positive disutility matrix of rank at most two
by its first entry, the rows coincide or lie on an affine line.
We use this geometry to find orders of agents and chores in which the disutility ratio of every pair of consecutive chores is \emph{monotone} across agents.

We can then search over agents and update the prices at every step.
For each pair of consecutive chores, we keep an agent index, initially set to the first agent.
At each iteration, we set the price ratio of each pair such that the selected agent is indifferent between the two chores, and then normalize the prices to make their sum equal the total earning requirement.
We update each index to the first agent whose cumulative earning requirement reaches the cumulative price of the chores up to the current pair of chores.
Because of the monotone structure, the search over agents only moves in one direction and must stop after polynomially many iterations.
When the indices stop changing, the common ratio order ensures that the recovered allocation is individually optimal.

\subsection{Related work}
\label{sec:related-work}

\paragraph{Competitive equilibrium with chores.}
Competitive equilibrium (and the corresponding disutility profiles) under equal incomes admit a characterization as critical points of the Nash product on the Pareto frontier, 
and equilibrium sets are known can be nonconvex and disconnected even with additive disutilities~\cite{bogomolnaia2017competitive,bogomolnaia2019dividing}.
In the general earning requirements case, all competitive disutility profiles can be enumerated in strongly polynomial time therefore yield polynomial algorithms for computing CE when either the number of agents or the number of chores is fixed~\cite{branzei2024algorithms}.
Our rank-two algorithm complements these results by restricting the disutility matrix while allowing both dimensions to grow.

The most of algorithms for computing a CE in chores Fisher market focus on developing algorithms for computing an approximate CE, including exterior-point method~\cite{boodaghians2022polynomial}, combinatorial algorithm~\cite{chaudhury2022competitive}, 
and greedy Frank-Wolfe algorithm~\cite{chaudhury2024competitive}.
These guarantees leave open the complexity of exact equilibrium computation.
For the broader exchange model, prior work establishes conditions guaranteeing equilibrium existence~\cite{chaudhury2022existence}
and $\mathsf{PPAD}$-hardness of computing an equilibrium to inverse-polynomial accuracy~\cite{chaudhury2022competitive}.
For exact equilibria in linear chores Fisher market, a complementary-pivot formulation gives membership in $\mathsf{PPAD}$~\cite{chaudhury2023complementary}, while a combinatorial local-search argument gives membership in $\mathsf{PLS}$~\cite{chaudhury2022competitive}.

\cite{chaudhury2026competitive} consider a labor market model which couples both goods and chores Fisher market, 
and show that there are polynomial-time algorithms that can compute a CE in that market model.

\paragraph{Continuous local search and equilibrium complexity.}
The class $\CLS$ was introduced to capture continuous local search problems combining fixed-point and potential-based existence arguments~\cite{daskalakis2011continuous}.
\cite{fearnley2022complexity} establishes $\CLS = \mathsf{PPAD}\cap\mathsf{PLS}$.
They prove that computing an approximate KKT point of a continuously differentiable function with a Lipschitz-continuous gradient is $\CLS$-complete.
\cite{fearnley2025complexity} shows that $\CLS$-completeness persists even for computing a KKT point of a quadratic objective over box constraints.

This complexity class also arises in equilibrium computation.
\cite{babichenko2021settling} proves that computing a possibly mixed Nash equilibrium in a congestion game to prescribed precision is $\mathsf{PPAD}\cap\mathsf{PLS}$-complete.
\cite{garg2024satiation} establishes $\CLS$-membership for exact equilibrium computation in Fisher markets with buyer utility limits and seller earning limits under a money-clearing condition.
Our results extend this line of work by providing completeness classifications both for exact linear chores equilibria and for finite-precision equilibrium price computation in CES chores Fisher markets.

\paragraph{Price potential functions and dynamics.}
\cite{eisenberg1959consensus} establishes the classical convex programming formulation for linear Fisher goods markets.
The most classic price dynamics, t\^atonnement, can be interpreted as gradient descent on a convex price potential function~\cite{cheung2013tatonnement}.
For linear chores markets, \cite{chaudhury2024competitive} develops a novel nonconvex dual Eisenberg-Gale (EG) formulation, which allows standard optimization algorithms to compute a CE in chores Fisher market.
Based on this new formulation, \cite{chen2024computing} establishes linear convergence to an exact equilibrium using a difference-of-convex formulation, but with a problem-dependent constant that might be arbitrarily large.
\cite{tao2026fisher,chaudhury2026tatonnement} extends the duality framework to continuous, convex, $1$-homogeneous (CCH) disutilities.
Convex CES disutilities are the most classic CCH disutility functions. 
For this class, \cite{chaudhury2026tatonnement} analyzes relative t\^atonnement, while \cite{chaudhury2026fully} develops fully distributed multiplicative t\^atonnement, both with convergence guarantees for computing an approximate CE.

\paragraph{Concurrent work on linear chores markets.}
The concurrent work~\cite{liu2026competitive} independently proves that computing an exact competitive equilibrium in linear chores Fisher markets is $\CLS$-complete.
Starts from computing an approximate KKT point of a quadratic program over the unit box, their reduction exploits the nonconvex dual Eisenberg-Gale formulation, using Taylor expansions of a logarithmic price potential and chains of maximum-minimum gadgets to encode the quadratic objective, which is different from our approach.

\paragraph{Organization.}
We present the model and rational linear-equilibrium certificates in~\cref{sec:model}.
We prove the linear completeness theorem in~\cref{sec:linear-completeness} and the CES completeness theorem in~\cref{sec:ces}.
We give the rank-two algorithm and its analysis in~\cref{sec:tractable}.
The technical CES hardness proofs appear in~\cref{app:ces-hardness}, and exact CES equilibrium computation is treated separately in~\cref{app:ces-exact}.

\section{Preliminaries}
\label{sec:model}

We consider Fisher markets with chores involving $n$ agents and $m$ divisible chores, each with a unit supply.
Each agent $i$ has earning requirement $B_i > 0$ and a disutility function $D_i:\R_{\ge0}^m \to \R_{\geq 0}$.
An allocation $\bx := (x_{ij})_{i \in [n], j \in [m]} \in \mathbb{R}^{n \times m}$ assigns $x_{ij} \geq 0$ units of chore $j$ to agent $i$, whose bundle is $\bx_i:=(x_{i1},\ldots,x_{im})$.
We define the earning ratio by $B:=\frac{\max_i B_i}{\min_i B_i}$.

\paragraph{CCH disutilities.}
A disutility function is \emph{continuous, convex, and homogeneous of degree one} (CCH) if it is continuous and convex on $\R_{\ge0}^m$ and satisfies $D_i(t\bx_i) = t D_i(\bx_i)$, $\forall\,t \geq 0,\ \bx_i\in\R_{\ge0}^m$.
For the CCH markets considered here, we additionally assume that every $D_i$ is strictly increasing in each coordinate.
In particular, without loss of generality we set $D_i(\mathbf0)=0$ and $D_i(\bx_i)>0$ for all nonzero bundles $\bx_i\ge0$.
The linear and convex CES disutilities are special cases of this class.

\paragraph{Linear disutilities.}
In a linear chores Fisher market,
$D_i(\bx_i) := \sum_{j=1}^{m}d_{ij}x_{ij}$, 
where $d_{ij}>0$ for all $i,j$.
For this class, 
define the global disutility ratio by $D:=\frac{\max_{i,j}d_{ij}}{\min_{i,j}d_{ij}}$.

\paragraph{Convex CES disutilities.}
In a convex CES chores Fisher market, agents share an exponent $\rho>1$ and have disutilities
$D_i^\rho(\bx_i):=\left(\sum_{j=1}^m a_{ij}x_{ij}^{\rho}\right)^{1/\rho}$ where $a_{ij}>0$ for all $i,j$.
For CES markets, we use the ratio of the smallest to the largest coefficient, $D:=a_{\min}/a_{\max}$.

\subsection{Competitive equilibrium}\label{sec:equilibrium-definition}

A competitive equilibrium (CE) consists of prices $\bp\in\R_{>0}^{m}$ and allocations $\bx \geq 0$ such that each chore is fully allocated 
\begin{equation*}
    \label{eq:ce-definition-1}
    \sum\nolimits_{i=1}^n x_{ij}=1 \ \forall\,j\in[m], 
\end{equation*}
and each agent receives an optimal bundle satisfying her earning requirement at prices $\bp$: 
\begin{equation*}
    \label{eq:ce-definition-2}
    \bx_i\in\argmin_{\by\ge0} \left\{D_i(\by):\ \sum\nolimits_{j=1}^m p_j y_j\ge B_i\right\} \ \forall\,i\in[n].
\end{equation*}
For CCH disutilities satisfying the preceding assumptions, every optimal bundle earns exactly $B_i$: any bundle earning more can be scaled down to earn $B_i$, strictly decreasing its disutility by homogeneity and positivity.
Consequently, market clearing implies
\begin{equation*}
    \label{eq:price-total}
    \sum\nolimits_{j=1}^mp_j = \sum\nolimits_{i=1}^nB_i.
\end{equation*}
Also, under the same assumptions, every chore has a positive equilibrium price: 
since disutility is strictly increasing in every coordinate, 
deleting a positive amount of a zero-price chore from a bundle would reduce disutility without changing earnings.
If we scale earning requirements and prices by the same factor, then equilibrium allocations preserve.


\subsection{Rationality of CE in chores Fisher markets}
\label{sec:certificates}

We now specialize to linear disutilities.
Let $b_{ij} := p_jx_{ij}$ denote the money that agent $i$ earns by performing $x_{ij}$ units of chore $j$ at price $p_j$ per unit, and 
let
\begin{equation*}
    \beta_i:=\max_{j \in [m]} \frac{p_j}{d_{ij}}.
\end{equation*}
For positive linear disutilities, individual optimality is equivalent to earning only from chores that maximize payment-per-unit-disutility, and earning exactly $B_i$.
Thus, in the linear case, the equilibrium definition in~\cref{sec:equilibrium-definition} is equivalent to
\begin{equation}\label{eq:ce-spending}
    \begin{aligned}
    &b_{ij}\ge0,\quad p_j>0,\quad \beta_i>0,\\
    &\sum_{j=1}^m b_{ij}=B_i,\quad \sum_{i=1}^n b_{ij}=p_j,\\
    &p_j\le d_{ij}\beta_i,\quad
    b_{ij}(d_{ij}\beta_i - p_j) = 0.
    \end{aligned}
    \tag{Equilibrium Conditions}
\end{equation}
It is known that positive finite linear chores Fisher markets admit an equilibrium by~\cite[Theorem 8]{chaudhury2022existence}.

\begin{lemma}
    \label{lem:rational-ce}
    An exact equilibrium exists for all positive rational linear Fisher chores markets.
    The price vector of any equilibrium has polynomial rational encoding length and admits a rational equilibrium allocation of polynomial encoding length.
    Given exact rational equilibrium prices, such an allocation can be found by a linear program.
\end{lemma}
\begin{proof}
We fix an equilibrium and consider its bipartite agent-chore graph, 
with an edge $ij$ whenever $b_{ij}>0$.
For each connected component, \
we use the equations $p_j = d_{ij}\beta_i$ along a spanning tree to express every price and $\bbeta$ coordinate as a multiple of one chosen price.
Each multiplier is a product and quotient of at most $n+m-1$ rational input coefficients and therefore has polynomial encoding length.
To determine the chosen price, we sum the row and column balance equations over the component and obtain
\begin{equation*}
\sum_{j\text{ in the component}}p_j
=
\sum_{i\text{ in the component}}B_i.
\end{equation*}
We divide the component's total earning requirement by the sum of its relative prices to obtain the chosen price.
Since both sums have polynomial rational encoding length, the chosen price does as well.
Thus, every equilibrium price vector and its associated $\bbeta$ vector have polynomial rational encoding length.

With these prices fixed, 
we consider the linear system with $\sum_{j=1}^m b_{ij} = B_i$, $\sum_{i=1}^n b_{ij} = p_j$, and earning flows allowed only on the edges of the support graph.
The original equilibrium earnings satisfy this system.
We can then choose a basic feasible solution, whose coordinates have polynomial rational encoding length.
Since equilibrium allocation can be computed by dividing each earning amount by the price of the chore, each equilibrium allocation is of polynomial encoding length.
Once we have equilibrium prices, we can compute each agent's optimal chores and solve the LP corresponding to the linear system described above, which can be done in polynomial time.
\end{proof}

\section{CLS-completeness for Linear Chores Fisher Markets}
\label{sec:linear-completeness}

\subsection{Proof overview}

\begin{figure}[htbp]
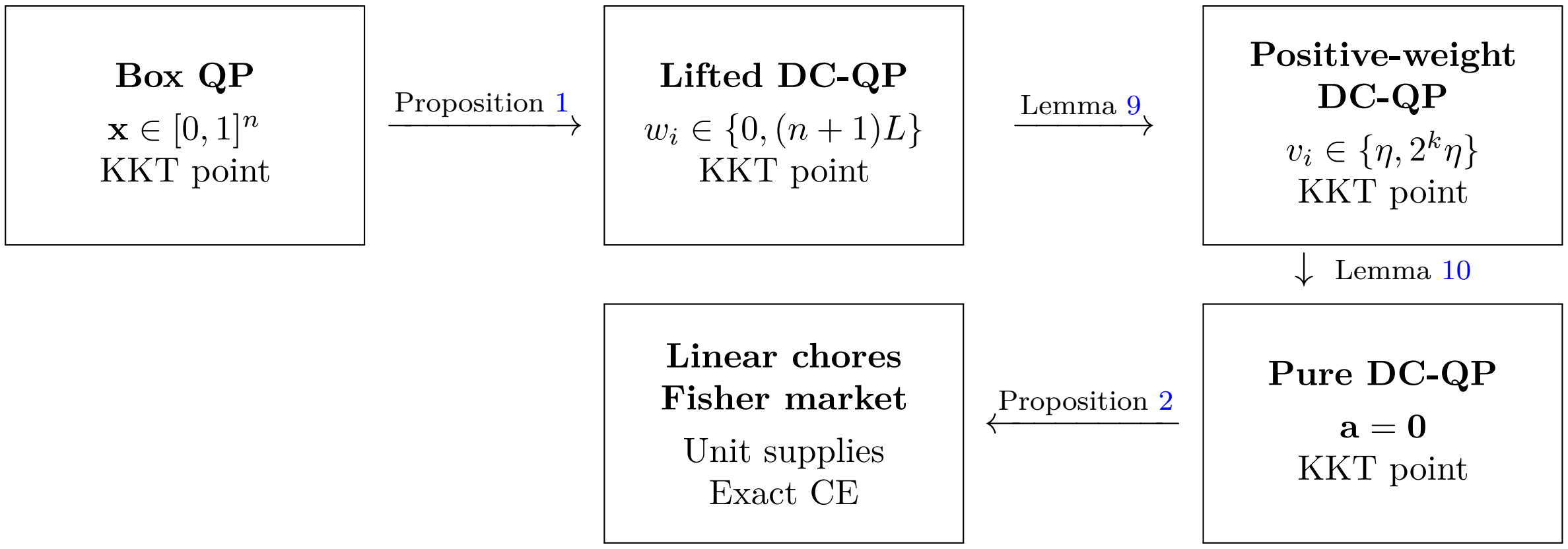

\centering
\begingroup
\setlength{\fboxsep}{4pt}
\newcommand{\reductionnode}[2]{%
  \fbox{\begin{minipage}[c][21mm][c]{0.20\linewidth}
    \centering\small
    \textbf{#1}\par\smallskip
    #2
  \end{minipage}}%
}
\begin{tabular}{@{}c@{\hspace{2mm}}c@{\hspace{2mm}}c@{\hspace{2mm}}c@{\hspace{2mm}}c@{}}
\reductionnode{Box QP}{$\bx\in[0,1]^n$\\KKT point}
& $\xrightarrow{\text{\scriptsize\cref{prop:lift}}}$
& \reductionnode{Lifted DC-QP}{$w_i\in\{0,(n+1)L\}$\\KKT point}
& $\xrightarrow{\text{\scriptsize\cref{lem:regularization}}}$
& \reductionnode{Positive-weight\\DC-QP}{$v_i\in\{\eta,2^k\eta\}$\\KKT point}
\\[1mm]
& & & & $\downarrow\;\text{\scriptsize\cref{lem:translation}}$
\\[1mm]
& & \reductionnode{Linear chores \\ Fisher market}{Unit supplies\\Exact CE}
& $\xleftarrow{\text{\scriptsize\cref{prop:market}}}$
& \reductionnode{Pure DC-QP}{$\ba=\mathbf0$\\KKT point}
\end{tabular}
\endgroup
\caption{Polynomial-time instance reductions from~\eqref{pg:box-qp} to the linear Fisher chores market.
From every exact equilibrium of the linear Fisher chores market, we recover a KKT point of the original Box QP
by decoding through all preceding stages in the reverse direction of the arrows.}
\label{fig:reduction-chain}
\end{figure}

To reduce from Box-constrained QP to the linear chores Fisher market, we follow the sequence of polynomial-time instance reductions illustrated in \cref{fig:reduction-chain}.
We first use a directed acyclic graph to construct a function that 
approximates $t^2/2$, which leads to a lifted QP with a feasible set involving the differences of its variables, 
which we call difference-of-variables constraints QP or difference constraints QP (DC-QP).
We can further reduce a DC-QP such that its objective function is a quadratic function with strictly negative diagonal entries without linear terms, which we call Pure DC-QP.
For each Pure DC-QP, we construct an equivalent $n$-agent-$n$-chore linear chores Fisher market instance, 
and show that every exact equilibrium of this market can be decoded to recover a KKT point of the Pure DC-QP, 
by using a directed acyclic graph with $n$ vertices, each represent a pair of agent and chore with the same index.

\subsection{Box-constrained QP}
\label{sec:gaps}

Since the reductions will cause small errors in the stationarity condition and complementary slackness, 
we first show that we can retrieve an exact KKT points from an approximate KKT points with a sufficiently small error, by solving a rational linear program in polynomial time.

We first introduce some notations. 
For a finite collection of rationals $\mathcal S$, define its coefficient height $\htn(\mathcal S)$ to be any integer $b\ge1$ such that every element in $\mathcal S$ can be written as $p/q$, where $p\in\Z$, $q\in\Z_{>0}$, and $|p|,q \leq 2^b$.
Such a height is computable directly from the binary encodings.
For integers $d,b \geq 1$, set
\begin{equation}
    \label{eq:gap}
    \gap(d,b) := 2^{-d(d+1)b-d\lceil\log_2(\max\{2,d\})\rceil-2}.
\end{equation}

\begin{lemma}
    \label{lem:gap}
    Let $\bA \in \mathbb{Q}^{m \times d}$ and $\bbeta \in \mathbb{Q}^m$ have all entries of
    height at most $b$, and suppose
    \[
        \rho^\star := \min\left\{ z_1 : \bA \bz \leq \bbeta,\ \bz \in \mathbb{R}^d_{\geq 0} \right\}
    \]
    is feasible. 
    Then, we have an optimal solution, and either $\rho^\star = 0$ or
    $\rho^\star \ge \mathfrak{g}(d,b)$. 
    Moreover, an optimal rational solution has encoding length polynomial in $d$ and $b$, 
    and can be computed in time polynomial in the encoding length of $(\bA, \bbeta)$.
\end{lemma}
\begin{proof}
    Let $P:=\{\bz\in\R^d_{\ge0}:\bA\bz\le\bbeta\}$.
    Since $P$ is nonempty, contains no line, and $z_1\ge0$ on $P$, an optimal vertex $\bz^\star$ exists.
    By clearing denominators in each row, 
    we obtain integer coefficients and right-hand sides bounded in absolute value by $H := 2^{(d+1)b}$.
    By the classical Cramer--Hadamard bounds underlying~\cite[Theorem~10.2 and Corollary~10.2a]{schrijver1998theory}, each coordinate of $\bz^\star$ is a ratio of integers whose absolute values are at most $\Delta:=d^{d/2}H^d$.
    Thus, if $\rho^\star>0$, we have
    \begin{equation*}
        \rho^\star=z_1^\star\ge\Delta^{-1}
        =d^{-d/2}2^{-d(d+1)b}\ge\gap(d,b),
    \end{equation*}
    where the last inequality follows from~\cref{eq:gap}.
    By these bounds, $\bz^\star$ also has encoding length polynomial in $d$ and $b$.
    An optimal rational solution can be computed in polynomial time using standard algorithms for linear programming with rational input~\cite[Chapter~13]{schrijver1998theory}.
\end{proof}

Let $\bQ \in \Q^{n \times n}$ be symmetric and $\bc \in \Q^n$.
Consider the following box-constrained QP:  
\begin{equation}
    \label{pg:box-qp}
    \begin{aligned}
        \min_{\bx} \quad & f(\bx) := \frac12\bx^\top \bQ\bx+\bc^\top\bx, \\ 
        \text{s.t.} \quad & \bx\in[0,1]^n.
    \end{aligned}
    \tag{Box-QP}
\end{equation}
The gradient is $\nabla f(\bx)=\bQ\bx+\bc$.
An exact KKT point of~\eqref{pg:box-qp} satisfies
\begin{equation}
    \label{eq:box-kkt}
    \nabla_i f(\bx) 
    \begin{cases}
        \geq 0 & \text{if }x_i=0, \\ 
        \leq 0 & \text{if }x_i=1, \\ 
        = 0 & \text{if }0<x_i<1.
    \end{cases}
\end{equation}

We call a point $\bx\in[0,1]^n$ a \emph{strong $\varepsilon$-KKT point} of~\eqref{pg:box-qp} if it satisfies:
\begin{equation*}
    \label{eq:box-approx-kkt}
    \nabla_i f(\bx) 
    \begin{cases}
        \geq - \varepsilon & \text{if }x_i=0, \\ 
        \leq \varepsilon & \text{if }x_i=1, \\ 
        \in [-\varepsilon, \varepsilon] & \text{if }0<x_i<1.
    \end{cases}
\end{equation*}

In the following lemma, we prove that, when $\varepsilon$ is sufficiently small, we can compute an exact KKT point of~\eqref{pg:box-qp} from a strong $\varepsilon$-KKT point in polynomial time.
We apply~\cref{lem:gap} directly in the proof.
\begin{lemma}
    \label{lem:box-round}
    If we have a rational strong $\varepsilon$-KKT point of \eqref{pg:box-qp} with $\varepsilon < \gap(n+1, \htn(\{0,1\}\cup\{Q_{ij}\}_{i,j}\cup\{c_i\}_i))$, 
    then we can compute an exact KKT point of \eqref{pg:box-qp} by solving a polynomial-time LP.
\end{lemma}
\begin{proof}
    Let $\hat{\bx} \in [0,1]^n$ denote the strong $\varepsilon$-KKT point.
    Set $b_{\rm box}:=\htn(\{0,1\}\cup\{Q_{ij}\}_{i,j}\cup\{c_i\}_i)$.
    Consider the following linear program
    \begin{equation}
        \label{eq:box-repair}
        \begin{aligned}
        \min_{\bx \in [0, 1]^n, \ \rho \geq 0} \quad &\rho\\
        \text{s.t.} \quad 
            & x_i=0 & \ \forall\, i : \hat{x}_i = 0 \\ 
            & x_i=1 & \ \forall\, i : \hat{x}_i = 1 \\
            & (\bQ \bx + \bc)_i \geq -\rho & \ \forall\, i : \hat{x}_i = 0 \\
            & (\bQ \bx + \bc)_i \leq \rho & \ \forall\, i : \hat{x}_i = 1 \\
            & -\rho \leq (\bQ \bx + \bc)_i \leq \rho & \ \forall\, i : 0 < \hat{x}_i < 1.
        \end{aligned}
    \end{equation}
    By the definition of the strong $\varepsilon$-KKT point, 
    $(\hat{\bx}, \varepsilon)$ is a feasible solution to~\cref{eq:box-repair}. 
    Hence, the optimal value $\rho^\star$ of~\cref{eq:box-repair} satisfies $\rho^\star \leq \varepsilon$.
    
    Moreover, we see that~\cref{eq:box-repair} satisfies the assumptions of~\cref{lem:gap}: after rearranging terms, coefficients and right-hand side lie in $\{0, \pm 1\} \cup \{\pm Q_{ij}\} \cup \{\pm c_i\}$, hence all of them have height at most $b_{\rm box}$. 
    By \cref{lem:gap}, we have that the optimal value of~\cref{eq:box-repair} satisfies either $\rho^\star = 0$ or $\rho^\star \ge \mathfrak{g}(n+1, b_{\rm box})$. 
    Since $\rho^\star \leq \varepsilon < \mathfrak{g}(n+1, b_{\rm box})$, $\rho^\star = 0$. 
    Furthermore, \cref{eq:box-repair} has an optimal
    solution $(\bx^\star, 0)$ of polynomial encoding length, 
    which can be computed in polynomial time.
    
    We now verify that $\bx^\star$ is a KKT point. Since $\rho^\star = 0$, the constraints of~\cref{eq:box-repair} impose the signs required by~\cref{eq:box-kkt} at every coordinate $i$ with $\hat{x}_i = 0$ or $\hat{x}_i = 1$.
    At every other coordinate, the last constraint imposes a zero derivative. A zero derivative satisfies the KKT condition whether $x^\star_i$ remains in the interior or moves to the boundary of $[0,1]$.
    Therefore, $\bx^\star$ is a KKT point, which completes the
    proof.
\end{proof}

\subsection{Difference-constraint QP}

Let $n \geq 2$, let $\bw \in \Q_{\geq 0}^n$ with $\one^\top \bw > 0$, $\ba \in \Q^n$ with $\one^\top\ba = 0$, 
and let $C_{ij} \in \Q$ for all ordered pairs $i \neq j$.
Consider the following optimization problem, which minimizes a quadratic subject to a feasible set mainly constraining the difference of two variables:  
\begin{equation}
    \begin{aligned}
        \min_{\bz \in \R^n} \quad & -\frac12\bz^\top \bW \bz+\ba^\top\bz \\ 
        \text{s.t.} \quad & P(\bw, \bC) := 
        \left\{ \bz \in \R^n: z_j - z_i \leq C_{ij}, \ \forall\, i \neq j, \,
        \bw^\top \bz = 0 \right\}.
    \end{aligned}
    \label{pg:difference-constraint-QP}
    \tag{DC-QP}
\end{equation}
where $\bW:=\diag(\bw)$.
We bound the magnitudes of the difference-constraint right-hand sides by
\begin{equation*}\label{eq:C-bar}
\bar{C} := \max\{1,\max_{i\ne j}|C_{ij}|\}.
\end{equation*}
Because $z_j - z_i \leq C_{ij}$ for all $i \neq j$, $\bw^\top \bz = 0$ and $\one^\top \bw > 0$, all $z_i$ has to be bounded within $[-\bar{C}, \bar{C}]$, otherwise all $z_i$ have to be all positive or all negative which contradicts $\bw^\top \bz = 0, \one^\top \bw > 0$.
Hence, the feasible set $P(\bw, \bC)$ is compact.

Let $\bA$ denote the incidence matrix with column $\e_j - \e_i$ for ordered pair $(i,j)$.
The KKT conditions of~\eqref{pg:difference-constraint-QP} can be written as 
\begin{equation}
    \label{eq:source-kkt}
    \begin{aligned}
    & \bA \bg = \bW \bz - \ba, \\ 
    & \bw^\top \bz = 0,\\
    & z_j - z_i \leq C_{ij}, \ \gamma_{ij} \geq 0, \ \gamma_{ij}(z_j-z_i-C_{ij}) = 0 \quad \forall\, i \neq j.
    \end{aligned}
    \tag{KKT Conditions}
\end{equation}

For rational input $\bw,\ba,C_{ij}$, fix a KKT multiplier support $\supp(\bg):=\{(i,j):i\ne j,\gamma_{ij}>0\}$ in~\eqref{eq:source-kkt}.
Imposing tightness for all $(i,j)\in\supp(\bg)$ and setting all other multipliers to zero reduces~\eqref{eq:source-kkt} to a rational linear feasibility system, with $\supp(\bg)$ held fixed.
Hence, a basic feasible solution of~\eqref{eq:source-kkt} with these support constraints provides a rational KKT certificate of polynomial encoding length.
Thus, the problem of finding a KKT point of~\eqref{pg:difference-constraint-QP} admits rational certificates of polynomial encoding length.

For $\varepsilon\ge0$, we call a primal point $\bz\in\R^n$ an \emph{$\varepsilon$-approximate KKT point} of~\eqref{pg:difference-constraint-QP} if there exists a nonnegative multiplier vector $\bg$ such that
\begin{equation}
    \label{eq:source-approx-kkt}
    \begin{aligned}
    &\bw^\top\bz=0,\quad \bg\ge0,\\
    &\|\bA\bg-\bW\bz+\ba\|_\infty\le\varepsilon,\\
    &z_j-z_i-C_{ij}\le\varepsilon &&\forall\,i\ne j,\\
    &|z_j-z_i-C_{ij}|\le\varepsilon &&\forall\,(i,j)\in\supp(\bg).
    \end{aligned}
    \tag{Approx-KKT Conditions}
\end{equation}

We define a small gap: 
\begin{equation}
    \label{eq:source-gap}
    \Delta_{\rm src} := \gap(2n+n(n-1)+1, \htn\bigl(\{0,1\}\cup\{w_i,a_i\}_i\cup\{C_{ij}\}_{i\ne j}\bigr)).
\end{equation}

In the next lemma, we prove that, from an approximate KKT point with error smaller than $\Delta_{\rm src}$, we can extract an exact KKT point of~\eqref{pg:difference-constraint-QP} in polynomial time.
\begin{lemma}
    \label{lem:source-round}
    Let $(\bz,\bg)$ be a rational tuple that satisfies $\bw^\top \bz = 0$ and $\bg \geq 0$.
    Suppose that $\bz$ is an $\varepsilon$-approximate KKT point of~\eqref{pg:difference-constraint-QP} with $\varepsilon<\Delta_{\rm src}$.
    Here, $\bg$ is the multiplier vector witnessing the approximate KKT conditions~\eqref{eq:source-approx-kkt}.
    Then, we have a polynomial-time algorithm for computing an exact KKT point of~\eqref{pg:difference-constraint-QP} by solving an LP.
\end{lemma}
\begin{proof}
    Fix $\supp(\bg)$, 
    and minimize $\rho\ge0$ over variables $(\bz',\bg',\rho)$ subject to
    \begin{equation}\label{eq:source-repair}
    \begin{aligned}
    &\bw^\top\bz'=0,\quad \bg'\ge0,\quad
    \gamma'_{ij}=0\quad\forall\,(i,j)\notin\supp(\bg),\\
    &z'_j-z'_i-C_{ij}\le\rho\quad\forall\,i\ne j,\\
    &|z'_j-z'_i-C_{ij}|\le\rho\quad\forall\,(i,j)\in\supp(\bg),\\
    &|(\bA\bg'-\bW\bz'+\ba)_i|\le\rho\quad\forall\,i\in[n].
    \end{aligned}
    \end{equation}
    Here, 
    each absolute-value inequality is equivalent to two linear inequalities.
    We can also replace each unrestricted variable $z'_i$ by the difference of two nonnegative variables, which doubles the number of variables but preserves linearity.
    Therefore, we can transform~\cref{eq:source-repair} into a linear program with $2n+n(n-1)+1$ nonnegative variables.
    Since the coefficients of the transformed LP belong to $\{0,\pm1\}\cup\{\pm w_i,\pm a_i\}_i\cup\{\pm C_{ij}\}_{i\ne j}$, its coefficient height is bounded by the height used in~\cref{eq:source-gap}, independently of $\supp(\bg)$.
    Since $(\bz,\bg,\varepsilon)$ is feasible for~\cref{eq:source-repair} with $\bz'=\bz$, $\bg'=\bg$, and $\rho=\varepsilon$, the optimal value is at most $\varepsilon < \Delta_{\text{src}}$.
    By~\cref{lem:gap}, any positive optimal value would be at least $\Delta_{\rm src}$. Thus, the optimal value must be zero.
    At zero residual, stationarity and the constraints defining the fixed multiplier support hold exactly, and all difference inequalities are feasible. Hence, an optimal solution satisfies~\eqref{eq:source-kkt}.

    Consider the following LP: 
    \begin{equation}\label{eq:source-exact-repair}
    \begin{aligned}
        \min\quad &0\\
        \text{s.t.}\quad
        &\bA\bg'=\bW\bz'-\ba,\quad \bw^\top\bz'=0, \quad \bg' \geq 0,\\
        &z'_j-z'_i\le C_{ij}
            &&\forall\,i\ne j,\\
        &z'_j-z'_i=C_{ij}
            &&\forall\,(i,j)\in\supp(\bg),\\
        &\gamma'_{ij}=0
            &&\forall\,(i,j)\notin\supp(\bg).
    \end{aligned}
    \end{equation}
    \cref{eq:source-exact-repair} is feasible by~\cref{lem:gap}.
    Let $(\bz^\star,\bg^\star)$ be a feasible solution to the linear program~\cref{eq:source-exact-repair}.
    For all $(i,j)\in\supp(\bg)$, the tightness constraint in~\cref{eq:source-exact-repair} imposes $z_j^\star-z_i^\star=C_{ij}$.
    For all $(i,j)\notin\supp(\bg)$, the support constraint in~\cref{eq:source-exact-repair} imposes $\gamma_{ij}^\star=0$.
    Hence, we have 
    \[
        \gamma_{ij}^\star(z_j^\star-z_i^\star-C_{ij})=0
        \quad\forall\,i\ne j.
    \]
    Combining this with other constraints of~\cref{eq:source-exact-repair}, 
    we see that $(\bz^\star,\bg^\star)$ satisfies all conditions in~\eqref{eq:source-kkt}, therefore $\bz^\star$ is an exact KKT point of~\eqref{pg:difference-constraint-QP}.
    Since \cref{eq:source-exact-repair} can be solved in polynomial time as a rational linear program, we can compute an exact KKT point of~\eqref{pg:difference-constraint-QP} in polynomial time.
\end{proof}

\subsection{From Box-constrained QP to Difference-constraint QP}
\label{sec:network}

\subsubsection{A function approximating $f(t) := \frac{1}{2} t^2$}

We next construct a function that corresponds to the optimal value of an linear program over some feasible set with difference constraints, such that the constructed function can approximate $t^2/2$ on $[-1,1]$, which can be used to reduce~\eqref{pg:box-qp} to~\eqref{pg:difference-constraint-QP}.
This is a known construction, we present it here for completeness.
By using $O(\log N)$ variables, we obtain subgradient error $1/N$ on this interval.
Since the number of variables and the coefficient bit lengths are polynomial in $\log N$, we can make the error $1/N$ exponentially small while keeping the encoding length polynomial.

\begin{figure}[t]
\centering
\begin{tikzpicture}[
  x=1cm,y=1cm,
  vertex/.style={circle,draw=black!65,fill=white,minimum size=7mm,inner sep=1pt,font=\small},
  flow arc/.style={->,draw=black!65,line width=0.7pt},
  added arc/.style={->,draw=black!60,line width=0.7pt,dashed},
  weight/.style={fill=white,inner sep=1.5pt,font=\small},
  annotation/.style={align=center,font=\small}
]
\node[annotation] at (3.05,6.7) {(a) Network for $N=2$};
\node[vertex] (source) at (1.35,5) {$s'$};
\node[vertex] (s) at (3.05,5.9) {$s$};
\node[vertex] (t) at (3.05,4.1) {$t$};
\node[vertex] (sink) at (4.75,5) {$t'$};
\draw[flow arc] (source) -- node[weight,sloped,above] {$(1,0)$} (s);
\draw[flow arc] (s) -- node[weight,left] {$(1,0)$} (t);
\draw[flow arc] (t) -- node[weight,sloped,below] {$(1,0)$} (sink);
\draw[flow arc] (source) -- node[weight,sloped,below] {$(1,\frac{1}{2})$} (t);
\draw[flow arc] (s) -- node[weight,sloped,above] {$(1,\frac{1}{2})$} (sink);

\node[annotation] at (9.9,6.7) {(b) Minimum cost for $N=2$};
\draw[->,draw=black!55] (8.3,4.1) -- (11.9,4.1) node[right,font=\small] {$F$};
\draw[->,draw=black!55] (8.3,4.1) -- (8.3,6.0) node[left,font=\small] {$g_2(F)$};
\draw[draw=black,line width=1pt] (8.3,4.1) -- (9.8,4.1) -- (11.3,5.6);
\foreach \xx/\labeltext in {8.3/0,9.8/1,11.3/2} {
  \draw[draw=black!55] (\xx,4.15) -- (\xx,4.05)
    node[below,font=\small] {$\labeltext$};
}
\draw[draw=black!55] (8.35,5.6) -- (8.25,5.6) node[left,font=\small] {$1$};
\draw[draw=black!25,densely dotted] (8.3,5.6) -- (11.3,5.6) -- (11.3,4.1);
\node[annotation] at (9.9,3.4) {$g_2(F)=(F-1)_+$};

\node[annotation] at (3.05,2.35) {(c) Network for $N=4$};
\node[vertex] (source-four) at (0,0) {$s''$};
\node[vertex] (source-inner) at (1.35,0) {$s'$};
\node[vertex] (s-inner) at (3.05,0.9) {$s$};
\node[vertex] (t-inner) at (3.05,-0.9) {$t$};
\node[vertex] (sink-inner) at (4.75,0) {$t'$};
\node[vertex] (sink-four) at (6.1,0) {$t''$};
\draw[flow arc] (source-inner) -- node[weight,sloped,above] {$(1,0)$} (s-inner);
\draw[flow arc] (s-inner) -- node[weight,left] {$(1,0)$} (t-inner);
\draw[flow arc] (t-inner) -- node[weight,sloped,below] {$(1,0)$} (sink-inner);
\draw[flow arc] (source-inner) -- node[weight,sloped,below] {$(1,\frac{1}{2})$} (t-inner);
\draw[flow arc] (s-inner) -- node[weight,sloped,above] {$(1,\frac{1}{2})$} (sink-inner);
\draw[added arc] (source-four) -- node[weight,above] {$(2,0)$} (source-inner);
\draw[added arc] (sink-inner) -- node[weight,above] {$(2,0)$} (sink-four);
\draw[added arc] (source-four.north) .. controls (0,2.15) and (4.75,2.15) ..
  node[weight,above,pos=0.5] {$(2,\frac{3}{2})$} (sink-inner.north);
\draw[added arc] (source-inner.south) .. controls (1.35,-2.15) and (6.1,-2.15) ..
  node[weight,below,pos=0.5] {$(2,\frac{3}{2})$} (sink-four.south);

\node[annotation] at (9.9,2.35) {(d) Minimum cost for $N=4$};
\draw[->,draw=black!55] (8.3,-1.25) -- (11.9,-1.25) node[right,font=\small] {$F$};
\draw[->,draw=black!55] (8.3,-1.25) -- (8.3,1.75) node[left,font=\small] {$g_4(F)$};
\draw[draw=black,line width=1pt]
  (8.3,-1.25) -- (9.08,-1.25) -- (9.86,-0.83) -- (10.64,0.01) -- (11.42,1.27);
\foreach \xx/\labeltext in {8.3/0,9.08/1,9.86/2,10.64/3,11.42/4} {
  \draw[draw=black!55] (\xx,-1.20) -- (\xx,-1.30)
    node[below,font=\small] {$\labeltext$};
}
\foreach \yy/\labeltext in {-0.83/1,0.01/3,1.27/6} {
  \draw[draw=black!55] (8.35,\yy) -- (8.25,\yy)
    node[left,font=\small] {$\labeltext$};
}
\draw[draw=black!25,densely dotted] (8.3,1.27) -- (11.42,1.27) -- (11.42,-1.25);
\node[annotation] at (9.9,-2.1) {$g_4(F)=\sum_{\ell=1}^3(F-\ell)_+$};
\end{tikzpicture}
\caption{We illustrate the unperturbed counting networks of~\cite[Section~3.1]{disser2018simplex} for $N=2$ and $N=4$ in the left panels, with each arc labeled by (capacity, unit cost).
The dashed arcs are the four arcs added in the second doubling step.
We plot the corresponding minimum-cost functions in the right panels.}
\label{fig:quadratic-cost-network}
\end{figure}
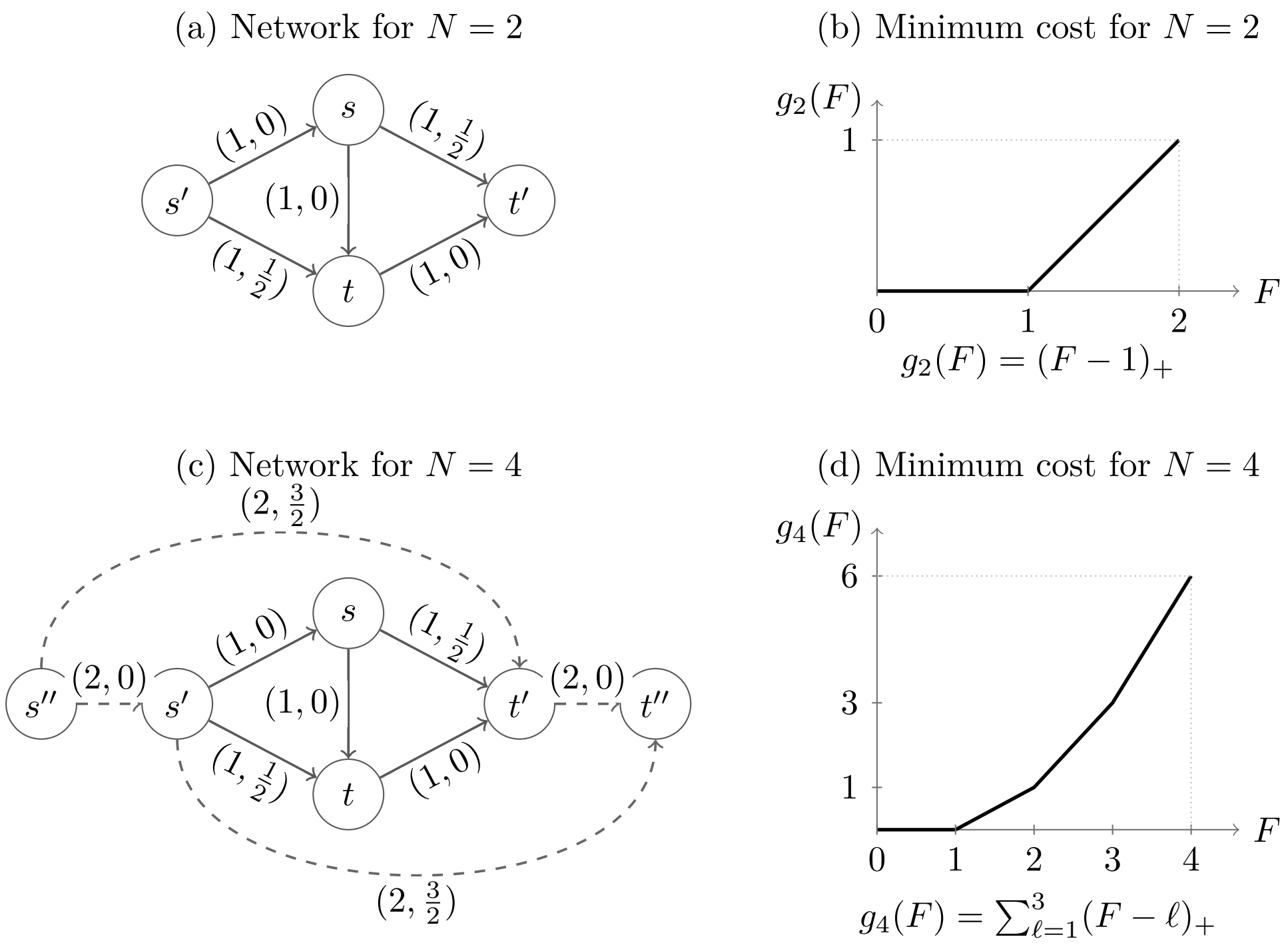

In the following lemma, we describe a network of size $O(\log N)$ whose minimum-cost function\footnote{For a network with arc set $E$, fixed arc capacities and unit costs $c_e$, $g_N(F)$ is the minimum of $\sum_{e\in E} c_e f_e$ over source-to-sink flows of total value $F\in[0,N]$ satisfying the capacity bounds and flow conservation at non-terminal vertices. The argument is $F$, while the arc flows $f_e$ are the optimization variables.} has $N$ linear segments.
In~\cref{fig:quadratic-cost-network}, we illustrate such networks for $N=2$ and $N=4$.

\begin{lemma}[\cite{disser2018simplex}]
    \label{lem:network}
    For all powers of two $N=2^q$, 
    we can construct a directed acyclic network with $2q + 2$ vertices and 
    $4q + 1$ arcs whose maximum flow is $N$ and 
    whose minimum-cost function $g_N(F): [0,N]\to\mathbb{R}$ is the linear interpolation of
    \begin{equation*}
        \label{eq:integer-parabola}
        g_N(j)=\frac12 j(j-1)\quad\forall\,j\in\{0,\ldots,N\}.
    \end{equation*}
    In this construction, all capacities and costs are nonnegative rationals of $O(q+1)$ bits.
\end{lemma}
\begin{proof}
    We follow the counting-gadget construction and the proof of Lemma~3.1 in~\cite[Section~3.1]{disser2018simplex}, setting all cost perturbations to zero.
    Their successive shortest path algorithm augments one unit of flow at each step, with successive path costs $0,1,\ldots,N-1$.
    Since shortest-path augmentation preserves minimum cost, the function $g_N$ has slope $j$ on $[j,j+1]$ for all $j\in\{0,\ldots,N-1\}$.
    By summing these marginal costs, we obtain $g_N(j)=\sum_{\ell=0}^{j-1}\ell=j(j-1)/2$.
    We start with one capacity-one, zero-cost arc and add two vertices and four arcs at each doubling step.
    The added capacities are powers of two at most $N$, and the added costs are either zero or $2^\ell-1/2$ for some $\ell\in\{0,\ldots,q-1\}$.
    Thus, the network has the stated size and coefficient encoding lengths.
\end{proof}

We then scale every capacity and cost by $1/N$, which leads to a scaled network where the maximum flow becomes one, and the minimum-cost function becomes $\bar g_N(F):=\frac{g_N(NF)}{N^2}$ for all $F\in[0,1]$.
In the scaled network, all scaled arc costs belong to $[0, 1/2)$.

Next, we define the convex conjugate $h_N$ of $\bar{g}_N$: 
\begin{equation*}\label{eq:hN}
    h_N(t)
    :=\max_{0\le F\le1}\{tF-\bar g_N(F)\} 
    =\max_{j=0,\ldots,N}
    \left\{\frac{j}{N}t-\frac{j(j-1)}{2N^2}\right\}
    =\frac1N\sum_{\ell=0}^{N-1}\pos{t-\frac{\ell}{N}}. 
\end{equation*}
In the following lemma, we represent $h_N(t)$ by an LP with $O(\log N)$ variables and difference-of-variable constraints.

\begin{lemma}
    \label{lem:dual}
    Let $G:=(V,E)$ be the scaled network defining $\bar g_N$, with arc capacities $u_e$ and costs $c_e$.
    Fix terminal potentials $p_s,p_t$ with $p_t-p_s=t$.
    Define the variable vectors $\bp:=(p_v)_{v\in V\setminus\{s,t\}}\in\R^{V\setminus\{s,t\}}$ and $\by:=(y_e)_{e\in E}\in\R^E$.
    Then, we have 
    \begin{equation}
        \label{eq:dual-lp}
        h_N(t)=\min_{\bp,\by}
        \left\{
        \sum_{e=(u,v)\in E}u_e(y_e-p_u):
        \begin{aligned}
        p_u-y_e&\le0,\\
        p_v-y_e&\le c_e
        \end{aligned}
        \quad\forall\,e=(u,v)\in E
        \right\}.
    \end{equation}
    The LP~\cref{eq:dual-lp} has $O(\log N)$ variables and difference inequalities.
    If $p_s,p_t\in[\alpha,\beta]$, we may also restrict all vertex potentials $p_v$ and arc variables $y_e$ to $[\alpha,\beta]$ without changing the optimal value in~\cref{eq:dual-lp}.
\end{lemma}
\begin{proof}
\begin{samepage}
Let $F\in[0,1]$ denote the $s$--$t$ flow value and $\mathbf f:=(f_e)_{e\in E}\in\R^E$ the arc-flow vector.
By the definition of $h_N$, we have
\begin{equation*}
h_N(t)=\max_{F,\mathbf f}
\left\{
\begin{aligned}
    &tF-\sum_{e\in E}c_ef_e: 
    \ 0\le f_e\le u_e
    &&\forall\,e\in E,\\
    &\sum\nolimits_{e=(s,v)\in E}f_e=F \text{ and }
    \sum\nolimits_{e=(v,t)\in E}f_e=F 
    &&\forall\,v\in V\setminus\{s,t\},\\
    &\sum\nolimits_{e=(v,w)\in E}f_e=\sum\nolimits_{e=(w,v)\in E}f_e
    &&\forall\,v\in V\setminus\{s,t\}
\end{aligned}
\right\}.
\end{equation*}
\end{samepage}

By eliminating the flow value $F$ and dualizing flow conservation at every non-terminal vertex, we obtain
\begin{equation*}
h_N(t)=\min_{\bp\in\R^{V\setminus\{s,t\}}}\sum_{e=(u,v)\in E}u_e\pos{p_v-p_u-c_e}.
\end{equation*}
For any capacity-feasible flow $(f_e)_e$ and vertex potentials $(p_v)_v$ with $p_t-p_s=t$, we have
$\sum_{e=(u,v)\in E} f_e(p_v-p_u)=tF$ and
$f_e(p_v-p_u-c_e)\le u_e\pos{p_v-p_u-c_e}$.
Thus, every feasible dual objective is at least $h_N(t)$, and equality follows from strong duality.
By introducing $y_e:=\max\{p_u,p_v-c_e\}$, we obtain the representation~\cref{eq:dual-lp}.

Let $\Pi$ denote clipping to $[\alpha,\beta]$.
Since $\Pi$ is nondecreasing and $1$-Lipschitz, we have, for all $a,b\in\R$ and $c\ge0$,
\begin{equation*}
\pos{\Pi(b)-\Pi(a)-c}\le\pos{b-a-c}.
\end{equation*}
If $b<a$, $\pos{\Pi(b)-\Pi(a)-c}=0$.
If $b\ge a$, $\Pi(b)-\Pi(a)\le b-a$.
By clipping each non-terminal vertex potential $p_v$ while keeping $p_s,p_t$ fixed and then setting
$y_e=\max\{p_u,p_v-c_e\}$, we obtain a feasible solution of~\cref{eq:dual-lp} with no larger objective value.
Since $c_e\ge0$, every such $y_e$ also belongs to $[\alpha,\beta]$.
Thus, restricting the non-terminal vertex potentials $p_v$ and arc variables $y_e$ to $[\alpha,\beta]$ preserves the optimal value in~\cref{eq:dual-lp}.
\end{proof}

We next present the function we mentioned in the beginning of this subsection, which approximates $t^2/2$ on $[-1,1]$.
Define the symmetrized function 
\begin{equation} 
    \psi_N(t) := h_N(t)+h_N(-t).
    \label{eq:psi}
\end{equation}
We use the following lemma to bound the difference between a subgradient of $\psi_N$ and the derivative of $\frac{1}{2} t^2$ by $1/N$, as needed to recover exact KKT points of~\eqref{pg:box-qp}.
\begin{lemma}
    \label{lem:square}
    For all $t\in[-1,1]$ and all $s\in\partial\psi_N(t)$,
    \begin{equation}\label{eq:square-error}
    |s-t|\le\frac1N.
    \end{equation}
\end{lemma}
\begin{proof}
For $t<0$, $h_N(t)=0$.
On an interval $(\ell/N,(\ell+1)/N)$ in $[0,1]$, the slope of $h_N$ is $(\ell+1)/N$.
At a breakpoint $t$, the subdifferential $\partial h_N(t)$ is the interval between the neighboring slopes of $h_N$.
At $t=0$, $\partial h_N(0)=[0,1/N]$, and at $t=1$, $\partial h_N(1)=\{1\}$.
By applying these slope bounds to the two terms in~\cref{eq:psi}, we obtain~\cref{eq:square-error}, including $t=0$ and the endpoints.
\end{proof}

\subsubsection{A lifted instance of~\eqref{pg:difference-constraint-QP}}

Using~\cref{lem:square}, 
we can encode~\eqref{pg:box-qp} as an instance of~\eqref{pg:difference-constraint-QP}.
Fix an instance of~\eqref{pg:box-qp}. 
We choose a coefficient bound $L$ and positive square weights $A_{ij}$ as follows: 
\begin{equation}
    \label{eq:hessian-weights}
    \begin{aligned}
        L &:= 1 + \sum\nolimits_{i,j=1}^n |Q_{ij}| \\
        A_{ij} &:= L - Q_{ij} > 0 \quad \forall\, 1 \leq i < j \leq n \\
        A_{i0} &:= L + \sum\nolimits_{j=1}^{n} Q_{ij} > 0 \quad \forall\, 1 \leq i \leq n.
    \end{aligned}
\end{equation}

We consider a set of auxiliary variables $z_0, z_1, \ldots, z_n$ with $\sum_{i=0}^{n} z_i = 0$.
We then consider variables $x_i := z_i - z_0$ for all $i \in [n]$ and $x_0 := 0$.
Define 
\begin{equation}
    \label{eq:native-g}
    g(\bx) 
    := - \frac{(n+1)L}{2} \sum_{i=1}^{n}x_i^2
    + \frac{L}{2}\left(\sum_{i=1}^{n}x_i\right)^2 = -\frac{(n+1)L}{2} \sum_{i=0}^{n}z_i^2.
\end{equation}

\begin{lemma}[Quadratic identity]
    \label{lem:hessian}
    For a sysmmetric matrix $\bQ \in \mathbb{Q}^{n \times n}$, we have 
    \begin{equation}
        \label{eq:hessian-identity}
        \frac12\bx^\top \bQ \bx
        =
        g(\bx)+
        \frac12\sum_{1\le i<j\le n}A_{ij}(x_i-x_j)^2+
        \frac12\sum_{i=1}^{n}A_{i0}x_i^2, 
    \end{equation}
    where $\bA$ is defined in~\cref{eq:hessian-weights} and $g(\bx)$ is defined in~\cref{eq:native-g}.
\end{lemma}
\begin{proof}
    The Hessian matrix of $g$ is $-(n+1)L \mathbf{I}_n + L \mathbf{J}_n$, where $\mathbf{I}_n$ is $n$-dimensional identity matrix and $\mathbf{J}_n$ is $n$-dimensional all-one matrix.
    The second term in~\cref{eq:hessian-identity} contributes off-diagonal entry $-A_{ij} = Q_{ij} - L$ to the overall Hessian matrix.
    The sum of the last two terms in~\cref{eq:hessian-identity} contributes diagonal entry $\sum_{j=1,\,j\ne i}^n A_{ij} + A_{i0} = \sum_{j=1,\,j\ne i}^n(L-Q_{ij}) + (L + \sum_{j=1}^n Q_{ij}) = nL + Q_{ii}$.
    Hence, the Hessian matrix of the right hand side of~\cref{eq:hessian-identity} is $\bQ$.
    Since both sides of~\cref{eq:hessian-identity} are homogeneous quadratics with the same Hessian, the identity holds.
\end{proof}

For a power of two $N\ge2$, we define the feasible set which is used to construct the lifted instance of~\eqref{pg:difference-constraint-QP}.
Let $\calG_N := (\calV_N, \calE_N)$ be the scaled network used in \cref{lem:dual}, with terminals $s,t$, arc capacities $u_e$, and arc costs $c_e$.
Define the index set 
\[
    \mathcal{I} := \{(i,j):1\le i<j\le n\}\cup\{(i,0):1\le i\le n\}.
\]
For all $(i,j) \in \mathcal{I}$, 
take two copies of this network, indexed by $\sigma\in\{+,-\}$.
For each copy, we define \emph{non-terminal vertex potential variables} $p_v^{ij,\sigma}$ for $v\in \calV_N \setminus \{s,t\}$ and 
\emph{arc variables} $y_e^{ij,\sigma}$ for $e\in \calE_N$.
In addition, we define the terminal potentials $(p_s^{ij,+},p_t^{ij,+}):=(z_j,z_i)$ and $(p_s^{ij,-},p_t^{ij,-}):=(z_i,z_j)$.

Let $r$ be the total number of main and auxiliary coordinates in~\eqref{eq:lifted-qp}.
For all $(i,j)\in\mathcal I$ and $\sigma\in\{+,-\}$, define the coordinate blocks
\begin{equation*}
    \bp^{ij,\sigma}:=(p_v^{ij,\sigma})_{v\in\calV_N\setminus\{s,t\}},
    \quad
    \by^{ij,\sigma}:=(y_e^{ij,\sigma})_{e\in\calE_N}.
\end{equation*}
We collect these coordinates into
\begin{equation*}
    \boldsymbol\zeta:=
    \left(\bz,
      (\bp^{ij,\sigma})_{(i,j)\in\mathcal I,\,\sigma\in\{+,-\}},
      (\by^{ij,\sigma})_{(i,j)\in\mathcal I,\,\sigma\in\{+,-\}}
    \right)\in\R^r,
    \quad \bz:=(z_0,z_1,\ldots,z_n).
\end{equation*}
Define the feasible set
\begin{equation}\label{eq:lifted-polytope}
\mathcal P_N:=
\left\{\boldsymbol\zeta\in\R^r:
    \begin{aligned}
        &\sum\nolimits_{i=0}^n z_i=0,\\
        &0\le\zeta_a-z_0\le1
            &&\forall\,a\in[r],\\
        &\begin{aligned}
           p_u^{ij,\sigma}-y_e^{ij,\sigma}&\le0,\\
           p_v^{ij,\sigma}-y_e^{ij,\sigma}&\le c_e
         \end{aligned}
            &&\begin{aligned}
               &\forall\,(i,j)\in\mathcal I,\\
               &\forall\,\sigma\in\{+,-\},\\
               &\forall\,e=(u,v)\in\calE_N.
              \end{aligned}
    \end{aligned}
\right\}.
\end{equation}
Next, by choosing a sufficiently large $N$ and the function $\psi_N(t)$ defined in~\cref{eq:psi}, we construct a lifted difference-of-variable constrained QP, such that from every exact KKT point of the lifted QP, we can compute an exact KKT point of the original box-constrained QP in polynomial time.

Let $M := \max_i (A_{i0}+\sum_{j=1,\,j\ne i}^nA_{ij} ) > 0$,
where $A_{ij}:=A_{ji}$ when needed.
Choose the smallest power of two $N\ge2$ satisfying
\begin{equation}
    \label{eq:N-choice}
    N>\frac{2M}{\Delta_{\rm box}}, 
\end{equation}
where $\Delta_{\rm box}:=\gap\!\left(n+1,\htn\bigl(\{0,1\}\cup\{Q_{ij}\}_{i,j}\cup\{c_i\}_i\bigr)\right)$.
Note that, with this chosen $N$, $\log(N)$ is polynomial in the encoding length of~\eqref{pg:box-qp}.


By~\cref{eq:psi,eq:dual-lp}, we obtain a quadratic objective in $\boldsymbol\zeta$.
Let $\ba$ encode all linear coefficients: 
\begin{equation*}
    \ba := 
    \begin{pmatrix}
    - \sum_{i=1}^n(c_i+A_{i0}) \\ 
    \left(
    c_i-A_{i0} - \sum\nolimits_{j=1,\,j\neq i}^nA_{ij}
    \right)_{i\in[n]} \\ 
    \left(
      \left(-A_{ij}\sum_{e=(v,w)\in\calE_N}u_e\right)_{v\in\calV_N\setminus\{s,t\}}
    \right)_{(i,j)\in\mathcal I,\,\sigma\in\{+,-\}} \\
    \left(
      \left(A_{ij}u_e\right)_{e\in\calE_N}
    \right)_{(i,j)\in\mathcal I,\,\sigma\in\{+,-\}}
    \end{pmatrix}.
\end{equation*}

One can verify $\one^\top\ba=0$.
The lifted QP has the following compact expression: 
\begin{equation}
    \label{eq:lifted-qp}
    \min_{\boldsymbol\zeta\in\mathcal P_N}
    \left\{-\frac12 \boldsymbol\zeta^\top \bW \boldsymbol\zeta+\ba^\top\boldsymbol\zeta\right\},
    \tag{Lifted-DC-QP}
\end{equation}
where $\bw$ has entries $(n+1)L$ on the coordinates $z_0,\ldots,z_n$ and zero on the coordinates $p_v^{ij,\sigma},y_e^{ij,\sigma}$, and $\bW=\diag(\bw)$.

In the following proposition, we prove that we can recover an exact KKT point of~\eqref{pg:box-qp} from every exact KKT point of~\eqref{eq:lifted-qp} in polynomial time.
\begin{proposition}
    \label{prop:lift}
    The construction of~\eqref{eq:lifted-qp} has polynomial encoding length.
    From every exact KKT point of~\eqref{eq:lifted-qp}, one can compute an exact KKT point of~\eqref{pg:box-qp} in polynomial time.
\end{proposition}
\begin{proof}
    Define $x_0=0$ and the convex function
    \begin{equation}\label{eq:H-global}
        H(\bx):=\bc^\top\bx+
        \sum_{(i,j)\in\mathcal I}A_{ij}\psi_N(x_i-x_j),
        \quad\forall\,\bx\in\R^n.
    \end{equation}
    For each $\bx\in[0,1]^n$, the variables $z_0,\ldots,z_n$ satisfy
    $z_0=-\sum_{i=1}^n x_i/(n+1)$ and $z_i=z_0+x_i$.
    Thus, we have $p_s^{ij,\sigma},p_t^{ij,\sigma}\in[z_0,z_0+1]$ for all $(i,j)\in\mathcal I$ and $\sigma\in\{+,-\}$.
    By~\cref{lem:dual,eq:psi} and the definition of $\ba$, we have
    \begin{equation*}
        H(\bx)=\min_{\boldsymbol\zeta\in\mathcal P_N}\left\{
        \ba^\top\boldsymbol\zeta:
        z_i-z_0=x_i\quad\forall\,i\in[n]
        \right\},
        \quad\forall\,\bx\in[0,1]^n.
    \end{equation*}
    By~\cref{lem:dual}, we may impose $z_0\le p_v^{ij,\sigma}\le z_0+1$ and $z_0\le y_e^{ij,\sigma}\le z_0+1$ without changing this minimum.

    Let $\boldsymbol\zeta$ be an exact KKT point of~\eqref{eq:lifted-qp}, and set $x_i=z_i-z_0$ for all $i\in[n]$.
    Since $\mathcal P_N$ is a polytope, we obtain from the KKT conditions
    $(\ba-\bW\boldsymbol\zeta)^\top(\boldsymbol\zeta'-\boldsymbol\zeta)\ge0$
    for all $\boldsymbol\zeta'\in\mathcal P_N$.
    By~\cref{eq:native-g}, this is equivalent to
    \begin{equation}\label{eq:joint-vi}
        \nabla g(\bx)^\top(\bx'-\bx)
        +\ba^\top(\boldsymbol\zeta'-\boldsymbol\zeta)\ge0,
        \quad x'_i:=z'_i-z'_0.
    \end{equation}
    When $\bx'=\bx$, the first term in~\cref{eq:joint-vi} vanishes. Hence, $\boldsymbol\zeta$ minimizes the linear objective among feasible points with these fixed main coordinates.
    Since this minimum is $H(\bx)$, we have $\ba^\top\boldsymbol\zeta=H(\bx)$.
    For each fixed $\bx'$, we minimize the left-hand side of~\cref{eq:joint-vi} over the variables $(p_v^{ij,\sigma})'$ and $(y_e^{ij,\sigma})'$ subject to $\boldsymbol\zeta'\in\mathcal P_N$ and obtain
    \begin{equation*}
        \nabla g(\bx)^\top(\bx'-\bx)
        +H(\bx')-H(\bx)\ge0,
        \quad\forall\,\bx'\in[0,1]^n.
    \end{equation*}
    Since $H$ is finite and convex on $\R^n$, we can apply the convex optimality conditions to obtain $\mathbf s\in\partial H(\bx)$ such that
    \begin{equation*}\label{eq:projected-kkt}
        \bigl(\nabla g(\bx)+\mathbf s\bigr)_i
        \begin{cases}
            \ge0 &\text{if }x_i=0,\\
            \le0 &\text{if }x_i=1,\\
            =0 &\text{if }0<x_i<1,
        \end{cases}
        \quad\forall\,i\in[n].
    \end{equation*}
    By the subdifferential sum and affine-composition rules applied to~\cref{eq:H-global}, there exist scalars $\alpha_{ij}\in\partial\psi_N(x_i-x_j)$ for all $(i,j)\in\mathcal I$ such that
    \begin{equation*}
        \mathbf s=\bc+
        \sum_{(i,j)\in\mathcal I}A_{ij}\alpha_{ij}(\e_i-\e_j),
    \end{equation*}
    where $\e_1,\ldots,\e_n$ are the standard unit vectors in $\R^n$ and $\e_0:=\mathbf0$.
    By~\cref{lem:square}, we have $|\alpha_{ij}-(x_i-x_j)|\le1/N$ for all $(i,j)\in\mathcal I$.
    By combining this bound with~\cref{eq:hessian-identity}, we obtain
    \begin{equation*}
        \left|
        \bigl(\nabla g(\bx)+\mathbf s
        -(\bQ\bx+\bc)\bigr)_i
        \right|
        \le\frac{A_{i0}+\sum_{j=1,\,j\ne i}^nA_{ij}}{N}
        \le\frac{M}{N}<\frac{\Delta_{\rm box}}2,
        \quad\forall\,i\in[n],
    \end{equation*}
    where the last inequality follows from~\cref{eq:N-choice}.
    The vector $\nabla g(\bx)+\mathbf s$ has the exact KKT signs, and, by the preceding bound, each coordinate differs from the corresponding derivative of the original quadratic by at most $M/N$.
    Therefore, $\bx$ is a strong $(M/N)$-KKT point of~\eqref{pg:box-qp}.
    Since $M/N<\Delta_{\rm box}/2$, we can apply~\cref{lem:box-round} to compute an exact KKT point of~\eqref{pg:box-qp} in polynomial time.

    We finally verify the encoding length of~\eqref{eq:lifted-qp}.
    By~\cref{eq:N-choice,lem:network,lem:dual}, the constructed QP has $O(n^2\log N)$ variables and inequalities, where $\log N$ is polynomial in the encoding length of $(\bQ,\bc)$.
    The coefficients of~\eqref{eq:lifted-qp} have polynomial binary encoding length: the capacities and costs have $O(\log N)$ bits, and each entry of $\ba$ is a signed sum of polynomially many terms of the form $A_{ij}u_e$ or $c_i$.
    The inequalities $0\le\zeta_a-z_0\le1$ for all $a\in[r]$ in~\cref{eq:lifted-polytope} imply $\zeta_b-\zeta_a\le1$ for all distinct $a,b\in[r]$.
    By adding these redundant inequalities and keeping the smallest right-hand side for each ordered pair, we obtain a complete difference-constraint matrix with at most $r(r-1)$ entries, each of polynomial encoding length.
    Thus, the construction of~\eqref{eq:lifted-qp} and the recovery of an exact KKT point of~\eqref{pg:box-qp} both take polynomial time.
\end{proof}


In the QP~\eqref{eq:lifted-qp}, the weights of the coordinates $p_v^{ij,\sigma}$ and $y_e^{ij,\sigma}$ are zero.
In the following lemma, we consider another QP where all weights are positive, 
and show how to recover an exact KKT point of~\eqref{eq:lifted-qp} 
from any KKT point of~\eqref{eq:C-baregularized}.
\begin{lemma}
    \label{lem:regularization}
    Consider an $r$-dimensional instance of~\eqref{pg:difference-constraint-QP}, and set
    $\bv := \bw + \eta\one$ with $\eta>0$.
    If
    \begin{equation}
        \label{eq:eta-bound}
        \eta(r+1)\bar{C} < \Delta_{\rm src},
    \end{equation}
    then every KKT point of
    \begin{equation}
        \label{eq:C-baregularized}
        \min\left\{-\frac12\bz^\top \bV \bz+\ba^\top\bz:
        z_j-z_i\le C_{ij},\ \bv^\top\bz=0\right\} 
        \tag{PosW-DC-QP}
    \end{equation}
    can be decoded into an exact KKT point of~\eqref{pg:difference-constraint-QP} by solving~\cref{eq:source-repair}.
\end{lemma}
\begin{proof}
Let $\widetilde{\bz}$ be an exact KKT point of~\eqref{eq:C-baregularized}.
If the multipliers $\bg$ are not supplied, we compute them by solving
\begin{equation*}
    \begin{aligned}
        \min_{\bg}\quad &0\\
        \text{s.t.}\quad
        &\bA\bg=\bV\widetilde{\bz}-\ba,\\
        &\gamma_{ij}\ge0
            &&\forall\,i\ne j,\\
        &\gamma_{ij}=0
            &&\forall\,i\ne j\text{ with }\widetilde z_j-\widetilde z_i<C_{ij}.
    \end{aligned}
\end{equation*}
Here, $\widetilde{\bz}$ is fixed and $\bV:=\diag(\bv)$.
Since $\widetilde{\bz}$ is a KKT point of~\eqref{eq:C-baregularized}, this LP is feasible, and we can compute $\bg$ in polynomial time.
From the gauge and difference constraints in~\eqref{eq:C-baregularized}, we obtain
$\|\widetilde{\bz}\|_\infty\le\bar{C}$.
Define
\begin{equation*}
h:=\frac{\bw^\top\widetilde{\bz}}{\sum_{i=1}^rw_i}
=-\frac{\eta}{\sum_{i=1}^rw_i}\one^\top\widetilde{\bz},
\quad
\bz:=\widetilde{\bz}-h\one.
\end{equation*}
Then, we have $\bw^\top\bz=0$.
Also, $z_j-z_i=\widetilde z_j-\widetilde z_i$ for all $i,j$. 
Therefore, for all $i\ne j$, the inequality $z_j-z_i\le C_{ij}$ in~\eqref{pg:difference-constraint-QP} is tight at $\bz$ if and only if the same inequality in~\eqref{eq:C-baregularized} is tight at $\widetilde{\bz}$.
Moreover, $\bA\bg - \bW\bz + \ba = \eta\widetilde{\bz} + h\bw$.
Since $w_i\le\sum_{j=1}^rw_j$ for all $i$ and $|h|\le\frac{\eta r\bar{C}}{\sum_{j=1}^rw_j}$, we have
\begin{equation*}
    \|\bA\bg-\bW\bz+\ba\|_\infty
    \le\eta(r+1)\bar{C}<\Delta_{\rm src}.
\end{equation*}
Since the uniform shift preserves all coordinate differences, the difference constraints and their tightness are unchanged.
The gauge holds by construction, and we keep the same multipliers. Thus, the pair $(\bz,\bg)$ satisfies primal feasibility and complementary slackness exactly.
Since the stationarity error is below $\Delta_{\rm src}$, we can apply~\cref{lem:source-round} to $(\bz,\bg)$ and compute an exact KKT point of~\eqref{pg:difference-constraint-QP} by solving an LP in polynomial time.
\end{proof}

In~\eqref{eq:lifted-qp}, 
positive weights are all equal to $(n+1)L$ before applying~\cref{lem:regularization}.
Choose the smallest integer $k\ge1$ such that 
\begin{equation}\label{eq:k-choice}
\eta:=\frac{(n+1)L}{2^k-1}
\le
\min\left\{1,\frac{\Delta_{\rm src}}{4(r+1)\bar{C}}\right\}.
\end{equation}
Then, we have~\cref{eq:eta-bound} which leads to a valid QP with all positive weights taking exactly two values.
In particular, the positive weights $v_i$ in the positive-weights QP satisfy 
$v_i \in \{\eta,(n+1)L+\eta\}=\{\eta,2^k\eta\}$.
The integer $k$ and the bit length of $\eta$ are polynomially bounded in the encoding length of~\eqref{eq:lifted-qp}.

We then remove the linear term in~\eqref{eq:C-baregularized} by a translation.

\begin{lemma}
    \label{lem:translation}
    Suppose $v_i>0$ for all $i$ and $\one^\top\ba=0$.
    By substituting
    \begin{equation}\label{eq:shift}
    z_i:=y_i+\frac{a_i}{v_i}\quad\forall\,i
    \end{equation}
    into~\eqref{eq:C-baregularized} and omitting the constant term in the objective, we can obtain the QP
    \begin{equation}\label{eq:pure-source}
        \min\left\{
        -\frac12\sum_{i=1}^r v_i y_i^2:
        y_j-y_i \leq C_{ij}+\frac{a_i}{v_i}-\frac{a_j}{v_j},\quad \bv^\top\by=0
        \right\}, 
        \tag{Pure-DC-QP}
    \end{equation}
    whose KKT points correspond bijectively to those of~\eqref{eq:C-baregularized}.
    Furthermore, the encoding length of~\eqref{eq:pure-source} is polynomial in that of~\eqref{eq:C-baregularized}.
\end{lemma}
\begin{proof}
    By substituting~\cref{eq:shift} into the objective of~\eqref{eq:C-baregularized}, we have
    \begin{equation*}
        -\frac12\bz^\top\bV\bz+\ba^\top\bz
        =-\frac12\sum_{i=1}^r v_i y_i^2+\frac12\sum_{i=1}^r\frac{a_i^2}{v_i}.
    \end{equation*}
    Since $\frac12\sum_{i=1}^r a_i^2/v_i$ is independent of $\by$,
    we can omit this term from the objective.
    Since $\sum_{i=1}^r a_i=0$, we have $\bv^\top\bz=\bv^\top\by+\sum_{i=1}^r a_i=\bv^\top\by$.
    For all $i\ne j$, we also have
    $z_j - z_i - C_{ij} = y_j - y_i - (C_{ij}+\frac{a_i}{v_i}-\frac{a_j}{v_j})$.
    Therefore, $\bz$ satisfies the constraints of~\eqref{eq:C-baregularized} if and only if $\by$ satisfies the constraints of~\eqref{eq:pure-source}.
    The inequalities $z_j-z_i\le C_{ij}$ and $y_j-y_i\le C_{ij}+a_i/v_i-a_j/v_j$ have equal slack for all $i\ne j$.
    Thus, any nonnegative multipliers $\bg$ satisfy complementary slackness for~\eqref{eq:C-baregularized} at $\bz$ if and only if they satisfy complementary slackness for~\eqref{eq:pure-source} at $\by$.

    By~\cref{eq:shift}, we have $\bV\bz-\ba=\bV\by$.
    Then, we have
    \begin{equation*}
        \bA\bg=\bV\bz-\ba
        \quad\Longleftrightarrow\quad
        \bA\bg=\bV\by.
    \end{equation*}
    Therefore, $(\bz,\bg)$ satisfies the KKT conditions of~\eqref{eq:C-baregularized} if and only if $(\by,\bg)$ satisfies the KKT conditions of~\eqref{eq:pure-source}.
    Since~\cref{eq:shift} has the inverse $y_i=z_i-a_i/v_i$ for all $i$, we obtain a bijection between the KKT points of the two QPs.

    Finally, the coefficients $v_i$ are unchanged, and we compute each bound $C_{ij}+a_i/v_i-a_j/v_j$ using a constant number of rational arithmetic operations on the coefficients of~\eqref{eq:C-baregularized}.
    The numerator and denominator of $1/v_i$ are the denominator and numerator of $v_i$, respectively.
    Therefore, the encoding length of~\eqref{eq:pure-source} is polynomial in that of~\eqref{eq:C-baregularized}.
\end{proof}

Thus, \eqref{eq:pure-source} minimizes a negative weighted norm over difference constraints, with strictly positive weights taking exactly two values in ratio $2^k$.

\subsection{From Difference-constraints QP to Linear chores Fisher market}
\label{sec:market}

In this subsection, we encode the difference constraints of~\eqref{eq:pure-source} in disutility ratios and use equilibrium prices and earning flows to obtain $\bA\bg=\bV\by$, the stationarity condition for~\eqref{eq:pure-source}.
We can therefore recover the KKT points of the difference-constraints QP from the equilibrium of the constructed market.
The main step is to show that every agent receives a positive earning flow of her corresponding chore at every equilibrium.

Consider a nonempty instance of~\eqref{pg:difference-constraint-QP} with
$n$ coordinates, $\ba=\mathbf0$, and strictly positive weights:
\begin{equation}
    \label{eq:market-source}
    \min_{\bz\in P(\bw,\bC)}-\frac12\bz^\top \bW \bz,
    \quad w_i>0.
    \tag{General-Pure-DC-QP}
\end{equation}
Compute $\Delta_{\rm src} = \gap(2n+n(n-1)+1, \htn\bigl(\{0,1\}\cup\{w_i,a_i\}_i\cup\{C_{ij}\}_{i\ne j}\bigr))$ and $\bar{C} = \max\{1,\max_{i\ne j}|C_{ij}|\}$.
Choose the following parameter 
\begin{equation}
    \label{eq:market-parameters}
    \delta:=\min\left\{
    \frac{\Delta_{\rm src}}{128 \bar{C}^2},\
    \frac{\min_iw_i}{64(\sum_{i=1}^nw_i)\bar{C}},\
    \frac1{8 \bar{C}}
    \right\}.
\end{equation}
Construct a linear chores Fisher market with $n$ agents and $n$ unit-supply chores:
\begin{equation}
    \label{eq:market-construction}
    B_i:=w_i,\quad
    d_{ii}:=w_i,\quad
    d_{ij}:=w_j\bigl[1+\delta(C_{ij}+\frac{\Delta_{\rm src}}{16})\bigr]\quad\forall\,i\ne j.
\end{equation}
That is, in the constructed market, each disutility coefficient $d_{ij}$ is a small multiplicative perturbation of $w_j$, and for $i\ne j$ the perturbation encodes the difference-constraint bound $C_{ij}$.
In the constructed market instance, all entries are strictly positive rationals, and all parameters have polynomial binary encoding length.

\begin{proposition}
    \label{prop:market}
    From any exact CE of the linear chores Fisher market with inputs in~\cref{eq:market-construction}, we can compute an exact KKT point of~\eqref{eq:market-source} by solving one LP in polynomial time.
    Moreover, $D < 2B$.
    \end{proposition}
\begin{proof}
Fix an instance of~\eqref{eq:market-source} and construct the corresponding market instance.
Let $(\bp,\bx)$ be any exact CE of this market. Then, $(\bp,\bx)$ satisfies~\eqref{eq:ce-spending}.
We normalize prices and disutility coefficients by the weights:
\begin{equation*}
q_j:=\frac{p_j}{w_j},\quad
r_{ii}:=1,\quad r_{ij}:=1+\delta(C_{ij}+\frac{\Delta_{\rm src}}{16})\quad\forall\,i\ne j.
\end{equation*}
Then, we have $d_{ij}=w_jr_{ij}$ for all pairs and $1-2\delta \bar{C} \leq r_{ij} \leq 1+2\delta \bar{C}$.

\paragraph{Prices are close to weights $(\bp \approx \bw)$.}
By~\eqref{eq:ce-spending}, we have $d_{ij} \beta_i \geq p_j$ for all agents $i$ and all chores $j$ thus $r_{ij} \beta_i \geq q_j$, which leads to $\beta_i \geq (\max_j q_j) /(1+2\delta \bar{C})$ for all agents $i$.
Since some agent earns a positive amount from a minimum-$q$ chore, we obtain from the complementary slackness condition $b_{ij}(d_{ij}\beta_i-p_j) = 0$ 
\begin{equation*}
    \min_j q_j \geq 
    \frac{1-2\delta \bar{C}}{1+2\delta \bar{C}}\,\max_j q_j.
\end{equation*}
By market clearing, we have $\sum_{j=1}^n w_j q_j = \sum_{j=1}^n p_j = \sum_{i=1}^n B_i = \sum_{i=1}^n w_i$, and therefore $\min_j q_j \leq 1 \leq \max_j q_j$.
By calculation, $|q_j - 1| \leq 8 \delta \bar{C}$ for all $j \in [n]$.

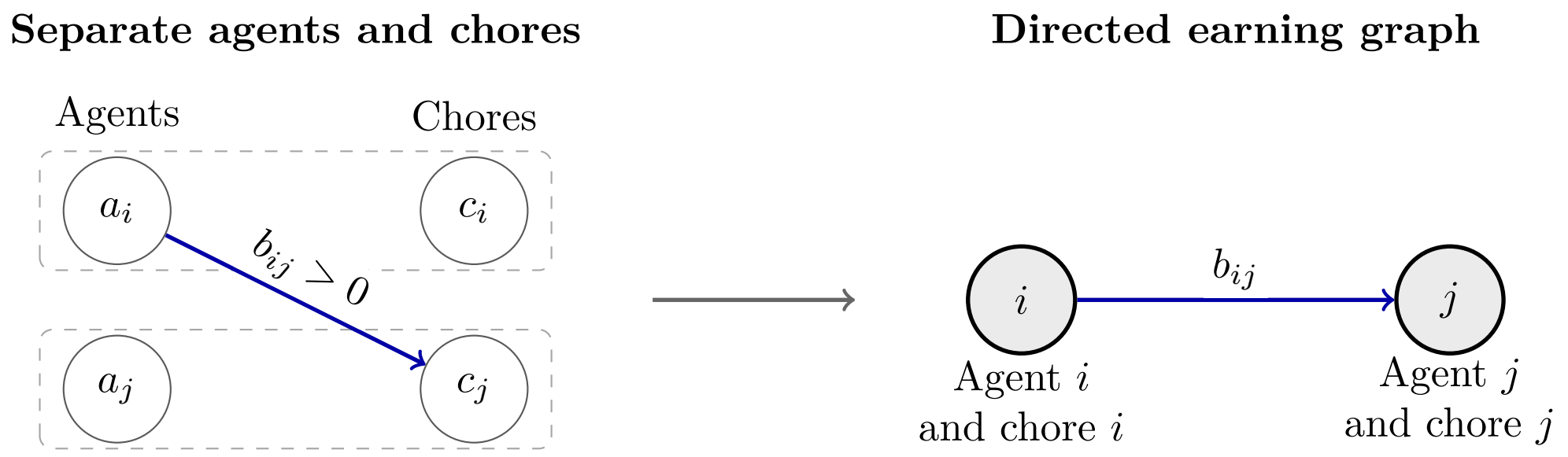
\begin{figure}[htbp]
    \centering
    \begin{tikzpicture}[
      x=1cm,y=1cm,
      vertex/.style={circle,draw=black!65,fill=white,minimum size=9mm,inner sep=1pt,font=\small},
      earning arc/.style={->,draw=blue!65!black,line width=1pt},
      weight/.style={fill=white,inner sep=2pt,font=\small},
      explanation/.style={align=center,font=\small}
    ]
    \node[font=\small\bfseries] at (1.5,2) {Separate agents and chores};
    \node[font=\small] at (0,1.3) {Agents};
    \node[font=\small] at (3,1.3) {Chores};
    \draw[draw=black!35,dashed,rounded corners=3pt] (-0.65,0) rectangle (3.65,1);
    \draw[draw=black!35,dashed,rounded corners=3pt] (-0.65,-1.5) rectangle (3.65,-0.5);
    \node[vertex] (ai) at (0,0.5) {$a_i$};
    \node[vertex] (ci) at (3,0.5) {$c_i$};
    \node[vertex] (aj) at (0,-1) {$a_j$};
    \node[vertex] (cj) at (3,-1) {$c_j$};
    \draw[earning arc] (ai) -- node[weight,sloped,above] {$b_{ij}>0$} (cj);
    
    \draw[->,line width=1pt,draw=black!60] (4.5,-0.25) -- (6.2,-0.25);
    \node[explanation,text width=2.4cm] at (5.1,0.5) {};
    
    \node[font=\small\bfseries] at (9.4,2) {Directed earning graph};
    \node[vertex,draw=black,line width=1pt,fill=black!8] (i) at (7.6,-0.25) {$i$};
    \node[vertex,draw=black,line width=1pt,fill=black!8] (j) at (11.2,-0.25) {$j$};
    \draw[earning arc] (i) -- node[weight,above] {$b_{ij}$} (j);
    \node[explanation] at (7.6,-1.1) {Agent $i$\\and chore $i$};
    \node[explanation] at (11.2,-1.1) {Agent $j$\\and chore $j$};
    \end{tikzpicture}
    \caption{From separate agent and chore vertices to the directed earning graph.
    Vertices with the same index are enclosed in dashed boxes. 
    The blue arc retains its weight $b_{ij}$ after grouping.
    Other arcs are omitted from this schematic.
    For each CE $(\bp, \bx)$, we have an earning graph where $\displaystyle b_{ij} = p_jx_{ij}$ is the money agent $i$ earns from chore $j$.
    In this graph, inner-vertex payments $b_{ii} = p_i x_{ii}$ are omitted.
    }
    \label{fig:earning-graph}
\end{figure}

\paragraph{Positive off-diagonal spending is acyclic.}
Consider a directed earning graph where each vertex corresponds to a pair of agent and chore with the same index, i.e., the vertex $i$ corresponds agent $i$ and chore $i$ for all $i \in [n]$.
There is an arc from vertex $i$ pointing to vertex $j$ if there is an earning flow from chore $j$ to agent $i$, with flow $b_{ij}$.
See~\cref{fig:earning-graph} for an illustration on such a graph.
Unlike this directed graph, a bipartite graph with separate agent and chore vertices is commonly used to represent earning flows in linear chores Fisher markets.

If $b_{ij}>0$ for $i\ne j$, then 
\begin{equation}
    q_j = r_{ij} \beta_i \geq r_{ij}q_i
    \label{eq:edge-ratio}
\end{equation}
because $b_{ij} (d_{ij} \beta_i - p_j) = 0$ and $w_i \beta_i \geq p_i$ for all $i \in [n], j \in [n]$.
Feasibility of~\eqref{eq:market-source} implies
$\sum_{(i,j)\in\mathcal C}C_{ij}\ge0$ for all directed cycles $\mathcal C$.
Therefore,
\begin{equation*}
    \label{eq:cycle}
    \begin{aligned}
    \log\left( \prod\nolimits_{(i,j)\in\mathcal C} r_{ij} \right) = \sum\nolimits_{(i,j)\in\mathcal C}\log r_{ij}
    &\ge
    \delta\sum_{(i,j)\in\mathcal C}(C_{ij}+\frac{\Delta_{\rm src}}{16})
    -\delta^2\sum_{(i,j)\in\mathcal C}(C_{ij}+\frac{\Delta_{\rm src}}{16})^2\\
    &\ge \frac{|\mathcal C|\delta\Delta_{\rm src}}{16}-
    4|\mathcal C|\delta^2 \bar{C}^2
    \ge\frac{|\mathcal C|\delta\Delta_{\rm src}}{32}>0, 
    \end{aligned}
\end{equation*}
For the first inequality, we use $\log(1+t) \geq t-t^2$ when $|t| \leq 1/2$.
Thus, $\prod_{(i,j)\in\mathcal C} r_{ij} > 1$ for all directed cycles $\mathcal{C}$.
However, if there is a cycle of positive earning flows, then, by~\cref{eq:edge-ratio}, $\prod\nolimits_{(i,j)\in\mathcal C} r_{ij} \leq \prod\nolimits_{(i,j)\in\mathcal C} \frac{q_j}{q_i} = 1$.
Hence, by contradiction, the positive off-diagonal earning graph is a Directed Acyclic Graph.

\begin{figure}[htbp]
\centering
\begin{tikzpicture}[
  x=1cm,y=1cm,
  vertex/.style={circle,draw=black!65,minimum size=11mm,inner sep=1pt,font=\small},
  positive/.style={fill=blue!8,draw=blue!65!black},
  negative/.style={fill=orange!10,draw=orange!75!black},
  flow arc/.style={->,draw=black!70,line width=1pt},
  path vertex/.style={circle,draw=black!60,fill=white,minimum size=5.8mm,inner sep=0.5pt,font=\small},
  path one/.style={->,draw=blue!65!black,line width=1pt},
  path two/.style={->,draw=orange!75!black,line width=1pt,dashed},
  path three/.style={->,draw=teal!75!black,line width=1pt},
  path four/.style={->,draw=violet!80!black,line width=1pt,densely dotted},
  weight/.style={fill=white,inner sep=2pt,font=\small},
  annotation/.style={align=center,font=\small},
  heading/.style={align=center,font=\small\bfseries}
]
\node[heading,text width=6.8cm] at (2.8,4.2) {(a) An acyclic earning flow};
\node[vertex,positive,label=below:{$s$}] (s) at (0,0.6) {$+12$};
\node[vertex,negative,label=above:{$u$}] (u) at (2.3,1.1) {$-4$};
\node[vertex,positive,label=below:{$v$}] (v) at (4.4,-0.65) {$+1$};
\node[vertex,negative,label=above:{$t$}] (t) at (5.15,2.25) {$-1$};
\node[vertex,negative,label=below:{$z$}] (z) at (1.5,-2.1) {$-8$};
\draw[flow arc] (s) to[bend left=12] node[weight,above] {$12$} (u);
\draw[flow arc] (u) to[bend right=12] node[weight,above right] {$8$} (v);
\draw[flow arc] (v) -- node[weight,left] {$9$} (t);
\draw[flow arc] (t.east) .. controls (7.2,2.25) and (7.2,-2.1) ..
  node[weight,right,pos=0.5] {$8$} (z.east);

\node[heading,text width=7.3cm] at (11.35,4.2) {(b) Decomposition into weighted paths};
\foreach \name/\xx in {s/8.5,u/9.6} {
  \node[path vertex] (p1-\name) at (\xx,2.4) {$\name$};
}
\draw[path one] (p1-s) -- (p1-u);
\node[annotation,text=blue!65!black] at (7.6,2.4) {$P_1$};
\node[annotation,text=blue!65!black] at (14.15,2.4) {$\lambda_1=4$};

\foreach \name/\xx in {s/8.5,u/9.6,v/10.7,t/11.8} {
  \node[path vertex] (p2-\name) at (\xx,1.1) {$\name$};
}
\foreach \from/\dest in {s/u,u/v,v/t} {
  \draw[path two] (p2-\from) -- (p2-\dest);
}
\node[annotation,text=orange!75!black] at (7.6,1.1) {$P_2$};
\node[annotation,text=orange!75!black] at (14.15,1.1) {$\lambda_2=1$};

\foreach \name/\xx in {s/8.5,u/9.6,v/10.7,t/11.8,z/12.9} {
  \node[path vertex] (p3-\name) at (\xx,-0.2) {$\name$};
}
\foreach \from/\dest in {s/u,u/v,v/t,t/z} {
  \draw[path three] (p3-\from) -- (p3-\dest);
}
\node[annotation,text=teal!75!black] at (7.6,-0.2) {$P_3$};
\node[annotation,text=teal!75!black] at (14.15,-0.2) {$\lambda_3=7$};

\foreach \name/\xx in {v/10.7,t/11.8,z/12.9} {
  \node[path vertex] (p4-\name) at (\xx,-1.5) {$\name$};
}
\foreach \from/\dest in {v/t,t/z} {
  \draw[path four] (p4-\from) -- (p4-\dest);
}
\node[annotation,text=violet!80!black] at (7.6,-1.5) {$P_4$};
\node[annotation,text=violet!80!black] at (14.15,-1.5) {$\lambda_4=1$};
\end{tikzpicture}
\caption{
    Numerical illustration of an acyclic earning flow and its path decomposition. 
    The labels on arcs denote flow amounts, and the signed number inside each vertex is its divergence. 
    The positive divergences sum to $13$.
    On the right, each arc flow equals the sum of the weights of the paths containing that arc.}
\label{fig:cross-flow-bound}
\end{figure}
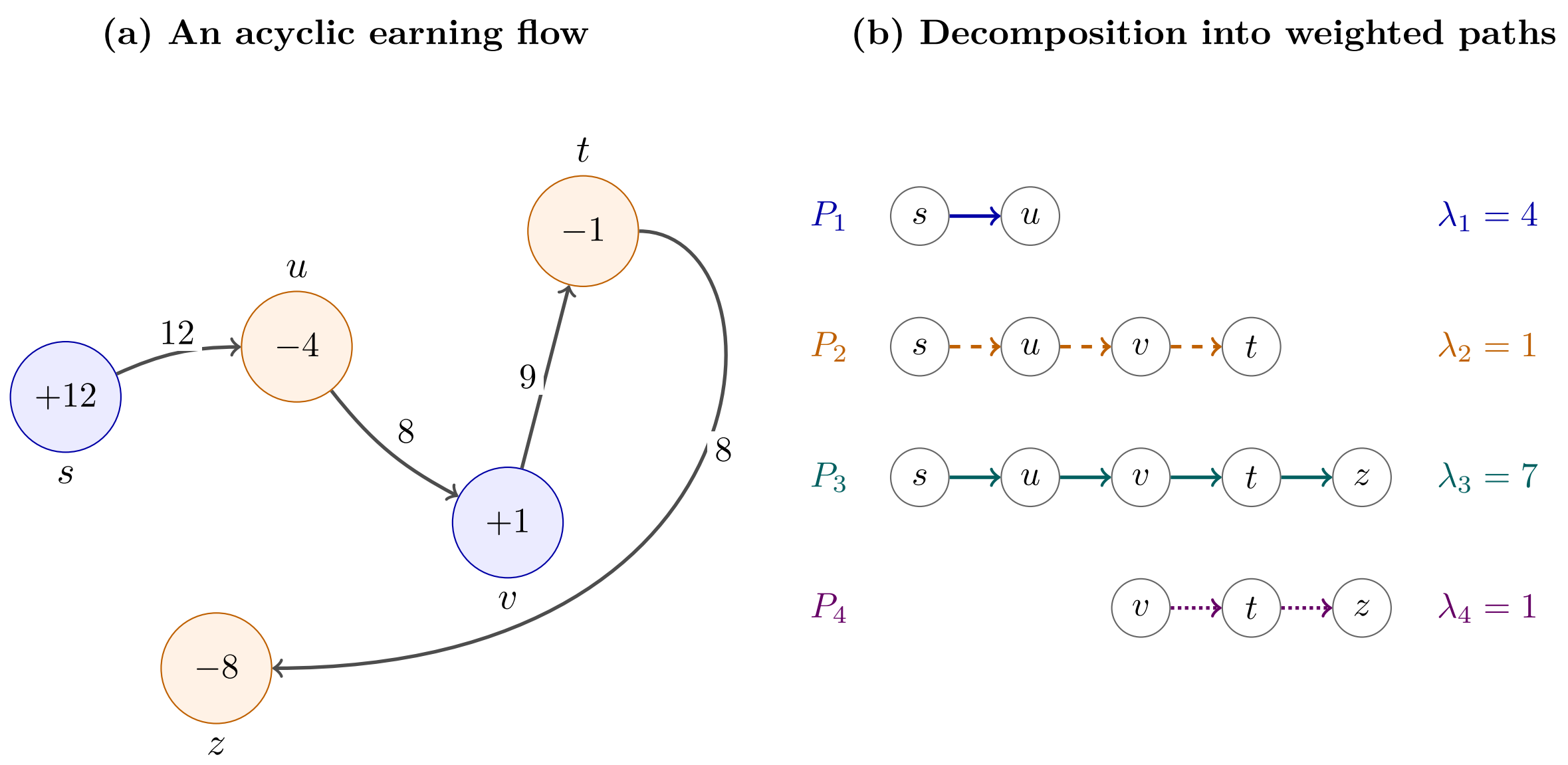

\paragraph{$b_{ii}>0$ for all $i$.}
We define the divergence of the vertex $i$ by the difference between the sum of earning flows from the $i$th agent and the sum of earning flows to the $i$th chore, which is 
\begin{equation*}
    \sum\nolimits_{j=1,\,j\neq i}^n b_{ij} - \sum\nolimits_{j=1,\,j\neq i}^n b_{ji} = w_i - p_i.
\end{equation*}
We say that a vertex is a positive divergence vertex if $\sum\nolimits_{j=1,\,j\neq i}^n b_{ij} - \sum\nolimits_{j=1,\,j\neq i}^n b_{ji} = w_i - p_i > 0$, and negative divergence vertex likewise.
By $|p_i / w_i - 1| \leq 8 \delta \bar{C}$,
the total positive divergence is at most 
\begin{equation*}
    \sum\nolimits_{i=1}^n (w_i - p_i)_+ \leq 8 \delta (\sum\nolimits_{i=1}^n w_i) \bar{C}.
\end{equation*}

A nonnegative acyclic earning flow can be decomposed into  directed paths from positive-divergence vertices to negative-divergence vertices, with total path weight $\sum\nolimits_{i=1}^n (w_i - p_i)_+$.
One can obtain this decomposition by extracting paths starting from positive-divergence vertices until the remaining imbalance is exhausted.
We illustrate such a process in~\cref{fig:cross-flow-bound}.

Since every vertex occurs at most once along a path, each path contributes at most its weight to the outgoing cross-flow at any vertex.
Summing over the paths, we bound this cross-flow by the total path weight, which equals the sum of positive divergences.
Combining this observation with the preceding bound on total positive divergence, we have 
\begin{equation*}
    \label{eq:diagonal}
    \sum_{j=1,\,j\neq i}^n b_{ij} \leq 8\delta (\sum_{i=1}^n w_i) \bar{C} \leq \frac{\min_i w_i}{8} \leq \frac{w_i}8.
\end{equation*}
Hence, $b_{ii} = w_i - \sum_{j=1,\,j\neq i}^n b_{ij} \geq \frac{7}{8} w_i > 0$.
Consequently, by complementary slackness, we have $\beta_i = q_i$.

\paragraph{Extract an exact KKT point of QP.}
By~\cref{lem:source-round}, to compute an exact KKT point of the QP, it suffices to compute an approximate KKT point with sufficiently small error.
Define a primal point $\bz^0$ and 
inequality multipliers $\bg$ by $z_i^0 := \frac{q_i-1}{\delta}$ for all $i \in [n]$ and 
$\gamma_{ij} := \frac{b_{ij}}{\delta}$ for all $i \neq j$.
Since $b_{ij}\ge0$ and $\delta>0$, we have $\bg\ge0$ and
$\supp(\bg)=\{(i,j):i\ne j,\ b_{ij}>0\}$.
By~\eqref{eq:ce-spending}, agent $i$ earns $w_i$ in total and chore $i$
has total payment $p_i$.
Therefore, from flow balance, we obtain
\begin{equation}
    \label{eq:exact-stationarity}
    \begin{aligned}
    (\bA\bg)_i
        &=\frac{\sum_{j=1,\,j\ne i}^nb_{ji}-\sum_{j=1,\,j\ne i}^nb_{ij}}{\delta}
          =\frac{p_i-w_i}{\delta}
          =w_i z_i^0
          \quad\forall\,i\in[n],\\
    \bw^\top\bz^0
        &=\frac{\sum_{i=1}^n p_i-\sum_{i=1}^nw_i}{\delta}=0.
    \end{aligned}
\end{equation}
Thus, the first three conditions in~\eqref{eq:source-approx-kkt} hold exactly with $\ba = \mathbf{0}$.
The bound $|q_i-1|\le8\delta\bar{C}$ also implies
$|z_i^0|\le8\bar{C}$ for all $i\in[n]$.

This certificate $(\bz^0,\bg)$ must also satisfy the approximate feasibility and complementary slackness conditions in \eqref{eq:source-approx-kkt} with $\ba=\mathbf0$.
Since $b_{ii}>0$, we obtain $p_i=d_{ii}\beta_i=w_i\beta_i$ by complementary slackness in~\eqref{eq:ce-spending}.
Hence, $\beta_i=p_i/w_i=q_i$ for all $i\in[n]$.

For each $i\ne j$, the inequality $p_j\le d_{ij}\beta_i$ in \eqref{eq:ce-spending}
holds regardless of whether $b_{ij}$ is positive.
Dividing by $w_j>0$ and substituting the disutilities from~\cref{eq:market-construction},
we obtain
\begin{equation}
    \label{eq:decoded-price-inequalities}
    \begin{aligned}
    q_j=\frac{p_j}{w_j}
        &\le\frac{d_{ij}\beta_i}{w_j}\\
        &=\bigl[1+\delta(C_{ij}+\frac{\Delta_{\rm src}}{16})\bigr]\beta_i\\
        &=\bigl[1+\delta(C_{ij}+\frac{\Delta_{\rm src}}{16})\bigr]q_i
        \quad\forall\,i\ne j.
    \end{aligned}
\end{equation}
For all $(i,j)\in\supp(\bg)$, we have $b_{ij}=\delta\gamma_{ij}>0$.
By \eqref{eq:ce-spending}, we have $b_{ij}(d_{ij}\beta_i-p_j)=0$. Then, we have
$p_j=d_{ij}\beta_i$.
Thus, equality holds in~\cref{eq:decoded-price-inequalities} on every such pair.
By substituting $q_i=1+\delta z_i^0$ into
\cref{eq:decoded-price-inequalities} and dividing by $\delta>0$, we obtain
\begin{equation}
    \label{eq:decoded-difference-inequalities}
    z_j^0-z_i^0-C_{ij}
    \le\frac{\Delta_{\rm src}}{16}+\delta(C_{ij}+\frac{\Delta_{\rm src}}{16})z_i^0
    \quad\forall\,i\ne j,
\end{equation}
again with equality for all $(i,j)\in\supp(\bg)$.
Since $0<\Delta_{\rm src}/16\le1\le\bar{C}$, we have
$|C_{ij}+\Delta_{\rm src}/16|\le2\bar{C}$.
Using this bound together with $|z_i^0|\le8\bar{C}$, we bound the absolute value of the
right-hand side of~\cref{eq:decoded-difference-inequalities} by
\begin{equation*}
    \label{eq:market-error}
    \varepsilon:=\frac{\Delta_{\rm src}}{16}+16\delta\bar{C}^2
    \le\frac{3\Delta_{\rm src}}{16}
    <\Delta_{\rm src},
\end{equation*}
where the inequalities follow from~\cref{eq:market-parameters}.
Since the decoded difference inequality holds for every pair and is an equality on the multiplier support, this error bounds every constraint violation and the absolute slack on that support.
Consequently,
\begin{equation}
    \label{eq:market-approx-feasibility}
    \begin{aligned}
    z_j^0-z_i^0-C_{ij}&\le\varepsilon
        &&\forall\,i\ne j,\\
    |z_j^0-z_i^0-C_{ij}|&\le\varepsilon
        &&\forall\,(i,j)\in\supp(\bg).
    \end{aligned}
\end{equation}
Combining~\cref{eq:exact-stationarity}, $\bg\ge0$, and
\cref{eq:market-approx-feasibility}, we see that $(\bz^0,\bg)$ satisfies
\eqref{eq:source-approx-kkt} for~\eqref{eq:market-source} with
$\varepsilon<\Delta_{\rm src}$.
By~\cref{lem:source-round}, solving the LP~\cref{eq:source-exact-repair} with
$\ba=\mathbf0$ and the fixed support $\supp(\bg)$ produces an exact KKT point
of~\eqref{eq:market-source}.
Also $(\bz^0, \bg)$ is computable from a rational equilibrium
certificate by polynomially many rational arithmetic operations, and all
parameters in~\cref{eq:market-parameters} have polynomial binary encoding
length. Hence, we decode the certificate by solving one polynomial-time LP.

By~\cref{eq:market-construction},
$1-2\delta\bar{C}\le d_{ij}/w_j\le1+2\delta\bar{C}$ for all $i,j$.
By~\cref{eq:market-parameters}, we have $2\delta\bar{C}\le1/4$ and obtain
\[
    D\le B\frac{1+2\delta\bar{C}}{1-2\delta\bar{C}}
      \le\frac53B<2B.
\]
This completes the proof.
\end{proof}

\subsection{Completing the reduction}\label{sec:composition}\label{sec:proof}

We combine the constructions in dependency order, verify that every equilibrium can be decoded, and bound the encoding length and numerical ranges.

\begin{proof}[Proof of \cref{thm:main}]
Membership in $\CLS$ follows from~\cite{chaudhury2022competitive}.
To prove $\CLS$-hardness, we start with a rational instance $(\bQ,\bc)$ of the $\CLS$-complete problem of finding a KKT point of~\eqref{pg:box-qp}~\cite{fearnley2025complexity}.
We construct a linear chores Fisher market as follows.
\begin{enumerate}
    \item We compute
    \begin{equation*}
        \Delta_{\rm box}
        =\gap\!\left(n+1,\htn\bigl(\{0,1\}\cup\{Q_{ij}\}_{i,j}\cup\{c_i\}_i\bigr)\right).
    \end{equation*}
    We choose the smallest power of two $N\ge2$ satisfying~\cref{eq:N-choice} and construct~\eqref{eq:lifted-qp}.
    This QP has $r$ variables and weights $w_i\in\{0,(n+1)L\}$.

    \item We compute $\Delta_{\rm src}$ using~\cref{eq:source-gap}, with the coefficients and bounds from~\eqref{eq:lifted-qp}.
    For these bounds $C_{ij}$, we set $\bar C=\max\{1,\max_{i\ne j}|C_{ij}|\}$.
    We choose $k,\eta$ according to~\cref{eq:k-choice} and set $\bv=\bw+\eta\one$ to obtain~\eqref{eq:C-baregularized}.
    Then, we have
    \begin{equation*}
        v_i\in\{\eta,(n+1)L+\eta\}=\{\eta,2^k\eta\}
        \quad\forall\,i\in[r].
    \end{equation*}

    \item By applying the substitution~\cref{eq:shift}, we remove the linear term from~\eqref{eq:C-baregularized} and obtain~\eqref{eq:pure-source}, with the same coefficients $v_i$ of the terms $-y_i^2/2$.

    \item We recompute $\Delta_{\rm src}$ from~\cref{eq:source-gap} for~\eqref{eq:pure-source}, using the weights $v_i$, the zero linear coefficient vector, and the bounds $C_{ij}+a_i/v_i-a_j/v_j$.
    We also compute
    \begin{equation*}
        \bar C=\max\left\{1,\max_{i\ne j}\left|C_{ij}+\frac{a_i}{v_i}-\frac{a_j}{v_j}\right|\right\}.
    \end{equation*}
    With these values of $\Delta_{\rm src}$ and $\bar C$, we choose $\delta$ according to~\cref{eq:market-parameters}.

    \item We apply~\cref{eq:market-construction} to~\eqref{eq:pure-source}, with earning requirement $v_i$ for agent $i$.
    We divide each earning requirement by $\eta$.
    Then, we have $B_i=v_i/\eta\in\{1,2^k\}$ for all $i\in[r]$.
    This scaling preserves equilibrium allocations and multiplies equilibrium prices by $1/\eta$.
\end{enumerate}

Given any exact equilibrium price vector $\bp$ for the market with earning requirements $v_i/\eta$, we use $\eta\bp$ as prices for the market with earning requirements $v_i$.
If an equilibrium allocation $(x_{ij})$ is not supplied, we compute one by applying~\cref{lem:rational-ce} at prices $\eta\bp$ and earning requirements $v_i$.
Then, we have earning flows $b_{ij}=\eta p_jx_{ij}$ for the market before scaling.
By applying~\cref{prop:market}, we solve~\cref{eq:source-exact-repair} for~\eqref{eq:pure-source} and obtain an exact KKT point together with nonnegative multipliers $\bg$ for its difference inequalities.
By applying~\cref{eq:shift}, we obtain an exact KKT point of~\eqref{eq:C-baregularized}.
By~\cref{lem:translation}, the same multipliers satisfy the KKT conditions of both QPs. We therefore retain $\bg$.
By applying~\cref{lem:regularization} to the KKT point of~\eqref{eq:C-baregularized} and $\bg$, we solve~\cref{eq:source-exact-repair} for~\eqref{eq:lifted-qp} and obtain an exact KKT point of~\eqref{eq:lifted-qp}.

From the variables $z_0,\ldots,z_n$ at this KKT point of~\eqref{eq:lifted-qp}, we define $x_i=z_i-z_0$ for all $i\in[n]$.
By the proof of~\cref{prop:lift}, we have a strong $(M/N)$-KKT point $\bx$ of~\eqref{pg:box-qp}.
By applying~\cref{lem:box-round}, we solve~\cref{eq:box-repair} to obtain an exact KKT point of~\eqref{pg:box-qp}.
Thus, we use one equilibrium computation, two instances of~\cref{eq:source-exact-repair}, and one instance of~\cref{eq:box-repair}.
When only equilibrium prices are returned, we also solve the LP in~\cref{lem:rational-ce} to obtain an allocation.

Let $\mathcal L$ denote the binary encoding length of $(\bQ,\bc)$.
By~\cref{eq:hessian-weights}, the values $L,A_{ij},A_{i0}$ have encoding length polynomial in $\mathcal L$.
This bound also holds for $(n+1)L$.
By~\cref{eq:gap} and~\cref{eq:N-choice}, we have polynomial bounds in $\mathcal L$ on $\log(1/\Delta_{\rm box})$ and $\log N$.
By~\cref{prop:lift}, the QP~\eqref{eq:lifted-qp} has $r=O(n^2\log N)$ variables and encoding length polynomial in $\mathcal L$.

For~\eqref{eq:lifted-qp}, the exponent in~\cref{eq:source-gap} is polynomial in $\mathcal L$.
By~\cref{eq:k-choice}, we have polynomial bounds in $\mathcal L$ on $k$ and the encoding length of $\eta=(n+1)L/(2^k-1)$.
The coefficients $a_i/v_i$ and $C_{ij}+a_i/v_i-a_j/v_j$ have polynomial encoding length by~\cref{lem:translation}.
We apply~\cref{eq:source-gap} to the coefficients of~\eqref{eq:pure-source} and then use~\cref{eq:market-parameters}.
Then, we have polynomial bounds in $\mathcal L$ on the encoding lengths of $\Delta_{\rm src},\bar C,\delta$ for~\eqref{eq:pure-source}.
Therefore, we can compute the rational disutilities $d_{ij}$ and earning requirements $B_i=v_i/\eta$ in polynomial time, and these coefficients have polynomial encoding length.

The agents corresponding to $z_0,\ldots,z_n$ have earning requirement $2^k$, and the agents corresponding to $p_v^{ij,\sigma}$ and $y_e^{ij,\sigma}$ have earning requirement $1$.
Therefore, both earning requirements occur.
Then, we have $B=2^k$ and, by~\cref{prop:market}, $D<2B=2^{k+1}$.
The market has $r$ agents, $r$ unit-supply chores, and positive rational disutilities.
By~\cref{lem:rational-ce}, the market admits an exact equilibrium with a rational certificate of polynomial encoding length.
We can compute an exact KKT point of~\eqref{pg:box-qp} from any equilibrium of this market in polynomial time.
Since finding a KKT point of~\eqref{pg:box-qp} is $\CLS$-complete, we obtain $\CLS$-hardness of exact equilibrium computation.
Together with membership in $\CLS$, we have proved the theorem.
\end{proof}

We also state the following corollary for running-time bounds expressed in terms of $D$ and $B$.
\begin{corollary}
    \label{cor:runtime}
    Unless every problem in $\CLS$ admits a polynomial-time algorithm, there is no algorithm that computes an exact CE for all positive rational linear Fisher chores markets in time
    \begin{equation}\label{eq:desired-runtime}
    \poly\left(n,m,1+\log D,1+\log B\right).
    \end{equation}
\end{corollary}
\begin{proof}[Proof of \cref{cor:runtime}]
Let $\mathcal L$ be the binary encoding length of a Box-QP instance.
For the market constructed from this instance in the proof of~\cref{thm:main}, the dimensions, total encoding length, $k=\log_2 B$, and $\log_2 D<k+1$ are all polynomial in $\mathcal L$.
An algorithm with the bit-operation bound \cref{eq:desired-runtime} would therefore run in time polynomial in $\mathcal L$ on this market.
Composing it with the equilibrium-to-KKT recovery procedure in that proof would solve the $\CLS$-complete box-QP KKT problem in polynomial time.
\end{proof}

\section{CLS-completeness for CES Chores Fisher Markets}
\label{sec:ces}

We use the convex CES disutilities defined in~\cref{sec:disutility-classes}, 
with a common exponent $\rho>1$ and unit supplies.
The coefficients $a_{ij}$ and earning requirements $B_i$ are positive rationals.
For simplicity, we rescale earning requirements to impose $\sum_{i=1}^n B_i=1$.
Then, we have $\sum_{j=1}^m p_j=1$ at equilibrium.
For positive prices $\bp$, agent $i$ has a unique optimal bundle $(x_{ij}(\bp))_j$.
Denote $\sigma:=\rho/(\rho-1)$.
Let $\omega_{ij}(\bp)$ denote the fraction of agent $i$'s earning requirement obtained from chore $j$.
Then, we have 
\begin{equation}\label{eq:ces-demand}
\omega_{ij}(\bp)=\frac{a_{ij}^{1-\sigma}p_j^\sigma}
{\sum_{k=1}^m a_{ik}^{1-\sigma}p_k^\sigma},\quad
x_{ij}(\bp)=\frac{B_i\omega_{ij}(\bp)}{p_j}.
\end{equation}
Since these shares are positive and sum to one for each agent, we have $x_{ij}(\bp)>0$ for all $i,j$ and $\sum_{j=1}^mp_jx_{ij}(\bp)=B_i$.
The total demand for chore $j$ is $g_j(\bp) := \sum_{i=1}^n x_{ij}(\bp)$.
We write $\bgg(\bp):=(g_j(\bp))_{j=1}^m$ for the aggregate-demand vector.
Since each chore has unit supply, we measure market clearing error by $\|\bgg(\bp)-\one\|_\infty$.
Note that $g_j(\bp)$ is not rational in general, even if the market data and prices are rational.

To formulate finite-precision equilibrium price computation, we first define an approximate equilibrium price vector.
\begin{definition}[Approximate equilibrium price vector]\label{def:ces-approximate-prices}
For $\epsilon>0$, we call a positive price vector $\bp$ an
\emph{$\epsilon$-approximate equilibrium price vector} if
\begin{equation*}
    \sum\nolimits_{j=1}^m p_j=1,\qquad \|\bgg(\bp)-\one\|_\infty\le\epsilon.
\end{equation*}
\end{definition}

Since demand values may be irrational even at rational prices, we use rational demand approximations to verify candidate prices.
For a rational exponent $\rho>1$, consider a CES market as above and a rational tolerance $0<\epsilon\le1$.

We define the disutility coefficient ratio $D=\min_{i,j}a_{ij}/\max_{i,j}a_{ij}$.
We consider price vectors whose coordinates sum to one and are each at least $D/(4m)$:
\begin{equation}
    \label{eq:ces-cutoff}
    \Delta_1:=\left\{\bp:p_j\ge\frac{D}{4m},\ \sum\nolimits_{j=1}^mp_j=1\right\}.
\end{equation}
We show in~\cref{sec:ces-membership} that every exact equilibrium has $p_j\ge D/m$. Thus, this price restriction excludes no exact equilibria.
To verify candidate prices using rational arithmetic, we fix a deterministic demand-evaluation algorithm.
Given the market data, the tolerance $\epsilon$, and rational prices $\bp\in\Delta_1$, 
the algorithm evaluates~\cref{eq:ces-demand} through logarithms and exponentials and returns a rational vector $\widehat{\bgg}(\bp)$ with
\begin{equation*}
\|\widehat{\bgg}(\bp)-\bgg(\bp)\|_\infty\le\frac{\epsilon}{4}
\quad\text{for }\bp\in\Delta_1.
\end{equation*}
We prove in~\cref{sec:ces-membership} that this evaluation can be performed in polynomial time.

With this price domain and evaluation algorithm, we define the following search problem.
\begin{definition}[Finite-precision equilibrium price computation]
    \label{def:ces-price-problem}
    An instance consists of a CES market as above and a rational tolerance $0<\epsilon\le1$.
    We encode these rational data in binary.
    Given these data, we seek a rational price vector $\bp\in\Delta_1$ satisfying
    \begin{equation}
        \label{eq:ces-certified-clearing}
        \|\widehat{\bgg}(\bp)-\one\|_\infty\le\frac{3\epsilon}{4}.
        \tag{Rational Approx-CE}
    \end{equation}
    We require the total number of bits used to write $\bp$ to be at most a fixed polynomial in the number of bits used to describe the market and $\epsilon$.
\end{definition}

Here we count the bits used for the numerators and denominators of the rational prices.
In~\cref{sec:ces-membership}, we prove that every instance has a solution within such a polynomial size bound.
We choose the fixed polynomial large enough for that construction, independently of the particular instance.

If a rational vector $\bp\in\Delta_1$ satisfies~\eqref{eq:ces-certified-clearing}, then the triangle inequality gives
\[
\|\bgg(\bp)-\one\|_\infty
\le\|\bgg(\bp)-\widehat{\bgg}(\bp)\|_\infty
 +\|\widehat{\bgg}(\bp)-\one\|_\infty
\le\frac{\epsilon}{4}+\frac{3\epsilon}{4}=\epsilon.
\]
Thus, every solution to finite-precision equilibrium price computation is an $\epsilon$-approximate equilibrium price vector.
Conversely, for any rational $\epsilon/2$-approximate equilibrium price vector $\bp\in\Delta_1$, we have
\[
\|\widehat{\bgg}(\bp)-\one\|_\infty
\le\|\widehat{\bgg}(\bp)-\bgg(\bp)\|_\infty
 +\|\bgg(\bp)-\one\|_\infty
\le\frac{\epsilon}{4}+\frac{\epsilon}{2}=\frac{3\epsilon}{4}.
\]
Hence, this vector satisfies~\eqref{eq:ces-certified-clearing}.

We now state the complexity classification for finite-precision equilibrium price computation.
\begin{theorem}
    \label{thm:ces}
    For each fixed rational exponent $\rho>1$, finite-precision equilibrium price computation in~\cref{def:ces-price-problem} is $\CLS$-complete.
\end{theorem}

\begin{remark}[Rational certificates and exact equilibria]\label{rem:ces-rational-exact}
For finite-precision equilibrium price computation in~\cref{def:ces-price-problem}, we use normalized rational prices to certify a market clearing error of at most $\epsilon$.
Exact equilibrium prices give zero market clearing error and may be irrational even for rational market data and $\rho=2$.
We give an example and a fixed-point representation for exact equilibrium computation in~\cref{app:ces-exact}.
\end{remark}

\subsection{Membership for all fixed convex CES exponents}\label{sec:ces-membership}

We use a price potential function to express aggregate demand as a gradient.
For positive prices $\bp$, we define
\begin{equation}\label{eq:ces-potential}
\Phi(\bp):=\sum_{i=1}^n\frac{B_i}{\sigma}
\log\left(\sum_{j=1}^m a_{ij}^{1-\sigma}p_j^\sigma\right).
\end{equation}
Then, we have $\nabla\Phi=\bgg$ and $\bp^\top \bgg(\bp)=1$.
The function $\Phi$ is the CES specialization of the equilibrium potential studied in~\cite{chaudhury2026tatonnement,tao2026fisher}.

At an equilibrium, 
the gradient of $\Phi$ equals $\one$, 
which is orthogonal to every direction whose coordinates sum to zero.
Thus, equilibrium prices are stationary for $\Phi$ along directions that preserve the price normalization.
To keep gradient steps within $\Delta_1$, we use Euclidean projection, denoted by $\operatorname{proj}_{\Delta_1}$: it sends a vector to its nearest point in $\Delta_1$.

We choose a step size for which the potential decreases whenever the projected step changes the prices.
We then use continuous local search to obtain a small projected step, and use demand ratios to show that the resulting prices have small market clearing error.
Finally, we control numerical evaluation and rounding errors to ensure that the returned rational prices satisfy~\cref{def:ces-price-problem}.

In the following theorem, we establish membership and the existence of solutions with the required encoding length.
\begin{theorem}[CES membership]\label{thm:ces-membership}
For each fixed rational exponent $\rho>1$, finite-precision equilibrium price computation in~\cref{def:ces-price-problem} belongs to $\CLS$.
The same holds for the variant in which the rational exponent $\rho>1$ is supplied as part of the instance and encoded in binary.
In both cases, every instance has a rational solution whose binary encoding length is polynomial in that of the instance.
\end{theorem}

\begin{proof}
Let $\mathcal L$ denote the binary encoding length of the market data and $\epsilon$, including $\rho$ when it is supplied as part of the instance.

\paragraph{Price bounds and projected descent.}
We use $\sigma-1=1/(\rho-1)>0$.
By~\cref{eq:ces-demand}, we have
\begin{equation}\label{eq:ces-small-price-demand}
\frac{x_{ij}(\bp)}{x_{ik}(\bp)}
=\left(\frac{p_j a_{ik}}{p_k a_{ij}}\right)^{\sigma-1}.
\end{equation}
For any equilibrium price vector $\bp$, we therefore have
\begin{equation*}\label{eq:ces-price-ratios}
\min_i\frac{a_{ij}}{a_{ik}}
\le\frac{p_j}{p_k}\le
\max_i\frac{a_{ij}}{a_{ik}}.
\end{equation*}
Indeed, if $p_j/p_k>\max_i a_{ij}/a_{ik}$, we have $x_{ij}(\bp)>x_{ik}(\bp)$ for all $i$, which contradicts $g_j(\bp)=g_k(\bp)=1$.
By interchanging $j,k$ in the same argument, we obtain the lower bound on $p_j/p_k$.
Thus, we have $p_j\ge D/m$ for all $j$ at any equilibrium, independently of $\rho$.

For positive prices with $\sum_{l=1}^m p_l=1$, if $p_j\le D/(2m)$ and $p_k=\max_l p_l$, then $p_k\ge1/m$ and $a_{ik}/a_{ij}\le1/D$.
By~\cref{eq:ces-small-price-demand}, we have $x_{ij}(\bp)/x_{ik}(\bp)\le2^{1-\sigma}$ for every agent, and hence $g_j(\bp)\le2^{1-\sigma}g_k(\bp)$.
We use the following lower bound on how far this demand-ratio factor is from one:
\begin{equation}\label{eq:ces-boundary-gap}
1-2^{1-\sigma}\ge\frac{1}{2}\min\{\sigma-1,1\}
\ge\frac{\sigma-1}{2\sigma}=\frac1{2\rho}.
\end{equation}
The function $s\mapsto1-2^{-s}$ is concave on $[0,1]$ and increasing on $[1,\infty)$.
From these properties, we obtain $1-2^{1-\sigma}\ge\frac12\min\{\sigma-1,1\}$.

For $\bp\in \Delta_1$, we compute the Hessian of the term $(B_i/\sigma)\log\sum_{j=1}^m a_{ij}^{1-\sigma}p_j^\sigma$ in~\cref{eq:ces-potential} as
\begin{equation*}
B_i\left[(\sigma-1)\diag\left(\frac{\omega_{ij}}{p_j^2}\right)
-\sigma\left(\frac{\omega_{ij}}{p_j}\right)_j
\left(\frac{\omega_{ij}}{p_j}\right)_j^{\!\top}\right].
\end{equation*}
Therefore, $\bgg$ is $L$-Lipschitz in Euclidean norm for $L:=32\sigma m^2/D^2$, and $\Phi$ is $(4m^2/D)$-Lipschitz.
These Lipschitz bounds have polynomial binary length.

We take a projected gradient step of size $1/L$ and choose a displacement tolerance $h$:
\begin{equation*}
\bG(\bp):=\operatorname{proj}_{\Delta_1}\left(\bp-\frac{\bgg(\bp)}L\right),\quad
h:=\min\left\{\frac{D}{4m},\frac{1}{8\rho L},\frac{\epsilon}{8L}\right\}.
\end{equation*}
Euclidean projection does not increase distances between vectors.
We can compute this projection in polynomial time by sorting the coordinates and finding a common shift such that, after replacing values below $D/(4m)$ by $D/(4m)$, the coordinates sum to one.
In particular, $\bG$ is $2$-Lipschitz in Euclidean norm.
By the descent lemma and the optimality conditions for $\operatorname{proj}_{\Delta_1}$, we have
\begin{equation}\label{eq:ces-descent}
\Phi(\bp)-\Phi(\bG(\bp))\ge\frac L2\|\bG(\bp)-\bp\|_2^2.
\end{equation}

\paragraph{Small steps and market clearing error.}
We now show that a small projected gradient step guarantees small market clearing error:
\begin{equation*}
\|\bG(\bp)-\bp\|_\infty\le h
\quad\Longrightarrow\quad
\|\bgg(\bp)-\one\|_\infty\le\frac{\epsilon}{4}.
\end{equation*}
By the optimality conditions for $\operatorname{proj}_{\Delta_1}$, there exists a multiplier $\lambda$ such that
\[
G_j(\bp)=\max\left\{\frac{D}{4m},p_j-\frac{g_j(\bp)-\lambda}{L}\right\}.
\]
For all $j$ with $G_j(\bp)>D/(4m)$, we have $|g_j(\bp)-\lambda|\le Lh$.
We choose $k$ such that $p_k=\max_jp_j$.
Then, we have $p_k\ge1/m$ and $G_k(\bp)\ge1/m-D/(4m)>D/(4m)$.

Suppose that $G_j(\bp)=D/(4m)$ for some $j$.
Then, we have $p_j\le D/(4m)+h\le D/(2m)$, and from the formula for $G_j(\bp)$ we obtain $\lambda\le g_j(\bp)$.
By~\cref{eq:ces-small-price-demand,eq:ces-boundary-gap}, we have
\[
\lambda\le g_j\le2^{1-\sigma}g_k
\le\left(1-\frac{1}{2\rho}\right)(\lambda+Lh).
\]
Therefore, we have $\lambda+Lh\le 2\rho Lh\le1/4$.
For all $j$ with $G_j(\bp)>D/(4m)$, we have $g_j\le\lambda+Lh$, and for all $j$ with $G_j(\bp)=D/(4m)$, we have $g_j\le(1-1/(2\rho))g_k\le\lambda+Lh$.
The inequalities $g_j\le1/4$ for all $j$ contradict $\bp^\top \bgg=1$ and $\sum_{j=1}^mp_j=1$.
Thus, we have $G_j(\bp)>D/(4m)$ for all $j$.

Since $\sum_{j=1}^mp_j=1$ and $\bp^\top\bgg=1$, the price-weighted average of $g_j-\lambda$ is $1-\lambda$.
Since each term has magnitude at most $Lh$, this average satisfies $|\lambda-1|\le Lh$.
By the triangle inequality, we then have
\begin{equation*}
|g_j-1|\le2Lh\le\min\left\{\frac{\epsilon}{4},\frac{1}{4\rho}\right\}
\quad\text{for all }j.
\end{equation*}

\paragraph{Numerical evaluation and continuous local search.}
We evaluate~\cref{eq:ces-demand} by defining
\[
z_{ij}:=\sigma\log p_j-(\sigma-1)\log a_{ij}.
\]
For each $i$, we subtract $\max_jz_{ij}$ from $z_{ij}$ before computing the exponentials.
We discard terms whose total contribution is below the requested error tolerance.
The remaining exponent arguments have magnitude polynomial in the input length and the number of requested accuracy bits.
We compute the logarithms with additional precision depending on $\log \sigma$, $\log(1/D)$, and $\log m$.
Since $p_j\ge D/(4m)$, we can therefore approximate $\bgg$, $\Phi$, and $\bG$ in time polynomial in the input length and the number of requested accuracy bits, even when $\sigma$ is numerically large.
In particular, we can verify~\eqref{eq:ces-certified-clearing} in polynomial time on $\Delta_1$.

Continuous local search on rational convex polytopes belongs to $\CLS$ by~\cite[Proposition~5.6 and Theorem~6.1]{fearnley2022complexity}.
We use $Lh^2/8$ as the threshold for potential decrease.
This threshold and its reciprocal have encoding length polynomial in $\mathcal L$.

By~\cite[Theorem~E.2 and Remark~4]{fearnley2022complexity}, we can construct polynomial-size linear arithmetic circuits using rational affine operations and min/max gates.
We approximate $\Phi$ uniformly within $Lh^2/32$ and $\bG$ within $DLh^2/(64m^3)$ in each coordinate.

We project the circuit approximation of $\bG$ back to $\Delta_1$.
Before projection, the Euclidean error is at most the sum of the coordinate errors, hence at most $DLh^2/(64m^2)$.
Since $\bG$ takes values in $\Delta_1$ and Euclidean projection is nonexpansive, the same bound holds after projection.
The sorting procedure described above gives a polynomial-size linear arithmetic circuit for this projection.

To use the domain $[0,1]^m$, we compose both circuit approximations with $\operatorname{proj}_{\Delta_1}$.
For any $\bz\in[0,1]^m$, we define $\bp:=\operatorname{proj}_{\Delta_1}(\bz)$.
Both circuits depend on $\bz$ only through $\bp$, and the circuit approximating $\bG$ takes values in $\Delta_1$.

We use a rational affine normalization to place the potential approximation in $[0,1]$ and scale the potential-decrease tolerance by the same factor.
This normalization has polynomial encoding length because $|\Phi|$ has a polynomial-bit bound and the approximation error is at most $Lh^2/32$.

We compute valid Lipschitz bounds from the linear circuits.
In the continuous local-search formulation of~\cite[Definition~5]{fearnley2022complexity}, called \emph{General-Continuous-Localopt}, a solution either gives a point with small potential decrease or exhibits a violation of a supplied Lipschitz bound.
Since we supply valid bounds, only the small-decrease alternative is possible.

Since $\Phi$ is $(4m^2/D)$-Lipschitz, the error in the projected approximation of $\bG$ changes the potential by at most $Lh^2/16$.
For a point returned with potential decrease at most $Lh^2/8$ before normalization, we therefore have
\[
\Phi(\bp)-\Phi(\bG(\bp))
\le\frac{Lh^2}{8}+2\left(\frac{Lh^2}{32}\right)+\frac{Lh^2}{16}=\frac{Lh^2}{4}.
\]
By~\cref{eq:ces-descent}, the potential decrease is at least $(L/2)\|\bG(\bp)-\bp\|_2^2$.
Since it is also at most $Lh^2/4$, we obtain $\|\bG(\bp)-\bp\|_2\le h$.
We can therefore apply the small-displacement implication proved above to obtain $\|\bgg(\bp)-\one\|_\infty\le\min\{\epsilon/4,1/(4\rho)\}$.

\paragraph{Rounding the prices.}
To bound the encoding length of the prices, we choose the smallest integer $k$ with
\begin{equation*}
2^k\ge\max\left\{\frac{8m^2}{D},\frac{8m^2L}{\epsilon}\right\}.
\end{equation*}
We round $p_1,\ldots,p_{m-1}$ down to multiples of $2^{-k}$ and choose the last price to keep the sum equal to one.

Before this rounding, we have $p_j>D/(2m)$ for all $j$.
Indeed, if $p_j\le D/(2m)$, we compare $p_j$ with a maximal price in~\cref{eq:ces-small-price-demand} and obtain
$g_j\le(1-1/(2\rho))(1+1/(4\rho))<1-1/(4\rho)$, contrary to $|g_j-1|\le1/(4\rho)$.

The rounded prices remain in $\Delta_1$ and differ from $\bp$ by at most $m^2 2^{-k}$ in Euclidean norm.
Since $\bgg$ is $L$-Lipschitz, rounding increases the market clearing error by at most $\epsilon/8$. Therefore, the rounded prices have market clearing error at most $\epsilon/4+\epsilon/8$.
Since the evaluation error is at most $\epsilon/4$, the evaluated residual is at most $5\epsilon/8<3\epsilon/4$.
Thus, the rounded prices satisfy~\eqref{eq:ces-certified-clearing} and have encoding length $O(mk)$.
We have thus proved membership in $\CLS$ and the existence of accepted outputs of polynomial encoding length.
\end{proof}

\subsection{Hardness for all fixed convex CES exponents}\label{sec:ces-fixed-proof}

\subsubsection{Proof overview}\label{sec:ces-proof-overview}

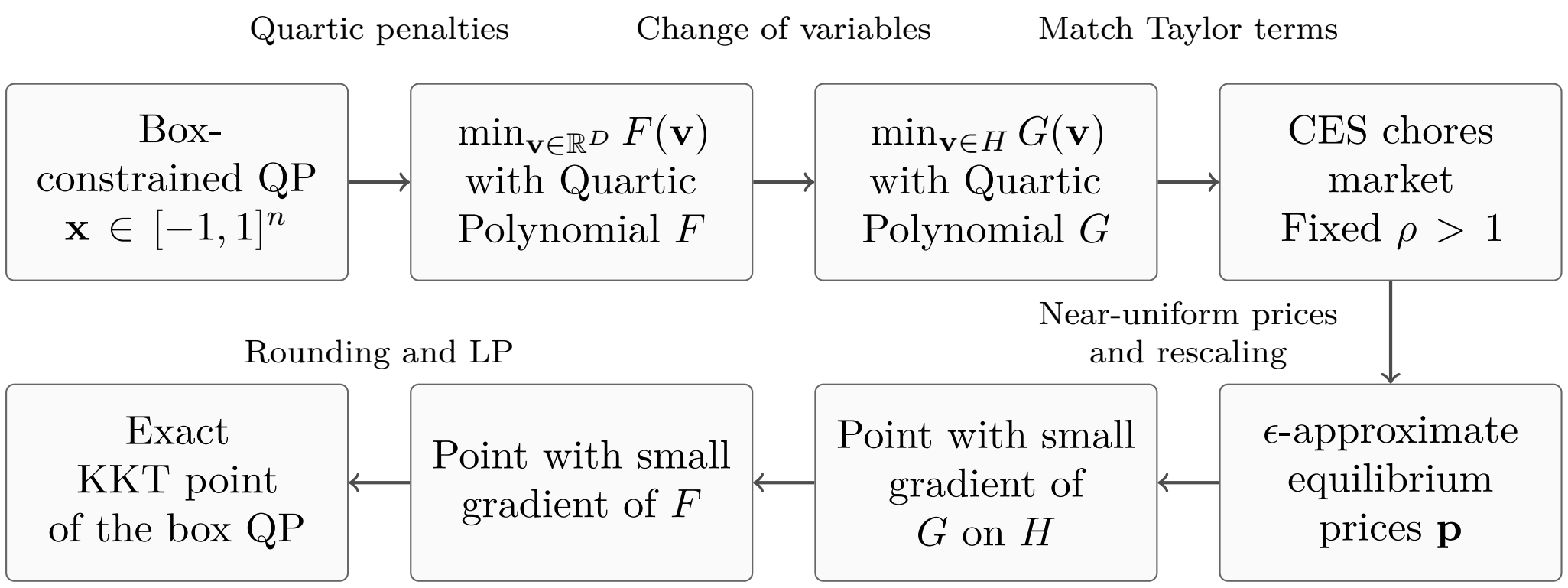
\begin{figure}[H]
\centering
\begin{tikzpicture}[
  stage/.style={draw=black!60,rounded corners=2pt,fill=black!2,
    text width=2.7cm,minimum height=17mm,align=center,font=\small},
  reductionarrow/.style={->,draw=black!70,line width=0.8pt},
  annotation/.style={font=\scriptsize,align=center,fill=white,inner sep=2pt}
]
\node[stage] (box) at (0,0)
  {Box-constrained QP\\$\bx\in[-1,1]^n$};
\node[stage] (quartic) at (3.5,0)
  {$\min_{\bv \in \mathbb{R}^D} F(\bv)$ \\ with Quartic Polynomial $F$};
\node[stage] (model) at (7,0)
  {$\min_{\bv \in H} G(\bv)$ \\ with Quartic Polynomial $G$};
\node[stage] (market) at (10.5,0)
  {CES chores market\\Fixed $\rho>1$};
\draw[reductionarrow] (box) -- node[annotation,above,yshift=11mm]
  {Quartic penalties} (quartic);
\draw[reductionarrow] (quartic) -- node[annotation,above,yshift=11mm]
  {Change of variables} (model);
\draw[reductionarrow] (model) -- node[annotation,above,yshift=11mm]
  {Match Taylor terms} (market);
\node[stage] (kkt) at (0,-2.6)
  {Exact KKT point\\of the box QP};
\node[stage] (stationary) at (3.5,-2.6)
  {Point with small\\gradient of $F$};
\node[stage] (modelstationary) at (7,-2.6)
  {Point with small\\gradient of $G$ on $H$};
\node[stage] (prices) at (10.5,-2.6)
  {$\epsilon$-approximate\\equilibrium prices $\bp$};
\draw[reductionarrow] (market) -- (prices);
\draw[reductionarrow] (prices) -- node[annotation,above,yshift=9mm]
  {Near-uniform prices\\and rescaling} (modelstationary);
\draw[reductionarrow] (modelstationary) -- (stationary);
\draw[reductionarrow] (stationary) -- node[annotation,above,yshift=9mm]
  {Rounding and LP} (kkt);
\end{tikzpicture}
\caption{Proof strategy for fixed-exponent CES hardness, with $H:=\{\by:\sum_{j=1}^my_j=0\}$.
We construct instances along the top row and decode every accepted price output along the bottom row.
}
\label{fig:ces-proof-strategy}
\end{figure}

We prove $\CLS$-hardness of finite-precision equilibrium price computation in~\cref{def:ces-price-problem} by a reduction from Box-constrained QP to a CES chores Fisher market with any fixed rational exponent $\rho>1$.
To do so, we follow the sequence of polynomial-time instance reductions illustrated in~\cref{fig:ces-proof-strategy}.

First, we construct a quartic polynomial $F$ such that, from a point where its gradient has sufficiently small infinity norm, we can compute an exact KKT point of the original Box QP.
We then express this stationarity problem using a quartic polynomial $G$ whose variables sum to zero. 
We can therefore use these variables to represent deviations of normalized prices from the uniform vector $\one/m$.
For each such polynomial $G$, 
we construct a CES market whose price potential function approximates $G$ \emph{near uniform prices}.

For prices satisfying $\sum_{j=1}^m p_j = 1$, we write $\bp=(\one+h\by)/m$, where $h>0$ is a small scale and $\by$ records the scaled price deviations, with $\sum_{j=1}^my_j=0$.
Before we round the disutility coefficients to rationals, the potential satisfies a fourth-order expansion of the form
\[
 \Phi\left(\frac{\one+h\by}{m}\right)
 =C_h+h^4G(\by)+O(h^5),
\]
where $C_h$ is independent of $\by$, and the approximation also holds for gradients when $\by$ is bounded.
We show that every approximate equilibrium price vector with sufficiently small error lies close enough to uniform prices for this expansion to apply.
Since the gradient of the price potential function is aggregate demand, small market clearing error means that this gradient is close to $\one$.
The vector $\one$ is orthogonal to every direction whose coordinates sum to zero.
Thus, the potential has small derivatives along these directions.
After rescaling and using the Taylor approximation, we obtain an approximate stationary point of $G$ and hence of $F$.
We can therefore decode every accepted price output to recover an exact KKT point of the original Box QP by rounding and solving a rational linear program.
The market data and tolerances have polynomial encoding length, and all construction and decoding steps take polynomial time.
We carry out these steps in the same order below, stating the technical results where they are needed.
For the simplicity of the presentation, we defer some proofs to~\cref{app:ces-hardness}.

\subsubsection{From Box-constrained QP to Unconstrained Quartic Program}
\label{ces2src:section}

We first replace the box constraints by quartic penalties to express approximate stationarity using the gradient of an unconstrained polynomial.
An affine change of variables maps $[0,1]^n$ to $[-1,1]^n$ and preserves KKT points.
We therefore consider a rational quadratic $f(\bx):=\frac12\bx^\top\bQ\bx+\bc_0^\top\bx$ with symmetric $\bQ$ and the problem
\begin{equation}\label{ces2src:box-qp}
 \min_{\bx\in[-1,1]^n} f(\bx).
 \tag{Auxiliary-Box-QP}
\end{equation}

For each coordinate $x_i$, we introduce a chain of $k$ auxiliary variables and penalize deviations from successive squarings.
A final quartic penalty on each chain keeps approximate stationary points close enough to the box for the sake of recovering KKT points reversely.
The resulting homogeneous degree-four part is a positive multiple of the sum of fourth powers.
To match the quartic term of a EG dual potential function, we also need a positive multiple of the fourth power of the Euclidean norm.
We add one further variable to obtain this radial term while keeping its effect on the derivatives of the original variables small.
Thus, the complete variable vector $\bw$ has dimension $D := n(k+1)+1$.
We write the additional term as $c\|\bw\|_2^4$ and call $c$ the \emph{radial coefficient}.

In the following proposition, we make this reduction precise and allow the radial coefficient to be chosen after the chain parameters.
The proof of~\cref{ces2src:hardness} is in~\cref{app:ces-quartic-source}.
\begin{proposition}
    \label{ces2src:hardness}
    From $f$, we can compute in polynomial time a positive rational penalty scale $M$ and an integer chain length $k\ge1$.
    For any positive rational radial coefficient $c$ of polynomial encoding length, we can then construct in polynomial time a quartic polynomial $F:\R^D\to\R$ and a positive rational tolerance $\eta$ of polynomial encoding length, 
    where the homogeneous degree-four part of $F$ is
    \begin{equation}
        \label{ces2src:quartic-form}
        M\sum\nolimits_{j=1}^D w_j^4+c\left(\sum\nolimits_{j=1}^D w_j^2\right)^2.
    \end{equation}
    From any rational point $\bw$ with $\|\nabla F(\bw)\|_\infty\le\eta$, we can compute an exact KKT point of~\eqref{ces2src:box-qp} in polynomial time.
\end{proposition}

\subsubsection{From Unconstrained Quartic Program to Zero-sum-constrained Quartic Program}
\label{cesq:paired-model}

We next represent the variables of the quartic source by deviations of normalized prices from the uniform vector.
For $m$ chores, these deviations lie in the subspace
\begin{equation}
    H := \{\by\in\R^m:\one^\top\by=0\}.
    \label{eq:def-H}
\end{equation}
We write $\nabla_H$ for the gradient within $H$, or equivalently the orthogonal projection of a gradient onto $H$.
We represent each variable of the source polynomial by the difference of two coordinates to avoid restricting it by the zero-sum constraint.

Choose the source parameters $M,k,D$ as in~\cref{ces2src:hardness} and set $m:=2D$.
For positive rational coefficients $A,B$, choose the radial coefficient $c:=2MB/A$ and complete the construction of $F$.
We denote its variables here by $\bv\in\R^D$ and rescale the polynomial to $\widetilde F := \frac{2A}{M} F$.
The homogeneous quartic part of the function $\widetilde{F}$ is then
\[
    2A\sum\nolimits_{i=1}^D v_i^4+4B\left(\sum\nolimits_{i=1}^D v_i^2\right)^2.
\]
We introduce $\bs\in\R^D$ and define the paired coordinates by
\begin{equation}
    \label{cesq:paired-coordinates}
    y_{2i-1}:=s_i+v_i,\qquad y_{2i}:=s_i-v_i
    \quad\forall\,i\in[D].
\end{equation}
Accordingly, $s_i = \frac{y_{2i - 1} + y_{2i}}{2}$ and $v_i = \frac{y_{2i-1}-y_{2i}}{2}$ for all $i \in [D]$.
Thus, we have $\by \in H$ if and only if $\sum\nolimits_{i=1}^D s_i=0$.
To obtain the quartic part required for the CES construction, we define
\begin{equation}
    \label{cesq:quartic-model}
    G(\bs,\bv) := \widetilde F(\bv)+\|\bs\|_2^2 + 2A\sum\nolimits_{i=1}^D(s_i^4+6s_i^2v_i^2) + 8B\|\bs\|_2^2\|\bv\|_2^2+4B\|\bs\|_2^4.
\end{equation}
Through~\cref{cesq:paired-coordinates}, we also regard $G$ as a polynomial in $\by$.
The degree-four part of $G$ is
\begin{equation}
    Q_4 := A\sum\nolimits_{j=1}^m y_j^4+B\left(\sum\nolimits_{j=1}^m y_j^2\right)^2.
    \label{eq:q4-g}
\end{equation}
Since their derivative in the direction $\bs$ is at least $2\|\bs\|_2^2$, a small gradient within $H$ forces $\bs$ to be small.
The resulting error in the gradient with respect to $\bv$ is quadratic in $\|\bs\|_2$.

For a radius $R\ge1$ and the source tolerance $\eta>0$, choose a positive rational
\begin{equation}\label{cesq:paired-tolerance}
 \xi\le\min\left\{1,
 \frac{A\eta}{100M[1+(24A+16B)R]}\right\}.
\end{equation}
In the following lemma, we quantify the stationarity implication needed for decoding.
We provide its proof in~\cref{app:ces-paired-decoding}.

\begin{lemma}
    \label{cesq:paired-decoding}
    If $\by\in H$ where $H$ is defined in~\cref{eq:def-H}, $\|\by\|_2\le R$, and $\|\nabla_HG(\by)\|_2\le\xi$, 
    then the coordinates
    $v_i = \frac{y_{2i-1}-y_{2i}}{2}$ satisfy $\|\nabla F(\bv)\|_\infty\le\eta$.
\end{lemma}

Let $P_d$ denote the rational homogeneous degree-$d$ part of $G$ for $d=1,2,3$.
After omitting the constant term, we obtain the quartic program
\begin{equation}
    \label{cesq:quartic-program}
    \begin{aligned}
        \min_{\by\in\R^m}\quad
        &G(\by) = A\sum\nolimits_{j=1}^m y_j^4 + B\left(\sum\nolimits_{j=1}^m y_j^2\right)^2
            +P_1(\by)+P_2(\by)+P_3(\by)\\
        \text{s.t.}\quad
        &\sum\nolimits_{j=1}^m y_j=0.
    \end{aligned}
    \tag{Zero-sum-constrained Quartic Program}
\end{equation}
By~\cref{cesq:paired-decoding,ces2src:hardness}, from any rational feasible point $\by$ satisfying $\|\by\|_2\le R$ and $\|\nabla_HG(\by)\|_2\le\xi$, we can compute an exact KKT point of~\eqref{ces2src:box-qp} in polynomial time.

\subsubsection{From Zero-sum-constrained Quartic Program to CES chores Fisher market}
\label{cesq:market-realization}

We now construct a CES market for which approximate equilibrium prices yield approximate stationary points of~\eqref{cesq:quartic-program}.
We highlight that, 
for a fixed rational $\sigma>1$, 
the quartic coefficients $A,B$ in this construction depend only on $\sigma$ and the number of chores $m$.
In our procedure, 
we first choose $A, B$ before computing the quartic polynomial in the unconstrained optimization problem.
For any $\sigma > 1$, we construct the corresponding~\eqref{cesq:quartic-program} with the computed $A, B$. 
From the~\eqref{cesq:quartic-program} instance, 
we construct the following CES chores Fisher market instance with $m$ chores and $O(m^3)$ agents.

In the constructed CES chores Fisher market, each chore has unit supply and each agent has common CES exponent $\rho=\sigma/(\sigma-1)$.
Agent $i$ has a positive rational earning requirement $B_i$, with $\sum_i B_i=1$. 
We specify agent $i$'s preferences over the $m$ chores by a vector of positive parameters $\bw_i:=(w_{i1},\ldots,w_{im})$. 
We set $\bw_i$ either to the uniform vector $\one/m$ or to a convex combination of $\one/m$ and one vector that assigns 
(1) its entire weight to one chore, 
(2) equal weights to two chores, 
(3) weights in the ratio $2:1$ to an ordered pair of chores, or 
(4) equal weights to three chores.
We then set disutility coefficients by rounding $w_{ij}^{-1/(\sigma-1)}$ to positive rationals.
We encode $P_1,P_2,P_3$ by making small changes to positive earning requirements that are initially equal within each family, while keeping their sum equal to one.

In the following proposition, we construct the market and obtain both the price localization and gradient bounds needed for decoding.
The proof is in~\cref{app:ces-baseline,app:ces-realization}.

\begin{proposition}
    \label{cesg:prop:embedding}
    For fixed rational $\sigma>1$ and $m\ge8$, 
    we can compute positive rational $A,B$ depending only on $\sigma,m$ in polynomial time.
    Given a rational quartic polynomial $G: H\to\R$ of the form in~\eqref{cesq:quartic-program} with these coefficients $A,B$, 
    we can compute a rational radius $R_* \geq 1$ independently of a target gradient tolerance $\xi$.
    
    Then, for any positive rational $\xi$, we can construct a CES chores Fisher market with exponent $\rho=\sigma/(\sigma-1)$, with positive rational data, such that, together with rational $h,\varepsilon>0$, every normalized positive $\varepsilon$-approximate equilibrium price vector $\bp$ satisfies
    \begin{equation}
        \label{cesq:rational-realization}
        \by := \frac{m\bp-\one}{h}\in H,\quad
        \|\by\|_2 \leq R_*,\quad \|\nabla_HG(\by)\|_2 \leq \xi.
    \end{equation}
    Furthermore, the construction takes polynomial time with polynomial binary encoding length.
\end{proposition}

To prove the connection between the stationarity conditions, we use Taylor expansion at uniform prices.
When we construct the market instance, we change earning requirements at scales $h^3,h^2,h$ to prescribe the terms $P_1,P_2,P_3$, respectively.
Before rounding disutility coefficients to rationals, we obtain
\begin{equation*}
    \Phi\left(\frac{\one+h\by}{m}\right)
    =C_h+h^4G(\by)+O(h^5),
\end{equation*}
where $C_h$ is independent of $\by$, and the expansion holds on bounded subsets of $H$.
Similarly, the gradient of $\Phi((\mathbf1+h\mathbf y)/m)$ with respect to $\mathbf y$ within bounded subsets of $H$ is approximated by $h^4 \nabla_H G(\mathbf y)$ with an $O(h^5)$ error in Euclidean norm.

We then show that every normalized positive price vector with sufficiently small market clearing error satisfies $\| m\bp - \one \|_2 \leq hR_*$.
Thus, by applying the above expansion and using $\nabla\Phi(\bp)=\bgg(\bp)$ and $\operatorname{proj}_H\one=\mathbf0$, we have 
\begin{equation}
    \|\nabla_HG(\by)\|_2
    \le \frac{\|\operatorname{proj}_H(\bgg(\bp)-\one)\|_2}{mh^3}+O(h)
    \le \frac{\varepsilon}{h^3}+O(h), 
    \label{eq:dual-EG-obj-approx-equivalent-polynomial}
\end{equation}
whenever $\|\bgg(\bp) - \one\|_\infty \leq \varepsilon$.
We can choose $h$ to control the error terms in~\cref{eq:dual-EG-obj-approx-equivalent-polynomial} and then $\varepsilon$ to make this bound at most $\xi$.
Therefore, $\by$ is a $\xi$-approximate KKT point of~\eqref{cesq:quartic-program}.
Finally, we round the disutility coefficients finely enough to control the resulting change in aggregate demand and preserve this decoding guarantee, which is~\cref{cesq:rational-realization} for the constructed rational market.
\subsubsection{Completing the reduction and decoding prices}
\label{cesq:composition}

We now combine the three preceding results in the order of the proof overview.
We first replace box constraints by quartic penalties, then form the zero-sum model $G$, and finally construct the CES market.
From every accepted price output, we reverse these steps to recover an exact KKT point.

\begin{theorem}\label{cesq:fixed-hardness}
For all fixed rational exponents $\rho>1$, finite-precision equilibrium price
computation in~\cref{def:ces-price-problem} is $\CLS$-hard.
\end{theorem}
\begin{proof}
\textbf{Constructing the quartic source.}
We start from a rational instance of~\eqref{pg:box-qp} on $[0,1]^n$, writing $\bz$ for its variables.
The affine change $\bx:=2\bz-\one$ gives an equivalent rational quadratic program on $[-1,1]^n$ with polynomial encoding length.
If $n<4$, we add variables with zero coefficients in the objective until $n=4$.
These variables do not affect the KKT conditions for the original coordinates.

By~\cref{ces2src:hardness}, we first choose $M,k$ and the source dimension $D=n(k+1)+1$, independently of the radial coefficient $c$.
We set $\sigma:=\rho/(\rho-1)$ and $m:=2D$.
By~\cref{cesg:prop:embedding}, we compute positive rational $A,B$ depending only on $\sigma,m$.
We choose $c:=2MB/A$ and complete the construction of $F$ and its gradient tolerance $\eta$ using~\cref{ces2src:hardness}.

\paragraph{Forming the model and constructing the market.}
We rescale $F$ to $\widetilde F=(2A/M)F$ and form $G$ by~\cref{cesq:paired-coordinates,cesq:quartic-model}.
Since its quartic part is $Q_4$, it has the form required by~\cref{cesg:prop:embedding}.
For its terms of degrees one through three, we compute the radius $R_*$ in that proposition.
We then choose a positive rational $\xi$ satisfying~\cref{cesq:paired-tolerance} with $R=R_*$.
The radius is independent of $\xi$. Therefore, these choices are not circular.
Applying~\cref{cesg:prop:embedding} with this target error gives a positive rational CES market, a scale $h>0$, and a market clearing tolerance $\varepsilon>0$.
We use this market and tolerance as an instance of~\cref{def:ces-price-problem}.

\paragraph{Decoding every accepted price output.}
Every accepted output $\bp$ is a normalized positive $\varepsilon$-approximate equilibrium price vector.
By~\cref{cesq:rational-realization}, the rescaled deviations $\by=(m\bp-\one)/h$ satisfy
\[
 \|\by\|_2\le R_*\qquad\text{and}\qquad
 \|\nabla_HG(\by)\|_2\le\xi.
\]
We recover the source coordinates by $v_i=(y_{2i-1}-y_{2i})/2$.
By~\cref{cesq:paired-decoding}, we have $\|\nabla F(\bv)\|_\infty\le\eta$.
We can therefore apply~\cref{ces2src:hardness}: we round the original coordinates of $\bv$ near the boundary of $[-1,1]^n$ and solve the affine version of the rational linear program in~\cref{eq:box-repair} on $[-1,1]^n$ to obtain an exact KKT point.
The affine substitution at the start of the proof gives an exact KKT point of the original instance of~\eqref{pg:box-qp}.

The dimensions and coefficient lengths of $F$ and $G$ are polynomial in the encoding length of the original quadratic program.
The scales, tolerances, and market data also have polynomial encoding length by~\cref{ces2src:hardness,cesg:prop:embedding}.
The construction and the decoding from an accepted rational price vector take polynomial time.
Since finding a KKT point of~\eqref{pg:box-qp} is $\CLS$-complete, finite-precision equilibrium price computation is $\CLS$-hard for this fixed exponent.
\end{proof}

We combine this hardness result with membership to obtain the following completeness classification.

\begin{corollary}\label{cor:ces-fixed-membership}
For all fixed rational exponents $\rho>1$, finite-precision equilibrium price computation in~\cref{def:ces-price-problem} is $\CLS$-complete.
\end{corollary}
\begin{proof}
The membership construction in~\cref{sec:ces-membership} also runs in polynomial time when a rational exponent $\rho>1$ is supplied with the market data and encoded in binary.
Thus, membership holds both for this variant and for each fixed rational exponent.
By~\cref{cesq:fixed-hardness}, finite-precision equilibrium price computation is $\CLS$-hard for each such constant.
Combining these results, we obtain $\CLS$-completeness.
Since hardness holds at $\rho=2$, the problem with a binary-encoded input exponent is also $\CLS$-complete, and we have proved~\cref{thm:ces}.
\end{proof}

\section{Strongly Polynomial-time Algorithm for Rank-two Linear Chores Fisher Markets}
\label{sec:tractable}

In this section, we provide a strongly polynomial algorithm for exact equilibrium computation in positive linear chores Fisher markets whose disutility matrix has ordinary rank at most two, with arbitrary positive rational earning requirements.

The key idea is to find orders of agents and chores 
in which the disutility ratio of each pair of consecutive chores is \emph{nondecreasing} across agents.
For each boundary between consecutive chores, we choose an agent index and set the corresponding price ratio equal to that agent's disutility ratio.
This makes the chosen agent indifferent between the two chores.
We normalize the prices to make their sum equal the total earning requirement.


Our algorithm use the standard procedure of repeatedly applying a monotone map from the smallest element of a finite product lattice~\cite[Section~2]{etessami2026tarski}.
To construct equilibrium earnings, 
we use the northwest-corner transportation rule, with earning requirements as row totals and chore prices as column totals~\cite{klinz2011northwest}.
Here, the rank-two condition supplies a common order for all chore boundaries.
We prove that the resulting price-index update is monotone and that its fixed points yield equilibria.

\subsection{Algorithm}
\label{sec:rank-two-algorithm}

We separate the computation into three oracles: 
The ordering oracle {\color{orange} \textsc{Order}} checks the rank condition and finds suitable agent and chore orders.
At each iteration, we use {\color{orange} \textsc{Update}} to construct prices, compute their cumulative sums, and choose the next indices.
When the indices stop changing, we use {\color{orange} \textsc{Recover}} to compute an allocation and restore the original agent and chore orders.

We first implement the ordering oracle, which uses row normalization to recognize rank at most two and then sorts the agents and chores.
The normalized coefficients are used only to find the permutations.
The price and allocation computations use the original coefficients in these orders.

\begin{algorithm}[H]
\caption{{\color{orange} \textsc{Order}}: Ordering oracle}
\label{alg:rank-two-ordering}
\small
\begin{algorithmic}[1]
\Require A positive rational disutility matrix $(d_{ij})$.
\Ensure Permutations $\pi,\tau$ making every adjacent disutility ratio nondecreasing across agents, or a report that the rank exceeds two.
\State $\widehat d_{ij}\gets d_{ij}/d_{i1}
       \quad\forall\,i\in[n],\,j\in[m]$.
\State Initialize $\pi$ and $\tau$ as the identity permutations.
\If{the normalized rows $\widehat{\bd}_i$ are not all identical}
    \State Choose $k$ such that $\widehat d_{ik}$ varies with $i$.
    \State Choose $i_-\in\argmin_i\widehat d_{ik}$ and $i_+\in\argmax_i\widehat d_{ik}$.
    \State $u_j\gets\widehat d_{i_-j}$, $v_j\gets\widehat d_{i_+j}\quad\forall\,j\in[m]$.
    \State $t_i\gets\frac{\widehat d_{ik}-u_k}{v_k-u_k}\quad\forall\,i\in[n]$.
    \If{$\widehat d_{ij}\ne(1-t_i)u_j+t_i v_j$ for some $i,j$}
        \State \Return \textnormal{``rank exceeds two''}.
    \EndIf
    \State Choose $\pi$ such that $t_{\pi(1)}\le\cdots\le t_{\pi(n)}$.
    \State Choose $\tau$ such that
           $v_{\tau(1)}/u_{\tau(1)}\ge\cdots\ge v_{\tau(m)}/u_{\tau(m)}$.
\EndIf
\State \Return $(\pi,\tau)$.
\end{algorithmic}
\end{algorithm}

After applying the permutations $\pi,\tau$, we use the original disutilities in these orders and permute the earning requirements with the agents.
The adjacent disutility ratios, cumulative earning requirements, and cumulative prices are
\begin{equation*}
    r_{ij} := \frac{d_{\pi(i),\tau(j)}}{d_{\pi(i),\tau(j+1)}} \ \forall\, j < m, \quad 
    S_i := \sum_{h=1}^iB_{\pi(h)}, \quad 
    P_j:=\sum_{k=1}^jp_k.
\end{equation*}
We set $S_0:=P_0:=0$ and collect the prefixes in $\mathbf S:=(S_0,\ldots,S_n)$ and $\bP:=(P_0,\ldots,P_m)$.
The update oracle constructs prices from the current indices and finds the first earning prefix that reaches each proper price prefix.

\begin{algorithm}[H]
\caption{{\color{orange} \textsc{Update}}: Price and index update oracle}
\label{alg:rank-two-update}
\small
\begin{algorithmic}[1]
\Require Ratios $r_{ij}>0$, earning prefixes $0=S_0<\cdots<S_n$, and indices $\ba\in[n]^{m-1}$.
\Ensure Prices $\bp$ with sum $S_n$, their prefixes $\bP$, and updated indices $\ba'$.
\State $q_m\gets1$.
\For{$j=m-1,m-2,\ldots,1$}
    \State $q_j\gets r_{a_j,j}q_{j+1}$.
\EndFor
\State $p_j\gets\frac{S_nq_j}{\sum_{k=1}^mq_k}\quad\forall\,j\in[m]$.
\State $P_0\gets0$, $P_j\gets\sum_{k=1}^jp_k\quad\forall\,j\in[m]$.
\State $\ba'\gets\ba$, $i\gets1$.
\For{$j=1,\ldots,m-1$}
    \While{$S_i<P_j$}
        \State $i\gets i+1$.
    \EndWhile
    \State $a'_j\gets i$.
\EndFor
\State \Return $(\bp,\bP,\ba')$.
\end{algorithmic}
\end{algorithm}

Once the indices are fixed, the recovery oracle computes payments from the overlaps of earning and price intervals.
We divide each payment by its chore price to obtain an allocation, then undo the permutations.

\begin{algorithm}[H]
\caption{{\color{orange} \textsc{Recover}}: Equilibrium recovery oracle }
\label{alg:rank-two-recovery}
\small
\begin{algorithmic}[1]
\Require Prices $\bp$ and prefixes $\bP$ from an update with unchanged indices for the ordered market, earning prefixes $\mathbf S$, and permutations $\pi,\tau$.
\Ensure Exact equilibrium prices and an allocation in the original agent and chore orders.
\ForAll{$i\in[n],\,j\in[m]$}
    \State $b_{ij}\gets\max\{0,\min(S_i,P_j)-\max(S_{i-1},P_{j-1})\}$.
    \State $x^{\mathrm{out}}_{\pi(i),\tau(j)}\gets b_{ij}/p_j$.
\EndFor
\State $p^{\mathrm{out}}_{\tau(j)}\gets p_j\quad\forall\,j\in[m]$.
\State \Return $(\bp^{\mathrm{out}},\bx^{\mathrm{out}})$.
\end{algorithmic}
\end{algorithm}

With these three oracles, we obtain the main algorithm in~\cref{alg:rank-two}.
We start with every boundary index equal to one and call the recovery oracle once an update leaves the indices unchanged.

\begin{algorithm}[H]
\caption{Exact equilibrium computation}
\label{alg:rank-two}
\small
\begin{algorithmic}[1]
\Require Positive rational disutilities $d_{ij}$, earning requirements $B_i$, and unit supplies.
\Ensure An exact CE, or a report that $\operatorname{rank}((d_{ij}))>2$.
\State Call \Call{\color{orange} Order}{$(d_{ij})$} to obtain $\pi,\tau$, or return its rank-failure report.
\State $r_{ij}\gets d_{\pi(i),\tau(j)}/d_{\pi(i),\tau(j+1)}
       \quad\forall\,i\in[n],\,j<m$.
\State $S_0\gets0$, $S_i\gets\sum_{h=1}^iB_{\pi(h)}\quad\forall\,i\in[n]$.
\State Set $\mathbf S\gets(S_0,\ldots,S_n)$ and initialize $\ba\in[n]^{m-1}$ by $a_j\gets1\quad\forall\,j<m$.
\Loop
    \State $(\bp,\bP,\ba') \gets \Call{\color{orange} Update}{(r_{ij}),\mathbf S,\ba}$.
    \If{$\ba'=\ba$}
        \State \Return \Call{\color{orange} Recover}{$\bp,\bP,\mathbf S,\pi,\tau$}.
    \EndIf
    \State $\ba\gets\ba'$.
\EndLoop
\end{algorithmic}
\end{algorithm}

\begin{figure}[H]
\centering
\small

\begin{tikzpicture}[
  stage/.style={
    draw=black!60,
    rounded corners=2pt,
    fill=blue!4,
    text width=3.4cm,
    minimum height=30mm,
    align=center,
    inner sep=5pt
  }
]
\node[stage] (start) at (0,0) {
  \textbf{Start: $a=(1,1)$}\\[4pt]
  $\mathbf p=(1,1,1)$\\[4pt]
  $(P_1,P_2)=(1,2)$\\[4pt]
  $a'=(1,2)$
};

\node[stage] (next) at (4.5,0) {
  \textbf{Next: $a=(1,2)$}\\[4pt]
  $\mathbf p=\left(\frac98,\frac98,\frac34\right)$\\[4pt]
  $(P_1,P_2)=\left(\frac98,\frac94\right)$\\[4pt]
  $a'=(2,3)$
};

\node[stage,fill=green!8] (fixed) at (9,0) {
  \textbf{Fixed point: $a=(2,3)$}\\[4pt]
  $\mathbf p=\left(\frac32,1,\frac12\right)$\\[4pt]
  $(P_1,P_2)=\left(\frac32,\frac52\right)$\\[4pt]
  $a'=(2,3)=a$
};

\draw[->,thick] (start.east) -- (next.west);
\draw[->,thick] (next.east) -- (fixed.west);
\end{tikzpicture}

\medskip
\textbf{At the fixed point: recover payments by interval overlap}

\smallskip
\begin{tikzpicture}[x=2.4cm,y=1cm,font=\small]
  \draw[fill=blue!12] (0,0) rectangle (1,0.5);
  \draw[fill=blue!24] (1,0) rectangle (2,0.5);
  \draw[fill=blue!36] (2,0) rectangle (3,0.5);

  \node at (0.5,0.25) {$A_1$};
  \node at (1.5,0.25) {$A_2$};
  \node at (2.5,0.25) {$A_3$};
  \node[anchor=east] at (-0.1,0.25) {Earnings};

  \draw[fill=orange!12] (0,-0.9) rectangle (1.5,-0.4);
  \draw[fill=orange!24] (1.5,-0.9) rectangle (2.5,-0.4);
  \draw[fill=orange!36] (2.5,-0.9) rectangle (3,-0.4);

  \node at (0.75,-0.65) {$c_1$};
  \node at (2,-0.65) {$c_2$};
  \node at (2.75,-0.65) {$c_3$};
  \node[anchor=east] at (-0.1,-0.65) {Prices};

  \foreach \t in {1,2}
    \draw[densely dashed,black!45] (\t,0) -- (\t,-0.9);
  \foreach \t in {1.5,2.5}
    \draw[densely dashed,black!45] (\t,0) -- (\t,-0.4);

  \foreach \t/\lab in {
    0/0,1/1,1.5/{\frac{3}{2}},
    2/2,2.5/{\frac{5}{2}},3/3
  } {
    \draw (\t,-0.9) -- (\t,-1);
    \node[below] at (\t,-1) {$\lab$};
  }
\end{tikzpicture}
\caption{A $3\times3$ rank-two market with unit chore supplies
and unit earning requirements.
The boundary indices increase from $(1,1)$ to $(1,2)$ and then to the fixed point $(2,3)$.}
\label{fig:rank-two-example}
\end{figure}
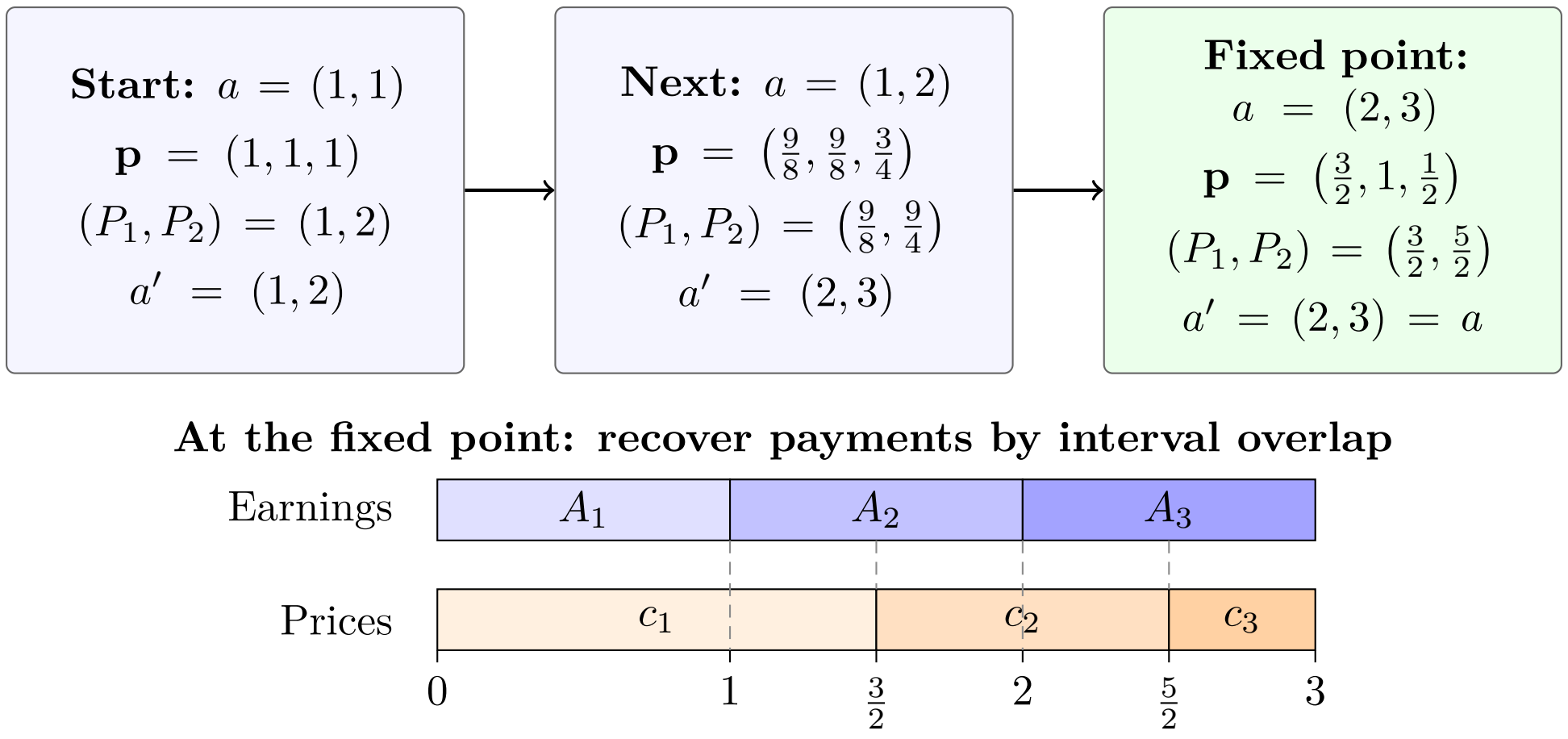

We provide an running example for~\cref{alg:rank-two} in~\cref{fig:rank-two-example}, where we consider a market instance with $3$ agents and $3$ chores, and the data: 
\[
    D:=\begin{pmatrix}
    1&1&1\\
    9&6&4\\
    7&4&2
    \end{pmatrix},
    \quad 
    (B_1,B_2,B_3):=(1,1,1),
    \quad 
    (S_1,S_2,S_3)=(1,2,3).
\]
In the current permutation, the adjacent disutility ratios increase across agents:
\[
    1<\frac32<\frac74
    \quad\text{for chores }1,2,
    \qquad
    1<\frac32<2
    \quad\text{for chores }2,3.
\]
For each boundary-index pair of agents $a=(a_1,a_2)$, we compute
\[
    \frac{p_j}{p_{j+1}}
      =\frac{d_{a_j,j}}{d_{a_j,j+1}},
    \quad 
    \sum\nolimits_{k=1}^{3}p_k=3, 
    \quad 
    P_j := \sum\nolimits_{k=1}^{j}p_k,
    \quad 
    a'_j := \min\{i:S_i \geq P_j\}
    \ \forall\, j = 1,2. 
\]

The computed equilibrium earnings and allocations are 
\[
    b=\begin{pmatrix}
    1&0&0\\
    \frac12&\frac12&0\\
    0&\frac12&\frac12
    \end{pmatrix}
    \quad \text{ and } \quad 
    x=\begin{pmatrix}
    \frac23&0&0\\
    \frac13&\frac12&0\\
    0&\frac12&1
    \end{pmatrix}.
\]

\begin{remark}[Rank-two and two-agent markets]
    One may think whether we can decompose the rank-two market into two-agent markets. 
    We comment that, the rank-two condition does not by itself justify directly applying an algorithm for two-agent markets.
    Indeed, after row normalization, we can express each disutility vector as a convex combination of two endpoint disutility vectors.
    However, replacing an agent by independently optimizing endpoint agents can change demand, even if we split its earning requirement using the same coefficients.

    For example, consider three unit-supply chores and three agents, each with earning requirement one, whose disutility vectors are
    \[
        \bu = (2,3,6),\qquad
        \bw = \frac{u+v}{2}=(4,3,4),\qquad
        \bv = (6,3,2).
    \]
    The disutility matrix has rank exactly two.
    Splitting the middle agent's earning requirement equally between the endpoint types gives two representative agents with earning requirements $3/2$ each.
    Their market has an equilibrium at
    \[
        \bp=\left(\frac35,\frac9{10},\frac32\right),
    \]
    where the $\bu$-agent performs all of chores $1$ and $2$, and the $\bv$-agent performs all of chore $3$.
    At these prices, however, the original $\bw$-agent and $\bv$-agent both uniquely minimize disutility per unit of payment on chore $3$, which lead to an aggregate demand that is larger than the supply on chore $3$. 
    Thus, these prices cannot support an equilibrium of the original market.
    Consequently, splitting earning requirements according to the rank factorization does not automatically preserve equilibria.
    
    Thus, using the fixed-agent algorithms, e.g., the algorithms in~\cite{branzei2024algorithms} would require an additional equilibrium-preserving reduction beyond the rank factorization.
\end{remark}

\subsection{Analysis}\label{sec:rank-two-analysis}

We prove correctness and a strongly polynomial complexity bound for~\cref{alg:rank-two}.

\begin{theorem}[Exact equilibrium for rank-two disutilities]\label{thm:rank-two}
Consider a positive rational linear Fisher chores market with $n$ agents, $m$ unit-supply chores, and arbitrary positive rational earning requirements.
If $\operatorname{rank}((d_{ij})_{i \in [n], j \in [m]}) \leq 2$, 
an exact equilibrium, including prices and an allocation, can be computed using 
\[
    O\!\left(nm(n+m)\right)
\]
rational arithmetic operations and comparisons.
The rank condition can be recognized within this bound.
The rational numbers computed by the algorithm have polynomial encoding length. 
Therefore, the algorithm is strongly polynomial and has polynomial bit complexity.
\end{theorem}

\begin{proof}
\textbf{Finding an order.}
We first verify the ordering oracle in~\cref{alg:rank-two-ordering}.
Normalize each row by its first entry, writing $\widehat d_{ij}=d_{ij}/d_{i1}$.
This operation preserves rank, and every normalized row has first coordinate one.
If all normalized rows are identical, the matrix has rank one and any row and column orders suffice.
Otherwise, under the rank-two assumption, the normalized rows lie on an affine line: their two-dimensional linear span intersects the hyperplane with first coordinate one in such a line.
Choose a coordinate $k$ that varies among the normalized rows, and let $\bu,\bv$ be rows with the smallest and largest $k$th coordinate.
Then, we have
\begin{equation}\label{eq:rank-two-affine}
t_i=\frac{\widehat d_{ik}-u_k}{v_k-u_k}\in[0,1],
\quad
\widehat d_{ij}=(1-t_i)u_j+t_i v_j\quad\forall\,i,j.
\end{equation}
We recognize the rank-two condition exactly by checking these identities: they hold for all rank-two inputs, and whenever they hold, all rows lie in the span of $\bu,\bv$.

In the rank-two case, order the agents by nondecreasing $t_i$ and the chores by nonincreasing $v_j/u_j$.
For $i<h$ and $j<k$ in these orders,
\[
\widehat d_{ij}\widehat d_{hk}-\widehat d_{ik}\widehat d_{hj}
=(t_h-t_i)(u_jv_k-v_ju_k)\le0.
\]
Restoring the positive row scales preserves the inequality.
In the rank-one case, these inequalities hold with equality for any orders.
Consequently the original coefficients, after permutation only, satisfy the multiplicative Monge condition
\begin{equation}\label{eq:rank-two-monge}
d_{ij}d_{hk}\le d_{ik}d_{hj}
\quad\forall\,i<h,\ j<k.
\end{equation}
Permute the earning requirements with the agents.
We can use the remaining construction for any matrix satisfying \cref{eq:rank-two-monge} in the supplied order, irrespective of its rank.

\paragraph{A monotone price construction.}
In the update oracle in~\cref{alg:rank-two-update}, we use adjacent disutility ratios $r_{ij}$ and cumulative earning requirements $S_i$, defined by
\[
r_{ij}=\frac{d_{ij}}{d_{i,j+1}},\quad
S_i=\sum_{h=1}^iB_h,\quad S_0=0.
\]
By condition~\cref{eq:rank-two-monge}, $r_{ij}$ is nondecreasing in $i$ for all $j<m$.
Given an index vector $\ba\in\{1,\ldots,n\}^{m-1}$, construct prices by
\begin{equation*}\label{eq:rank-two-prices}
\begin{aligned}
q_m&=1,\\
q_j&=r_{a_j,j}q_{j+1}
    \quad\text{for }j=m-1,\ldots,1,\\
p_j(\ba)&:=\frac{S_n q_j}{\sum_{k=1}^mq_k}
    \quad\forall\,j\in[m].
\end{aligned}
\end{equation*}
We define the cumulative prices by $P_j(\ba):=\sum_{k=1}^jp_k(\ba)$, with $P_0(\ba):=0$ and $P_m(\ba)=S_n$.
Increasing one adjacent price ratio multiplies all unnormalized prices to the left of that boundary by a common factor at least one.
After normalization, every proper price prefix weakly increases: a prefix to the left is scaled in its entirety, while a prefix to the right has an unscaled complement whose normalized total weakly decreases.
Thus, $\ba\mapsto \bP(\ba)$ is componentwise nondecreasing.

Start with $a_j=1$ and repeatedly construct $\bp(\ba)$ and update
\begin{equation}\label{eq:rank-two-update}
a'_j=\min\{i:S_i\ge P_j(\ba)\}\quad\forall\,j<m.
\end{equation}
The index selection in~\cref{eq:rank-two-update} is well defined because $0<P_j(\ba)<S_n$.
Since every selected index is at least one, the first update is componentwise no smaller than the initial vector.
If the indices do not decrease in one update, the price prefixes cannot decrease.
Since the selection rule is also nondecreasing, the next update cannot decrease the indices either.
By induction, every update is nondecreasing.
Each strict update increases $\sum_{j=1}^{m-1}a_j$ by at least one.
There are therefore at most $(n-1)(m-1)$ strict updates and at most $1+(n-1)(m-1)$ price constructions before a fixed index vector is reached.

\paragraph{Recovering an equilibrium.}
At a fixed index vector, the recovery oracle in~\cref{alg:rank-two-recovery} computes the payments
\begin{equation*}\label{eq:rank-two-payments}
b_{ij}=\max\{0,\min(S_i,P_j)-\max(S_{i-1},P_{j-1})\}.
\end{equation*}
The payments $b_{ij}$ are the lengths of intersections of the intervals $(S_{i-1},S_i)$ and $(P_{j-1},P_j)$.
Since each list of intervals partitions $(0,S_n)$ up to endpoints, summing these intersection lengths recovers each interval's length.
Therefore, $\sum_{j=1}^mb_{ij}=B_i$ and $\sum_{i=1}^nb_{ij}=p_j$.
If $b_{ij}>0$, then $P_{j-1}<S_i$ and $P_j>S_{i-1}$.
For $k<j$, we have $P_k\le P_{j-1}<S_i$. Thus, the selector in~\cref{eq:rank-two-update} satisfies $a_k\le i$.
For $k\ge j$ with $k<m$, we have $P_k\ge P_j>S_{i-1}$. Hence, the selected index satisfies $a_k\ge i$.
Since
\[
\frac{p_k/d_{ik}}{p_{k+1}/d_{i,k+1}}
=\frac{r_{a_k,k}}{r_{ik}},
\]
agent $i$'s payment per unit disutility weakly increases up to chore $j$ and weakly decreases after it.
Thus, we have $j\in\argmax_k p_k/d_{ik}$ whenever $b_{ij}>0$.
By taking $\beta_i=\max_jp_j/d_{ij}$ and $x_{ij}=b_{ij}/p_j$, we obtain \eqref{eq:ce-spending}.
We undo the row and column permutations to obtain an equilibrium of the original market.
For $m=1$, the index vector is empty, we set $p_1=S_n$, and we use the same payment formula.

\paragraph{Complexity.}
The ordering oracle uses $O(nm)$ rational operations and comparisons for row normalization, endpoint selection, and verification of~\cref{eq:rank-two-affine}.
Sorting takes $O(n\log n+m\log m)$ comparisons.
Each call to the update oracle uses $O(m)$ rational operations to construct prices and $O(n+m)$ comparisons to scan the increasing earning and price prefixes.
The recovery oracle uses $O(nm)$ operations for payments and allocations.
Combining these costs with the iteration bound gives $O(nm(n+m))$ operations overall.
In each update call, we recompute each unnormalized price from at most $m-1$ input ratios. Therefore, products do not accumulate across iterations.
Row normalization, affine parameters, budget sums, price normalization, and payment recovery all have polynomial intermediate encoding lengths.
We have thus proved both the arithmetic and bit-complexity assertions.
\end{proof}

We use ordinary linear-algebraic rank and do not require a supplied factorization or ordering.
This rank restriction does not imply uniqueness: two equal-earning agents with disutility rows $(1,2)$ and $(2,1)$ have the three equilibrium price vectors
\[
\left(\frac{2}{3},\frac{4}{3}\right),\quad(1,1),\quad\left(\frac{4}{3},\frac{2}{3}\right).
\]
Thus, we can compute exact equilibria efficiently for rank-two disutility matrices even when equilibrium prices are not unique.

\paragraph{Role of the rank-two assumption.}
We use the rank bound only in the ordering oracle.
After row normalization, the rows either coincide or lie on an affine line. We can therefore find a common order for every adjacent disutility ratio.
Once an order satisfying~\cref{eq:rank-two-monge} is available, the price iteration and payment recovery work regardless of rank.
At rank three, such an order may not exist.
For example, the rank-three disutility matrix
\[
 \begin{pmatrix}
 1&2&2\\
 2&1&2\\
 2&2&1
 \end{pmatrix}
\]
has no row and column permutations satisfying~\cref{eq:rank-two-monge}.
For any chore order, the agent with coefficient $1$ on the middle chore has the largest ratio at the first boundary and the smallest at the second. Thus, the two boundaries require incompatible agent orders.
With earning requirements $B_i=1$, unit prices and the identity allocation are nevertheless an equilibrium.
Without a common order, the preceding arguments no longer guarantee monotone index updates or optimality of the recovered allocation.
Extending the result to all rank-three markets would therefore require a way to handle these incompatible ratio orders.

\begingroup
\small
\bibliographystyle{alpha}
\bibliography{cls}

@inproceedings{boodaghians2022polynomial,
  title={Polynomial Time Algorithms to Find an Approximate Competitive Equilibrium for Chores},
  author={Boodaghians, Shant and Chaudhury, Bhaskar Ray and Mehta, Ruta},
  booktitle={Proceedings of the 2022 Annual ACM-SIAM Symposium on Discrete Algorithms (SODA)},
  pages={2285--2302},
  year={2022},
  organization={SIAM}
}

@article{bogomolnaia2017competitive,
  title={Competitive division of a mixed manna},
  author={Bogomolnaia, Anna and Moulin, Herv{\'e} and Sandomirskiy, Fedor and Yanovskaya, Elena},
  journal={Econometrica},
  volume={85},
  number={6},
  pages={1847--1871},
  year={2017},
  publisher={Wiley Online Library}
}

@article{bogomolnaia2019dividing,
  title={Dividing bads under additive utilities},
  author={Bogomolnaia, Anna and Moulin, Herv{\'e} and Sandomirskiy, Fedor and Yanovskaia, Elena},
  journal={Social Choice and Welfare},
  volume={52},
  number={3},
  pages={395--417},
  year={2019},
  publisher={Springer}
}

@article{branzei2024algorithms,
  title={Algorithms for competitive division of chores},
  author={Br{\^a}nzei, Simina and Sandomirskiy, Fedor},
  journal={Mathematics of Operations Research},
  volume={49},
  number={1},
  pages={398--429},
  year={2024},
  publisher={INFORMS}
}

@inproceedings{chaudhury2022existence,
  title={On the existence of competitive equilibrium with chores},
  author={Chaudhury, Bhaskar Ray and Garg, Jugal and McGlaughlin, Peter and Mehta, Ruta},
  booktitle={13th Innovations in Theoretical Computer Science Conference (ITCS 2022)},
  year={2022}
}

@article{chaudhury2022competitive,
  title={Competitive equilibrium with chores: Combinatorial algorithm and hardness},
  author={Chaudhury, Bhaskar Ray and Garg, Jugal and McGlaughlin, Peter and Mehta, Ruta},
  journal={arXiv preprint arXiv:2205.11363},
  year={2022}
}

@article{chaudhury2023complementary,
  title={A complementary pivot algorithm for competitive allocation of a mixed manna},
  author={Chaudhury, Bhaskar Ray and Garg, Jugal and McGlaughlin, Peter and Mehta, Ruta},
  journal={Mathematics of Operations Research},
  volume={48},
  number={3},
  pages={1630--1656},
  year={2023},
  publisher={INFORMS}
}

@inproceedings{daskalakis2011continuous,
  title={Continuous local search},
  author={Daskalakis, Constantinos and Papadimitriou, Christos},
  booktitle={Proceedings of the twenty-second annual ACM-SIAM symposium on Discrete Algorithms},
  pages={790--804},
  year={2011},
  organization={SIAM}
}

@article{fearnley2022complexity,
  title={The complexity of gradient descent: CLS $=$ PPAD $\cap$ PLS},
  author={Fearnley, John and Goldberg, Paul and Hollender, Alexandros and Savani, Rahul},
  journal={Journal of the ACM},
  volume={70},
  number={1},
  pages={1--74},
  year={2022},
  publisher={ACM, New York, NY}
}

@article{fearnley2025complexity,
  title={The complexity of computing KKT solutions of quadratic programs},
  author={Fearnley, John and Goldberg, Paul and Hollender, Alexandros and Savani, Rahul},
  journal={Journal of the ACM},
  volume={72},
  number={5},
  pages={1--48},
  year={2025},
  publisher={ACM, New York, NY}
}

@inproceedings{babichenko2021settling,
  title={Settling the complexity of Nash equilibrium in congestion games},
  author={Babichenko, Yakov and Rubinstein, Aviad},
  booktitle={Proceedings of the 53rd annual ACM SIGACT symposium on theory of computing},
  pages={1426--1437},
  year={2021}
}

@article{garg2024satiation,
  title={Satiation in Fisher markets and approximation of Nash social welfare},
  author={Garg, Jugal and Hoefer, Martin and Mehlhorn, Kurt},
  journal={Mathematics of Operations Research},
  volume={49},
  number={2},
  pages={1109--1139},
  year={2024},
  publisher={INFORMS}
}

@article{eisenberg1959consensus,
  title={Consensus of subjective probabilities: The pari-mutuel method},
  author={Eisenberg, Edmund and Gale, David},
  journal={The Annals of Mathematical Statistics},
  volume={30},
  number={1},
  pages={165--168},
  year={1959},
  publisher={JSTOR}
}

@inproceedings{cheung2013tatonnement,
  title={Tatonnement beyond gross substitutes? Gradient descent to the rescue},
  author={Cheung, Yun Kuen and Cole, Richard and Devanur, Nikhil},
  booktitle={Proceedings of the forty-fifth annual ACM symposium on Theory of computing},
  pages={191--200},
  year={2013}
}

@article{chaudhury2024competitive,
  title={Competitive equilibrium for chores: from dual Eisenberg-Gale to a fast, greedy, LP-based algorithm},
  author={Chaudhury, Bhaskar Ray and Kroer, Christian and Mehta, Ruta and Nan, Tianlong},
  journal={arXiv preprint arXiv:2402.10439},
  year={2024}
}

@inproceedings{chaudhury2026tatonnement,
  title={T{\^a}tonnement dynamics for fisher markets with chores},
  author={Chaudhury, Bhaskar Ray and Kroer, Christian and Mehta, Ruta and Nan, Tianlong},
  booktitle={Proceedings of the 58th Annual ACM Symposium on Theory of Computing},
  pages={386--397},
  year={2026}
}

@article{chen2024computing,
  title={Computing competitive equilibrium for chores: Linear convergence and lightweight iteration},
  author={Chen, He and Jiang, Chonghe and So, Anthony Man-Cho},
  journal={arXiv preprint arXiv:2410.04036},
  year={2024}
}

@article{chaudhury2026fully,
  title={Fully Distributed T{\^a}tonnement for Chores Markets},
  author={Chaudhury, Bhaskar Ray and Kroer, Christian and Mehta, Ruta and Nan, Tianlong},
  journal={arXiv preprint arXiv:2607.00300},
  year={2026}
}

@inproceedings{tao2026fisher,
  title={Fisher Meets Lindahl: A Unified Duality Framework for Market Equilibrium},
  author={Tao, Yixin and Zheng, Weiqiang},
  booktitle={Proceedings of the 58th Annual ACM Symposium on Theory of Computing},
  pages={286--297},
  year={2026}
}

@article{liu2026competitive,
  title={Competitive equilibrium for linear chores is {CLS}-complete},
  author={Liu, Qian and Ma, Shinuo and Tao, Yixin and Yang, Minji and Zheng, Weiqiang},
  journal={arXiv preprint arXiv:2611.06589},
  year={2026}
}

@article{disser2018simplex,
  title={The simplex algorithm is NP-mighty},
  author={Disser, Yann and Skutella, Martin},
  journal={ACM Transactions on Algorithms (TALG)},
  volume={15},
  number={1},
  pages={1--19},
  year={2018},
  publisher={ACM, New York, NY, USA}
}

@article{etessami2010complexity,
  title={On the complexity of Nash equilibria and other fixed points},
  author={Etessami, Kousha and Yannakakis, Mihalis},
  journal={SIAM Journal on Computing},
  volume={39},
  number={6},
  pages={2531--2597},
  year={2010},
  publisher={SIAM}
}

@book{schrijver1998theory,
  title={Theory of linear and integer programming},
  author={Schrijver, Alexander},
  year={1998},
  publisher={John Wiley \& Sons}
}

@article{etessami2026tarski,
  title={Tarski’s theorem, supermodular games, and the complexity of equilibria},
  author={Etessami, Kousha and Papadimitriou, Christos H and Rubinstein, Aviad and Yannakakis, Mihalis},
  journal={SIAM Journal on Computing},
  volume={55},
  number={6},
  pages={1185--1228},
  year={2026},
  publisher={SIAM}
}

@article{klinz2011northwest,
  title={The Northwest corner rule revisited},
  author={Klinz, Bettina and Woeginger, Gerhard J},
  journal={Discrete Applied Mathematics},
  volume={159},
  number={12},
  pages={1284--1289},
  year={2011},
  publisher={Elsevier}
}

@article{devanur2008market,
	title={Market equilibrium via a primal-dual algorithm for a convex program},
	author={Devanur, Nikhil R and Papadimitriou, Christos H and Saberi, Amin and Vazirani, Vijay V},
	journal={Journal of the ACM (JACM)},
	volume={55},
	number={5},
	pages={1--18},
	year={2008},
	publisher={ACM New York, NY, USA},
    doi={10.1145/1411509.1411512}
}

@inproceedings{orlin2010improved,
    author = {Orlin, James B.},
    title = {Improved algorithms for computing fisher's market clearing prices},
    year = {2010},
    isbn = {9781450300506},
    publisher = {Association for Computing Machinery},
    booktitle = {Proceedings of the Forty-Second ACM Symposium on Theory of Computing},
    pages = {291–300},
}

@book{walras1874elements,
    author    = {Walras, L{\'e}on},
    title     = {{\'E}l{\'e}ments d'{\'e}conomie politique pure
               ou Th{\'e}orie de la richesse sociale},
    publisher = {L. Corbaz; Guillaumin et Cie; H. Georg},
    address   = {Lausanne; Paris; B{\^a}le},
    year      = {1874--1877},
}

@article{chaudhury2026competitive,
  title={Competitive Equilibrium in Labor Economies through the Lens of Goods and Chores Fisher Markets},
  author={Chaudhury, Bhaskar Ray and Kroer, Christian and Mehta, Ruta and Nan, Tianlong and Yang, Zongjun},
  journal={arXiv preprint arXiv:2606.15060},
  year={2026}
}
\endgroup

\appendix

\addtocontents{toc}{\protect\setcounter{tocdepth}{-1}}

\section{Omitted proofs for CES hardness}\label{app:ces-hardness}

We give the technical proofs omitted from the hardness walkthrough in~\cref{sec:ces-fixed-proof}.
We first establish the quartic-source proposition and the paired-coordinate decoding lemma.
We then construct the reference CES market and prove the realization proposition, including localization of approximate equilibrium prices and rounding of disutility coefficients.

\subsection{Proof of the quartic-source proposition}
\label{app:ces-quartic-source}

We first establish the penalty estimates needed for the construction in~\cref{ces2src:hardness}.

For a penalty scale $M>0$, an integer chain length $k\ge1$, and $\lambda\ge0$,
let $\by:=(y_1,\ldots,y_k)\in\R^k$ and set $y_0:=x$.
We define the chain penalty and its minimized value by
\begin{equation*}
\begin{aligned}
 J_\lambda(x,\by)&:=
 M\sum_{l=1}^k(y_l-y_{l-1}^2)^2+M y_k^4
       +\lambda\left(x^2+\sum_{l=1}^k y_l^2\right),\\
 P_\lambda(x)&:=\min_{\by\in\R^k} J_\lambda(x,\by).
\end{aligned}
\end{equation*}
In the following lemma, we bound the derivative of the minimized penalty inside and outside the interval.

\begin{lemma}[Squaring-chain penalty]\label{ces2src:chain}
For every $x\in\R$, the function $J_\lambda(x,\cdot)$ has a unique stationary point $\by$, which is its global minimizer.
The function $P_\lambda$ is differentiable, convex and even, satisfies
$P_\lambda(0)=0$, and $P_\lambda'(x)/x$ is nondecreasing for $x>0$.
At the minimizer, we have
\begin{equation}\label{ces2src:chain-signs}
 0\le y_l\le y_{l-1}^2,\quad y_l\le |x|^{2^l}
 \quad\forall\,l\in[k].
\end{equation}
For $|x|\le1$, we have
\begin{equation}\label{ces2src:inside-bound}
 |P_\lambda'(x)|\le M\,2^{k+2}|x|^{2^{k+2}-1}+\lambda\,2^{k+2}.
\end{equation}
For $0<\zeta\le1$, we have
\begin{equation*}\label{ces2src:outside-bound}
 \frac{P_\lambda'(1+\zeta)}{1+\zeta}\ge \frac{M\zeta^2}{16}.
\end{equation*}
If $|x|<2$, the minimizing vector $\by$ satisfies $|y_l|<3$ for all $l\in[k]$.
\end{lemma}

\begin{proof}
We eliminate $y_k,y_{k-1},\ldots,y_1$ in this order.
At each step, the reduced objective has the form
$M(y-s)^2+\lambda y^2+V(y^2)$ with $V$ convex and even, starting with $V(u)=Mu^2$.
Since $V$ is nondecreasing on $[0,\infty)$, this objective is $2M$-strongly convex in $y$, and its minimized value is convex and even in $s$.
Repeating this argument, the stationarity equations have a unique solution, which is the global minimizer.

After eliminating all chain variables, we have $P_\lambda(x)=\lambda x^2+V_1(x^2)$, where $V_1$ is convex
and nondecreasing on $[0,\infty)$.
From this representation, we see that $P_\lambda$ is convex and even,
$P_\lambda(0)=0$, and $P_\lambda'(x)/x$ is nondecreasing for $x>0$.
By uniqueness of the minimizer, we also obtain differentiability of $P_\lambda$.

For any chain vector, we write $r_l:=y_l-y_{l-1}^2$ for the error at link $l$.
From $\partial J_\lambda/\partial y_l=0$ for all $l\in[k]$, we obtain
\[
 r_k=-2y_k^3-\frac{\lambda}{M}y_k,
 \quad
 r_l=2y_l r_{l+1}-\frac{\lambda}{M}y_l\quad\forall\,l<k.
\]
The equation for $r_k$ is equivalent to
$(1+\lambda/M)y_k+2y_k^3=y_{k-1}^2$.
Then, we have $y_k\ge0$ and $r_k\le0$.
For $l=k-1,\ldots,1$, we have
$y_l(1+\lambda/M-2r_{l+1})=y_{l-1}^2$.
By induction, we obtain the inequalities in~\cref{ces2src:chain-signs}.

By differentiating $P_\lambda(x)=\min_{\by}J_\lambda(x,\by)$ with respect to $x$, we have
$P_\lambda'(x)=2\lambda x-4Mxr_1$.
We expand the recurrence for $r_1$ and use~\cref{ces2src:chain-signs}.
The term originating from $-2y_k^3$ has absolute value at most
$M\,2^{k+2}|x|^{2^{k+2}-1}$.
For $|x|\le1$, the sum of the absolute values of the terms proportional
to $\lambda$, including $2\lambda x$, is at most
$2\lambda+4\lambda\sum_{l=1}^k2^{l-1}\le\lambda\,2^{k+2}$.
Therefore, we obtain~\cref{ces2src:inside-bound}.

We prove $P_\lambda(1+\zeta)\ge M\zeta^2/4$ first for $\lambda=0$.
If $J_0(1+\zeta,\by)<M\zeta^2/4$, we have $|r_l|<\zeta/2$ for all
$l\in[k]$ and $y_k^4<\zeta^2/4$.
Since $y_0=1+\zeta$ and $y_l=y_{l-1}^2+r_l$, we obtain inductively
$y_l\ge(1+\zeta)^2-\zeta/2>1+\zeta$ for all $l\in[k]$,
which contradicts $y_k^4<\zeta^2/4$.
Thus, we have $P_\lambda(1+\zeta)\ge P_0(1+\zeta)\ge M\zeta^2/4$.
Since $P_\lambda$ is convex, $P_\lambda(0)=0$, and $1+\zeta\le2$,
we also have
\begin{equation*}
 \frac{P_\lambda'(1+\zeta)}{1+\zeta}\ge \frac{P_\lambda(1+\zeta)}{(1+\zeta)^2}\ge \frac{M\zeta^2}{16}.
\end{equation*}

Finally, for fixed $|x|<2$, we compare the minimum with the value at
$y_1=\cdots=y_k=0$.
Then, we have $P_\lambda(x)-\lambda x^2\le Mx^4<16M$.
At the minimizing vector $\by$, we obtain $|r_l|<4$ for
all $l\in[k]$ and $y_k<2$ from this inequality.
By the identities $y_{l-1}^2=y_l-r_l$, we obtain $|y_l|<3$ for all
$l\in[k]$.
\end{proof}

The preceding bounds apply when the chain variables minimize the penalty exactly.
At an approximate stationary point of the quartic polynomial, the derivatives with respect to these variables may only be small.
We next bound the change in the derivative with respect to $x$ when we replace such chain variables by their exact minimizing values.

\begin{lemma}[Approximate elimination]\label{ces2src:approx-elimination}
Suppose $|x|,|y_l|\le H$ and $|\partial J_\lambda/\partial y_l|\le\theta$
for all $l\in[k]$, where $H\ge4$ and $\lambda\ge0$ is fixed.
Let $\by^*$ be the minimizer of $J_\lambda(x,\cdot)$.
If, in addition, $|f'(x)+\partial J_\lambda/\partial x|\le\theta$ for a differentiable function $f$, then
\begin{equation}\label{ces2src:effective-error}
 |f'(x)+P_\lambda'(x)|\le\theta(1+2H)^{k+1}.
\end{equation}
If $|x|<2$, we also have
\begin{equation}\label{ces2src:aux-distance}
 |y_l-y_l^*|\le\frac{\theta(1+2H)^{k+1}(H+4)^{k+1}}{2M}
 \quad\forall\,l\in[k].
\end{equation}
The same conclusions hold for each $x_i$, with $\partial f/\partial x_i$ in place of $f'$, when $f$ depends on
$x_1,\ldots,x_n$ and each $x_i$ has variables $y_{i1},\ldots,y_{ik}$.
\end{lemma}

\begin{proof}
We use $e_l$ to bound the absolute derivative with respect to $y_l$
after we minimize over $y_{l+1},\ldots,y_k$, with $y_{l-1}$ fixed at its
current value.
For $l=k$, we have $e_k\le\theta$.
By strong convexity, the distance from the current $y_{l+1}$ to its
minimizing value, with $y_l$ fixed, is at most $e_{l+1}/(2M)$.
The partial derivative with respect to $y_l$ changes by $4M|y_l|$
times this distance, where $4M|y_l|\le4MH$.
Then, we have
\[
 e_l\le\theta+2He_{l+1}.
\]
By applying the same estimate to the partial derivative with respect to $x$, we obtain
\begin{equation*}
 |f'(x)+P_\lambda'(x)|\le\theta+2He_1\le\theta(1+2H)^{k+1}.
\end{equation*}
We have proved~\cref{ces2src:effective-error}.

After we minimize over $y_{l+1},\ldots,y_k$, the objective as a function
of $y_l$ is $M(y_l-y_{l-1}^2)^2$ plus a convex function.
For an objective of this form, changing $y_{l-1}^2$ changes the minimizing value of $y_l$ by at most the same amount.
For $|x|<2$, the minimizer $\by^*$ of $J_\lambda(x,\cdot)$ satisfies
$|y_l^*|<3$ for all $l\in[k]$ by~\cref{ces2src:chain}.
We define $y_0^*:=x$, $d_l:=|y_l-y_l^*|$, and $d_0:=0$.
Since $|y_{l-1}^2-(y_{l-1}^*)^2|\le(H+3)d_{l-1}$, we have
\[
 d_l\le\frac{\theta(1+2H)^{k+1}}{2M}+(H+3)d_{l-1},\quad\text{with }d_0=0.
\]
By iterating this inequality and bounding the geometric sum by $(H+4)^{k+1}$,
we obtain~\cref{ces2src:aux-distance}.
\end{proof}

We now apply the two penalty estimates to construct the quartic polynomial in~\cref{ces2src:hardness} and verify its decoding guarantee.

\begin{proof}[Proof of \cref{ces2src:hardness}]
We bound the gradient of the source quadratic by setting
\[
 G:=1+\max_i\left(|(\bc_0)_i|+\sum_{j=1}^n|Q_{ij}|\right).
\]
Then, we have $|\nabla_i f(\bx)|\le G\max\{1,\|\bx\|_\infty\}$.
By the affine version of~\cref{lem:box-round} on $[-1,1]^n$, we compute $0<\Delta\le1$ such that any rational strong $\varepsilon$-KKT point with $\varepsilon<\Delta$ can be converted to an exact KKT point of~\eqref{ces2src:box-qp} in polynomial time.
We choose a boundary margin $\zeta$ and a penalty scale $M$ to keep changes in the quadratic gradient small across the margin and make the penalty gradient dominate beyond it:
\[
 \zeta:=\min\left\{\frac{1}{8},\frac{\Delta}{8G}\right\},
 \quad M\ge\frac{64G}{\zeta^2},
\]
where $M$ is the least power of two satisfying $M\ge64G/\zeta^2$.
We choose the chain length to make the first term on the right-hand side of~\cref{ces2src:inside-bound} small for $|x|\le1-\zeta$.
We choose the least integer
\[
 b\ge8+\left\lceil\log_2\left(\frac{8M}{\Delta}\right)\right\rceil
          +\left\lceil\log_2\left(\frac{1}{\zeta}\right)\right\rceil
\]
and a power of two $N$ satisfying $16b^2/\zeta\le N<32b^2/\zeta$.
We then set $k:=\log_2 N-2$, for which $N=2^{k+2}$.
Then, we have
\begin{equation}\label{ces2src:tiny-interior}
 MN(1-\zeta)^{N-1}\le\frac{\Delta}{8}.
\end{equation}
Indeed, $\log_2(8MN/\Delta)\le2b$, whereas
$\zeta(N-1)\ge16b^2-1$.

We now apply a separate squaring chain to each original variable.
For each $x_i$, we introduce the variables $y_{i1},\ldots,y_{ik}$ and define
$\bv:=(\bx,\by)$, $d:=n(k+1)$, and $S:=\sum_{j=1}^d v_j^2$.
The parameters $M,k,d$ are independent of $c$.
We define
\[
 F_0(\bv):=f(\bx)+M\sum_{i=1}^n\sum_{l=1}^k
              (y_{il}-y_{i,l-1}^2)^2+M\sum_{i=1}^n y_{ik}^4,
 \quad y_{i0}:=x_i.
\]
The coefficient of $v_j^4$ is $M$ for all $j\in[d]$.
We next add a radial correction while keeping its effect on the gradient small.
We use $S_0:=16d$ as a target bound on $S$, and prove below that every sufficiently accurate stationary point satisfies this bound.
To choose the scale of the correction on this region, we define
\[
 K_0:=6cS_0^2+4(c+M)S_0^3.
\]
We choose $R$ to be the least power of two satisfying
\begin{equation}\label{ces2src:R-choice}
 R\ge\max\left\{1,\frac{1}{c},\frac{16NK_0}{\Delta}\right\}.
\end{equation}
We introduce one more variable $z$ and define
\begin{align}
 E(S,z)&:=c(S-Rz)^2+2cSz^2+(c+M)z^4,\label{ces2src:E}\\
 F(\bv,z)&:=F_0(\bv)+E(S,z).\label{ces2src:F}
\end{align}
By expanding $E(S,z)$, we obtain~\cref{ces2src:quartic-form} for $\bw:=(\bv,z)$.
The terms of degree less than four in $E(S,z)$ are $cR^2z^2$ and $-2cRzS$.

For fixed $S\ge0$, the derivative
\[
 E_z=2cR^2z-2cRS+4cSz+4(c+M)z^3
\]
is strictly increasing in $z$, since $E_{zz}\ge2cR^2$.
This derivative $E_z$ has a unique zero $z_*(S)\in[0,S/R]$.
We define $\lambda(S):=E_S(S,z_*(S))$.
Since $E_S=2c(S-Rz+z^2)$ and $z_*(S)\le S/R$, we have $\lambda(S)\ge0$.
We also have the upper bound
\begin{equation}\label{ces2src:lambda-bound}
 0\le\lambda(S)
 \le \frac{6cS^2}{R^2}+\frac{4(c+M)S^3}{R^4}.
\end{equation}
To verify the upper bound in~\cref{ces2src:lambda-bound}, we use
$E_z(S,z_*)=0$ to obtain
\begin{equation*}
 S-Rz_*=\frac{2Sz_*}{R}+\frac{2\left(1+\frac{M}{c}\right)z_*^3}{R}.
\end{equation*}
By substituting $z_*\le S/R$ into $E_S=2c(S-Rz_*+z_*^2)$, we obtain
the stated upper bound on $\lambda(S)$.

We first bound $\|\bw\|_\infty$ for points satisfying $\|\nabla F(\bw)\|_\infty\le1$.
Let $C$ be one plus the sum of the absolute values of the coefficients of $F$ in degrees one through three, and set $D:=d+1$.
For $s:=\|\bw\|_\infty\ge1$, Euler's identity and the quartic part in~\cref{ces2src:quartic-form} give
\[
 4Ms^4\le3Cs^3+Ds\le(3C+D)s^3.
\]
Thus, $\|\bw\|_\infty\le H$ for
\[
 H:=\max\left\{4,1+\frac{3C+D}{M}\right\},
\]
which has polynomial binary encoding length.
To bound the increase in stationarity error when we eliminate $z$, we set
\[
 D_0:=1+2H(R+2H+1).
\]
We choose
\begin{equation}\label{ces2src:eta-choice}
 0<\eta\le
 \min\left\{1,\frac{\Delta}{16D_0(1+2H)^{k+1}},
                      \frac{M}{D_0(1+2H)^{k+1}(H+4)^{k+1}}\right\}.
\end{equation}
The logarithms of all inverse tolerances remain polynomial in the
encoding length of $f$ and in the encoding length of $c$.

We consider any rational point $(\bv,z)$ with $\|\nabla F(\bv,z)\|_\infty\le\eta$
and set
$z_*:=z_*(S)$, $\lambda:=\lambda(S)\ge0$.
Since $E_{zz}\ge2cR^2$, we have
$|z-z_*|\le\eta/(2cR^2)\le\eta/2$,
where the last inequality follows from \cref{ces2src:R-choice}.
In particular, we have $|z_*|\le H+1/2$.
If we replace $z$ by $z_*$, the partial derivative with respect to $v_j$
changes by at most
\[
 4cH|z-z_*|(R+|z|+|z_*|)
 \le \frac{2H\eta(R+2H+1)}{R^2}.
\]
After replacing $z$ by $z_*$, the radial correction contributes $2\lambda v_j$ to the derivative with respect to $v_j$.
Therefore, with $\theta:=D_0\eta$ and $\lambda$ treated as a fixed
nonnegative constant in the following gradient, we have
\begin{equation*}
 \|\nabla_{\bv}(F_0(\bv)+\lambda\sum_{j=1}^d v_j^2)\|_\infty\le\theta.
\end{equation*}

By applying~\cref{ces2src:approx-elimination} to the variables
$y_{i1},\ldots,y_{ik}$ for each $i\in[n]$, we obtain
\begin{equation*}
 |\nabla_i f(\bx)+P_\lambda'(x_i)|\le\theta(1+2H)^{k+1}\le\frac{\Delta}{16}\le1.
\end{equation*}
If $s=\|\bx\|_\infty\ge1+\zeta$, we choose $i$ with $|x_i|=s$.
By~\cref{ces2src:chain}, we have $|P_\lambda'(x_i)|\ge4G|x_i|$,
whereas $|\nabla_i f(\bx)|\le G|x_i|$.
Then, we have $|\nabla_i f(\bx)+P_\lambda'(x_i)|\ge3G|x_i|>1$,
which is a contradiction.
Therefore, we have $\|\bx\|_\infty<1+\zeta<2$.

For each $i$, the minimizer $(y_{i1}^*,\ldots,y_{ik}^*)$ of
$J_\lambda(x_i,\cdot)$ satisfies $|y_{il}^*|<3$ for all $l\in[k]$.
By~\cref{ces2src:aux-distance,ces2src:eta-choice}, we also have
$|y_{il}-y_{il}^*|\le1/2$ for all $i\in[n]$ and $l\in[k]$.
In particular, we have $|v_j|<4$ for all $j\in[d]$ and $S\le S_0=16d$.
By~\cref{ces2src:lambda-bound,ces2src:R-choice}, we obtain
\begin{equation}\label{ces2src:lambda-small}
 \lambda N\le \frac{NK_0}{R^2}\le\frac{\Delta}{16}.
\end{equation}

We now move coordinates near the boundary of the box to that boundary and leave the other coordinates unchanged.
We define $\bar{\bx}\in[-1,1]^n$ by setting $\bar x_i:=1$ if
$x_i\ge1-\zeta$, $\bar x_i:=-1$ if $x_i\le-1+\zeta$, and
$\bar x_i:=x_i$ otherwise.
Then, we have $\|\bar{\bx}-\bx\|_\infty\le\zeta$.
For $\bar x_i=1$, we use $P_\lambda'(x_i)\ge0$ to obtain
$\nabla_i f(\bx)\le\Delta/16$.
For $\bar x_i=-1$, we similarly have $\nabla_i f(\bx)\ge-\Delta/16$.
For $-1<\bar x_i<1$, we combine
\cref{ces2src:inside-bound,ces2src:tiny-interior,ces2src:lambda-small}
with $|\nabla_i f(\bx)+P_\lambda'(x_i)|\le\Delta/16$ and obtain
$|\nabla_i f(\bx)|\le\Delta/8+\Delta/16+\Delta/16=\Delta/4$.
Since $\|\bar{\bx}-\bx\|_\infty\le\zeta$, we also have
\begin{equation*}
 |\nabla_i f(\bar{\bx})-\nabla_i f(\bx)|\le G\zeta\le\frac{\Delta}{8}
 \quad\text{for all }i\in[n].
\end{equation*}
Therefore, $\bar{\bx}$ is a strong $(3\Delta/8)$-KKT point of~\eqref{ces2src:box-qp}.
Since $3\Delta/8<\Delta$, we apply the affine version of~\cref{lem:box-round} to obtain an exact KKT point of~\eqref{ces2src:box-qp}.

Finally, the positive definite quartic part in~\cref{ces2src:quartic-form} makes $F$ coercive. Therefore, $F$ attains a global minimum at a stationary point $\bw^*$.
The preceding bound gives $\|\bw^*\|_\infty\le H$.

On $[-H-1,H+1]^D$, the polynomial $\nabla F$ has a Lipschitz constant whose logarithm is polynomial in the encoding lengths of $f$ and $c$.
Since $\log(1/\eta)$ is also polynomial, sufficiently fine dyadic rounding of $\bw^*$ gives a rational point $\bw$ of polynomial encoding length with $\|\nabla F(\bw)\|_\infty\le\eta$.
The construction of $F$ and $\eta$, and the decoding of any such rational point, take polynomial time.
\end{proof}

\subsection{Proof of the paired-coordinate decoding lemma}
\label{app:ces-paired-decoding}

We bound the auxiliary coordinates $\bs$ in~\cref{cesq:quartic-model} and then their contribution to the derivatives in $\bv$.

\begin{proof}[Proof of \cref{cesq:paired-decoding}]
The change from $(\bs,\bv)$ to $\by$ multiplies direction norms by $\sqrt2$.
Since $\sum_{i=1}^D s_i=0$, changing $\bs$ in the direction $\bs$ preserves $H$, as does changing $\bv$ in any direction.
By differentiating~\cref{cesq:quartic-model} and using $\|\nabla_HG(\by)\|_2\le\xi$, we obtain
\[
 2\|\bs\|_2^2\le\bs^\top\nabla_{\bs}G\le\sqrt2\xi\|\bs\|_2,
 \qquad \|\nabla_{\bv}G\|_2\le\sqrt2\xi.
\]

Thus, $\|\bs\|_2\le\xi/\sqrt2$.
The additional terms in $G$ contribute at most $(24A+16B)\|\bs\|_2^2\|\bv\|_2$ to the norm of the gradient in $\bv$.
Since $\|\bv\|_2\le\|\by\|_2\le R$, we have
\[
 \|\nabla\widetilde F(\bv)\|_2
 \le\sqrt2\xi+\frac{(24A+16B)R\xi^2}{2}
 \le\frac{2A\eta}{M},
\]
where the last inequality follows from~\cref{cesq:paired-tolerance} and $\xi\le1$.
Because $\nabla\widetilde F=(2A/M)\nabla F$, we obtain
$\|\nabla F(\bv)\|_\infty\le\|\nabla F(\bv)\|_2\le\eta$.
\end{proof}

\subsection{A globally confining CES potential}
\label{cesq:baseline-section}\label{app:ces-baseline}

We fix a rational $\sigma>1$ and construct a symmetric CES market with $m$ chores.
Its potential will have no Taylor terms of degrees one through three at uniform prices.
We also prove a global bound that will keep approximate equilibrium prices near uniform after we modify the market.

We describe each agent by a positive probability vector, called its coefficient row, and an earning requirement.
We first introduce $m$ agents whose total earning requirement is
\[
 \beta:=1-\frac1{(2\sigma-1)^2}=\frac{4\sigma(\sigma-1)}{(2\sigma-1)^2}.
\]
For each $j\in[m]$, agent $j$ has earning requirement $\beta/m$ and coefficient row
\begin{equation*}
 \bw^{(j)}:=\left(1-\frac{1}{2\sigma}\right)\e_j+\frac{\one}{2\sigma m}.
\end{equation*}
We introduce one further agent with coefficient row $\one/m$ and earning
requirement $1-\beta$.
For each agent $i$, let $w_{ij}$ be coordinate $j$ of its coefficient row and let $B_i$ be its earning requirement.
The corresponding disutility coefficients are $a_{ij}:=w_{ij}^{-1/(\sigma-1)}$.
In this construction, we allow these coefficients to be real.
We define the price potential function
\[
 \Phi_0(\bp):=\sum_{i=1}^{m+1}\frac{B_i}{\sigma}\log\left(\sum_{j=1}^m w_{ij}p_j^\sigma\right).
\]

For the scaled price deviations $\bt:=m\bp-\one$, we write the potential as $\phi_0(\bt):=\Phi_0((\one+\bt)/m)$.
We call $\bt^\top\nabla\phi_0(\bt)$ the radial derivative, since it measures the derivative in the direction of the deviation vector $\bt$.
In the following lemma, we compute the quartic Taylor term and prove a global bound that we will use to show that approximate equilibrium prices are near uniform.

\begin{lemma}\label{cesq:global-coercivity}
There is an explicitly computable positive rational $c_{\sigma}(m)$ with
$\log(1/c_{\sigma}(m))=O_{\sigma}(\log m)$ such that, for every $\bp\geq0$ with $\sum_{j=1}^m p_j=1$,
\[
 \bt^\top\nabla\phi_0(\bt)\ge c_{\sigma}(m)\|\bt\|_2^4.
\]
The Taylor expansion of $\phi_0$ at $\bt=0$ on $\sum_{j=1}^m t_j=0$ has no linear,
quadratic, or cubic terms and has quartic term
\[
 \frac{(\sigma-1)(2\sigma-1)}{48m}\sum_{j=1}^m t_j^4
 +\frac{(\sigma-1)^2}{8m^2}\left(\sum_{j=1}^m t_j^2\right)^2.
\]
\end{lemma}
\begin{proof}
We set $z_j:=mp_j$ and $M:=m^{-1}\sum_{j=1}^m z_j^\sigma$.
We regard $z$ as taking the values $z_j$, $j\in[m]$, with equal weights $1/m$, and write $\mathbb E$ for this uniform average.
Then, we have $\mathbb Ez=1$.
By differentiating $\Phi_0$, we obtain $g_j(\bp)=H(z_j)$, where
\[
 \begin{aligned}
 H(z)&:=z^{\sigma-1}\left(\frac{\beta(2\sigma-1)}{(2\sigma-1)z^\sigma+M}+K\right),\\
 K&:=\beta\mathbb E\frac1{(2\sigma-1)z^\sigma+M}+\frac{1-\beta}{M}.
 \end{aligned}
\]
By Jensen's inequality, we have
\[
 K\ge\frac{\beta}{2\sigma M}+\frac1{(2\sigma-1)^2M}=\frac1{(2\sigma-1)M}.
\]
We set $d:=K-1/((2\sigma-1)M)$.
By the preceding lower bound, we have $d\geq0$.
By differentiating $H$ with $M,K$ held fixed, we obtain
\[
 H'(z)=z^{\sigma-2}\left[(\sigma-1)K+
 \beta(2\sigma-1)\frac{(\sigma-1)M-(2\sigma-1)z^\sigma}{((2\sigma-1)z^\sigma+M)^2}\right].
\]
The minimum of $((\sigma-1)-v)/(1+v)^2$ for $v\ge0$ is $-1/(4\sigma)$,
attained at $v=2\sigma-1$.
Since $\beta(2\sigma-1)/(4\sigma)=(\sigma-1)/(2\sigma-1)$, we have
$H'(z)\ge(\sigma-1)d z^{\sigma-2}$ for $z>0$.
Therefore, $H(z)-d z^{\sigma-1}$ is nondecreasing for $z>0$.
By continuity, this function is also nondecreasing on $[0,m]$.

We have $1\le M\le m^\sigma$ and $0\le z_j\le m$ for all $j$.
Then, we have $(2\sigma-1)z_j^\sigma+M\le2\sigma m^\sigma$ for all $j$.
The second derivative of $v\mapsto1/v$ is at least $1/(4\sigma^3m^{3\sigma})$ for $0<v\le2\sigma m^\sigma$.
By applying this lower bound to Jensen's inequality, we obtain
\[
 d\ge\frac{\sigma-1}{2\sigma^2m^{3\sigma}}\operatorname{Var}(z^\sigma).
\]
By the pointwise inequality $(z-1)(z^\sigma-1)\ge(z-1)^2$ and the
Cauchy--Schwarz inequality, we have
$\operatorname{Var}(z^\sigma)\ge\operatorname{Var}(z)$.
We also have
\[
 \operatorname{Cov}(z,z^{\sigma-1})\ge c\operatorname{Var}(z),\quad
 \text{where }c:=\begin{cases}1,&\sigma\ge2,\\(\sigma-1)m^{\sigma-2},&1<\sigma<2.\end{cases}
\]
For $\sigma\ge2$, we obtain this covariance bound from
$(z-1)(z^{\sigma-1}-1)\ge(z-1)^2$.
For $1<\sigma<2$, we use the lower bound $(\sigma-1)m^{\sigma-2}$ on the derivative
of $z^{\sigma-1}$ for $0<z\le m$.
Since $H(z)-d z^{\sigma-1}$ is nondecreasing, its covariance with $z$ is nonnegative.
Together with $\mathbb Ez=1$, this gives
\[
 \bt^\top\nabla\phi_0(\bt)
 =\mathbb E[(z-1)H(z)]
 \ge d\operatorname{Cov}(z,z^{\sigma-1})
 \ge\frac{(\sigma-1)c}{2\sigma^2m^{3\sigma+2}}\|\bt\|_2^4.
\]
We can choose a positive rational lower bound $c_{\sigma}(m)$ for
$(\sigma-1)c/(2\sigma^2m^{3\sigma+2})$ with $O_{\sigma}(\log m)$ bits.
For example, we replace $m^{3\sigma+2}$ by $m^{\lceil3\sigma+2\rceil}$ and,
when $\sigma<2$, replace $c$ by $(\sigma-1)/m$.

We now compute the Taylor expansion at uniform prices.
We write $T_r:=m^{-1}\sum_{j=1}^m t_j^r$ and expand
$(1/\sigma)\log\sum_{j=1}^m w_j(1+t_j)^\sigma$ through degree four for each coefficient row $\bw$.
We then average these expansions with the agents' earning requirements as weights.
The resulting weighted sum differs from $\phi_0$ only by a constant.
By the identity $\beta((2\sigma-1)/(2\sigma))^2=(\sigma-1)/\sigma$, we have
zero coefficients for the quadratic and cubic terms.
The quartic term is
$(\sigma-1)(2\sigma-1)T_4/48+(\sigma-1)^2T_2^2/8$.
The coefficients in these expansions are rational for fixed rational $\sigma$.
\end{proof}


\subsection{Realizing quartic models by CES markets}
\label{cesg:sec:embedding}\label{app:ces-realization}

We modify the market in~\cref{cesq:baseline-section} to prescribe the first three Taylor terms while retaining the radial bound needed to localize approximate equilibrium prices.

We fix a rational $\sigma>1$ and $m\geq8$.
As in~\cref{cesq:global-coercivity}, we use the scaled price deviations $\bt=m\bp-\one$.
For normalized prices, these deviations lie in the zero-sum subspace $H:=\{\bt:\one^T\bt=0\}$.
Normalized nonnegative prices correspond to the domain $\mathcal T:=\{\bt\in H:t_j\geq-1\}$.
We take Taylor expansions at $\bt=0$, which corresponds to uniform prices.
For an agent with coefficient row $\bw$, a positive probability vector, we define
\[
 \phi_{\bw}(\bt):=\frac1\sigma\log\sum_{j=1}^m w_j(1+t_j)^\sigma.
\]
Up to an additive constant, this is the agent's contribution to the potential in scaled coordinates per unit of earning requirement.
As above, the row $\bw$ determines the disutility coefficients $a_j=w_j^{-1/(\sigma-1)}$.
We choose rational earning requirements.

\paragraph{An auxiliary market for prescribing Taylor terms.}
We first construct an auxiliary market whose quadratic and cubic Taylor terms cancel.
We choose its row mixtures and earning requirements using the parameters
\[
 b_0:=1-\frac1{2\sigma},\quad b_1:=1-\frac1{4\sigma},\quad
 T_2:=\frac{\sigma-1}{\sigma b_1^2},\quad
 T_3:=\frac{(\sigma-1)b_0}{\sigma b_1^3}<T_2.
\]
We take $\bw^0$ from three permutation-invariant families: the singleton vectors $\e_j$ for $j\in[m]$, the pair vectors $(\e_j+\e_k)/2$ for $1\le j<k\le m$, and the uniform vector $\one/m$.
For each vector, we introduce one agent with coefficient row
$\bw:=b_1\bw^0+(1-b_1)\one/m$.
We assign total earning requirements $\beta_1$, $\beta_2$, and $\beta_u$ to the singleton, pair, and uniform families, respectively, where
\begin{equation*}
\begin{aligned}
 \beta_1&:=T_2-\frac{2(T_2-T_3)(m-2)}{m},\\
 \beta_2&:=\frac{4(T_2-T_3)(m-1)}{m},\\
 \beta_u&:=1-3T_2+2T_3.
\end{aligned}
\end{equation*}
We divide each family's total earning requirement equally among its agents throughout this auxiliary construction.
These requirements are positive because $\beta_1\geq T_2(2b_0/b_1-1)>0$, $\beta_2>0$, and $\beta_u\geq1/(2\sigma-1)^2>0$.

To obtain coefficient rows for interpolating polynomials of degree at most three, we add two more families, $\bw^0:=(2\e_j+\e_k)/3$, $j\ne k$, and
$\bw^0:=(\e_j+\e_k+\e_l)/3$, $j<k<l$, each with total earning requirement $\gamma:=\min\{\beta_1,\beta_2,\beta_u\}/100$.
We record the second and third moments of these families to choose compensating changes in the original families' earning requirements.
For $\bt\in H$, we define $A_2,A_3$ for each family by
\begin{equation*}
 \mathbb E(\bw^0\cdot \bt)^d=A_d\,m^{-1}\sum_{j=1}^m t_j^d,\qquad d=2,3,
\end{equation*}
where $\mathbb E$ denotes the uniform average over the rows in that family.
For the rows $(\e_j+\e_k)/2$, $(2\e_j+\e_k)/3$, and $(\e_j+\e_k+\e_l)/3$, respectively, we have
\begin{equation*}
\begin{aligned}
 A_2^{\mathrm{pair}}&=\frac{m-2}{2(m-1)},
 & A_3^{\mathrm{pair}}&=\frac{m-4}{4(m-1)},\\[3pt]
 A_2^{\mathrm{ordered}}&=\frac{5m-9}{9(m-1)},
 & A_3^{\mathrm{ordered}}&=\frac{m-3}{3(m-1)},\\[3pt]
 A_2^{\mathrm{triple}}&=\frac{m-3}{3(m-1)},
 & A_3^{\mathrm{triple}}&=\frac{(m-3)(m-6)}{9(m-1)(m-2)}.
\end{aligned}
\end{equation*}
The two added families change the second and third moments of the coefficient rows weighted by earning requirements.
To restore their original values, we define $U_d:=A_d^{\mathrm{ordered}}+A_d^{\mathrm{triple}}$ and replace
$\beta_1,\beta_2,\beta_u$ by $\beta_1+d_1,\beta_2+d_2,\beta_u+d_u$, where
\begin{align*}
 d_2&:=-\gamma\frac{U_2-U_3}{A_2^{\mathrm{pair}}-A_3^{\mathrm{pair}}},\\
 d_1&:=-\gamma U_2-A_2^{\mathrm{pair}}d_2,\\
 d_u&:=-2\gamma-d_1-d_2.
\end{align*}
Since $A_2^{\mathrm{pair}}-A_3^{\mathrm{pair}}=m/[4(m-1)]\geq1/4$
and the six coefficients $A_d^{\mathrm{pair}},A_d^{\mathrm{ordered}},A_d^{\mathrm{triple}}$
lie in $[0,1]$ for $m\geq8$, we have
$\max\{|d_1|,|d_2|,|d_u|\}\leq16\gamma$.
Therefore, the earning requirements $\beta_1+d_1,\beta_2+d_2,\beta_u+d_u$ remain positive.
Since $d_1+d_2+d_u=-2\gamma$, the total earning requirement remains one after adding the two families.
We let $F_1$ denote this market's potential in scaled coordinates, obtained by weighting each $\phi_{\bw}$ by the corresponding agent's earning requirement.

For a nonnegative integer vector $\boldsymbol{\alpha}$, write $|\boldsymbol{\alpha}|:=\sum_{j=1}^m\alpha_j$ and define the interpolation rows
\[
 \bw^{\boldsymbol{\alpha}}:=\frac{b_1\boldsymbol{\alpha}}{3}+\frac{(1-b_1)\one}{m}
 \quad\forall\,\boldsymbol{\alpha}\in\mathbb Z_{\geq0}^m\text{ with }|\boldsymbol{\alpha}|=3.
\]
In the following lemma, we verify the Taylor cancellation and the availability of these rows with positive earning requirements.

\begin{lemma}\label{cesg:lem:auxiliary}
The Taylor terms of $F_1$ of degrees one, two, and three vanish on $H$.
The market contains every row $\bw^{\boldsymbol{\alpha}}$ with a strictly positive earning requirement, has $O(m^3)$ agents, and satisfies $w_j\geq1/(4\sigma m)$ for every row and coordinate.
\end{lemma}
\begin{proof}
We now use $\mathbb E$ to denote the average over all coefficient rows $\bw$, with each agent's earning requirement as its weight.
These weights sum to one.
By the definitions of $\beta_1,\beta_2,\beta_u,d_1,d_2,d_u$, we have, for all $\bt\in H$,
\[
 \mathbb E(\bw\cdot \bt)^2=\frac{\sigma-1}{\sigma m}\sum_{j=1}^m t_j^2,
 \quad
 \mathbb E(\bw\cdot \bt)^3=\frac{(\sigma-1)b_0}{\sigma m}\sum_{j=1}^m t_j^3.
\]
By permutation invariance, we have $\mathbb E\bw=\one/m$.
We write $\bt^{\circ r}$ for the vector with coordinates $t_j^r$.
The quadratic-moment identity also determines mixed second moments.
Applying it to $\bt$ and the component of $\bt^{\circ2}$ whose coordinates sum to zero, we obtain
\begin{equation*}
 \mathbb E[(\bw\cdot \bt)(\bw\cdot \bt^{\circ2})]=\frac{\sigma-1}{\sigma m}\sum_{j=1}^m t_j^3.
\end{equation*}
For a fixed row $\bw$, we write $\mu_r:=\bw\cdot \bt^{\circ r}$ for the weighted coordinate moments of $\bt$.
The Taylor terms $J_{1,\bw},J_{2,\bw},J_{3,\bw}$ of $\phi_{\bw}$ of degrees one, two, and three are
\begin{align*}
 J_{1,\bw}&:=\mu_1,\\
 J_{2,\bw}&:=\frac{\sigma-1}{2}\mu_2-\frac \sigma2\mu_1^2,\\
 J_{3,\bw}&:=\frac{(\sigma-1)(\sigma-2)}6\mu_3
   -\frac{\sigma(\sigma-1)}2\mu_1\mu_2+\frac{\sigma^2}3\mu_1^3.
\end{align*}
By substituting the identities for the moments of $\bw$ into these expressions, we obtain
$\mathbb E J_{1,\bw}=\mathbb E J_{2,\bw}=\mathbb E J_{3,\bw}=0$ for all $\bt\in H$.
The listed singleton, ordered-pair, and triple families contain every row $\bw^{\boldsymbol{\alpha}}$, each with a positive earning requirement.
\end{proof}

\paragraph{Bounds on derivatives of $\phi_{\bw}$.}
Choose a computable rational $C_{\sigma}\geq1$, depending only on the fixed exponent $\sigma$, such that
\[
 \|D^k\phi_{\bw}(\bt)\|\leq C_{\sigma}
 \quad\text{for }1\leq k\leq5,\quad \|\bt\|_2\leq\frac12,
\]
uniformly over probability rows $\bw$, where the norm is the multilinear Euclidean operator norm.
Such a bound follows by differentiating the logarithm: its argument is at least $2^{-\lceil\sigma\rceil}$, and the first five derivatives of that argument are bounded on unit directions by constants depending only on $\sigma$.

The positive uniform component of each coefficient row also gives a gradient bound on the entire domain $\mathcal T$.
For each constructed row $\bw$ and all $\bt\in\mathcal T$, we have
$S:=\sum_{j=1}^m w_j(1+t_j)^\sigma\geq1/(4\sigma)$.
By H\"older's inequality, we have
\begin{equation*}\label{cesg:eq:global-atom-gradient}
 \|\nabla_{\bt}\phi_{\bw}(\bt)\|_1
 =\frac{\sum_{j=1}^m w_j(1+t_j)^{\sigma-1}}S
 \leq S^{-1/\sigma}\leq4\sigma.
\end{equation*}
Since the Taylor terms of $F_1$ of degrees one, two, and three vanish on $H$,
we obtain $|\bt^T\nabla F_1(\bt)|\leq (C_{\sigma}+32\sigma)\|\bt\|_2^4$ for $\|\bt\|_2\leq1/2$ by Taylor's theorem.
For $\|\bt\|_2>1/2$, we have
$|\bt^T\nabla F_1(\bt)|\leq4\sigma\|\bt\|_2\leq32\sigma\|\bt\|_2^4$.
More generally, we consider a signed sum of functions $\phi_{\bw}$ whose coefficients
have total absolute value $\Delta$.
After removing the Taylor terms of degrees one, two, and three from each radial derivative $\bt^T\nabla\phi_{\bw}(\bt)$, the corresponding signed sum of the remainders has absolute value at most
\begin{equation*}\label{cesg:eq:radial-remainder}
 (16C_{\sigma}+32\sigma)\Delta\|\bt\|_2^4
 \quad\forall\,\bt\in\mathcal T.
\end{equation*}
To obtain this bound for $r:=\|\bt\|_2>1/2$, we bound the sum of the first three
terms by $C_{\sigma}(r+r^2+r^3/2)\leq13C_{\sigma}r^4$ and use
$|\bt^T\nabla\phi_{\bw}(\bt)|\leq4\sigma r$.
For $r\leq1/2$, we obtain the same bound from Taylor's theorem.

We now combine the auxiliary market with the symmetric market from~\cref{cesq:global-coercivity} to retain the global radial bound.
We use the potential $\phi_0$ of that symmetric market.
By~\cref{cesq:global-coercivity}, we have $\bt^T\nabla \phi_0(\bt)\geq c_0\|\bt\|_2^4$ for all $\bt\in\mathcal T$, where
\[
 c_0:=\frac{(\sigma-1)\min\{1,\sigma-1\}}
 {2\sigma^2m^{3\lceil\sigma\rceil+3}}.
\]
In the basis $\sum_{j=1}^m t_j^4,(\sum_{j=1}^m t_j^2)^2$, the potential $\phi_0$ has quartic coefficients
\begin{equation*}
 A_0:=\frac{(\sigma-1)(2\sigma-1)}{48m},\qquad B_0:=\frac{(\sigma-1)^2}{8m^2}.
\end{equation*}
We compute the coefficients $A_1,B_1$ of
$\sum_{j=1}^m t_j^4$ and $(\sum_{j=1}^m t_j^2)^2$ in $F_1$ by exact rational Taylor expansion.
We combine the markets by multiplying the symmetric market's earning requirements by $1-\lambda$ and the auxiliary market's by $\lambda$.
We choose a mixing weight small enough to preserve the radial bound and positive quartic coefficients:
\begin{equation*}\label{cesg:eq:mixing}
 \lambda:=\min\left\{
 \frac14,\frac{c_0}{4(C_{\sigma}+32\sigma+c_0)},
 \frac{A_0}{4(A_0+|A_1|)},\frac{B_0}{4(B_0+|B_1|)}
 \right\},
 \quad F_*:=(1-\lambda)\phi_0+\lambda F_1.
\end{equation*}
The mixed potential retains a global radial bound:
\begin{equation}\label{cesg:eq:coercivity}
 \bt^T\nabla F_*(\bt)\geq \frac{c_0}{2}\|\bt\|_2^4
 \quad\forall\,\bt\in\mathcal T.
\end{equation}
The degree-four Taylor term of $F_*$ is
\begin{equation}\label{cesg:eq:quartic}
 Q_4(\bt):=A\sum_{j=1}^m t_j^4+B\left(\sum_{j=1}^m t_j^2\right)^2,
 \quad
 \frac34A_0\leq A\leq\frac54A_0,
 \quad\frac34B_0\leq B\leq\frac54B_0.
\end{equation}
The Taylor terms of $F_*$ of degrees one, two, and three vanish on $H$.
The constants and earning requirements have polynomial encoding length.

\paragraph{Prescribing Taylor terms on $H$.}
We next allow signed changes in the earnings of the agents with rows $\bw^{\boldsymbol{\alpha}}$, while requiring these changes to sum to zero.
We first choose changes that prescribe the Taylor terms of degrees one, two, and three on $H$.
We then scale them to keep every earning requirement positive.
To choose the changes, we use the following interpolation polynomials in a coefficient vector $\bv$:
\[
 L_{\boldsymbol{\alpha}}(\bv):=\prod_j\frac{\prod_{r=0}^{\alpha_j-1}(3v_j-r)}{\alpha_j!},
 \quad\text{for }|\boldsymbol{\alpha}|=3,
\]
For every nonnegative integer vector $\boldsymbol{\beta}$ with $|\boldsymbol{\beta}|=3$, they satisfy $L_{\boldsymbol{\alpha}}(\boldsymbol{\beta}/3)=\mathbf1_{\boldsymbol{\alpha}=\boldsymbol{\beta}}$.
Therefore, we can uniquely recover a polynomial
of degree at most three on the coefficient simplex $\{\bv\in\R_{\geq0}^m:\one^T\bv=1\}$ from its values at the nodes $\boldsymbol{\alpha}/3$.
The same conclusion holds at the nodes $\bw^{\boldsymbol{\alpha}}$ under the affine change $\bw:=b_1\bv+(1-b_1)\one/m$.
To relate interpolation to the potential, write each coefficient row as $\bw=\one/m+\bz$.
For $\bt\in H$, the coefficients of
$(\bz\cdot \bt)$, $(\bz\cdot \bt)^2$, and $(\bz\cdot \bt)^3$ in the Taylor terms
of degrees one, two, and three, respectively, are $1,-\sigma/2,\sigma^2/3$.
These coefficients are nonzero, and the remaining terms involve moments
of lower degree in $\bz$.
Then, we have an invertible block triangular transformation from the moments
of signed changes in earnings to the sum of these changes and their contributions
to the Taylor terms of degrees one, two, and three on $H$.
We can therefore prescribe these three Taylor terms while requiring the
sum of the earnings changes to be zero.

For homogeneous rational polynomials $P_d$ on $H$, $d=1,2,3$, we compute
signed changes $\bd^{(d)}$ in the earnings of the agents with rows $\bw^{\boldsymbol{\alpha}}$.
We require these changes to sum to zero and to contribute $P_d$ as the only
nonzero Taylor term through degree three.
By exact rational interpolation, we can compute $\bd^{(1)},\bd^{(2)},\bd^{(3)}$
in polynomial time, with polynomial binary encoding length.
To bound the total size of the earnings changes, we define $M_J:=1+\sum_{d=1}^3\|\bd^{(d)}\|_1$ and let $b_*>0$ be the smallest
earning requirement among the agents with rows $\bw^{\boldsymbol{\alpha}}$ in the mixed market with potential $F_*$.
For $h\in(0,1]$, we change these agents' earnings by
\[
 \boldsymbol{\delta}(h):=h^3\bd^{(1)}+h^2\bd^{(2)}+h \bd^{(3)},
 \quad\Delta:=\|\boldsymbol{\delta}(h)\|_1\leq hM_J.
\]
We choose $K\geq1$ with $|P_d(\bt)|\leq K\|\bt\|_2^d$ for all $\bt\in H$ and $d\in\{1,2,3\}$, and define
$R_* :=\max\{1,32K/c_0\}$ as a radius for localizing the rescaled price deviations.
We require $h$ to satisfy
\begin{equation}\label{cesg:eq:scale-conditions}
 hM_J\leq\min\left\{\frac{b_*}{2},\frac{c_0}{64(C_{\sigma}+2\sigma)}\right\},
 \quad hR_*\leq\frac{1}{2}.
\end{equation}
The bound involving $b_*$ keeps the earning requirements positive.
The remaining bounds control the radial remainder and the Taylor neighborhood.

We write $\nabla_H$ for the gradient within $H$, equivalently the orthogonal projection of an ambient gradient onto $H$.
For the prescribed homogeneous terms $P_1,P_2,P_3$, we define the quartic polynomial
\begin{equation*}
 G(\by):=Q_4(\by)+P_1(\by)+P_2(\by)+P_3(\by),\qquad \by\in H.
\end{equation*}
For a normalized price vector $\bp$, we use the rescaled deviations $\by:=(m\bp-\one)/h$.
In the following lemma, we show that prices with small market clearing error are near uniform and explain how to recover an approximate stationary point of $G$ from them.

\begin{lemma}\label{cesg:lem:unrounded}
Suppose $\bp$ is a normalized positive price vector of the mixed market with earnings
changed by $\boldsymbol{\delta}(h)$, and $\|\bgg(\bp)-\one\|_\infty\leq\varepsilon\leq h^3$.
Then, we have
\[
 \|m\bp-\one\|_2\leq R_*h.
\]
We also have
\begin{equation}\label{cesg:eq:decoding-error}
 \|\nabla_HG(\by)\|_2
 \leq\frac{\varepsilon}{h^3}
      +C_{\sigma}h\left(R_*^4+M_JR_*^3\right).
\end{equation}
\end{lemma}
\begin{proof}
We let $F$ denote the potential after the change $\boldsymbol{\delta}(h)$ in earnings,
and define $\bt=m\bp-\one$ and $r=\|\bt\|_2$.
By~\cref{cesg:eq:coercivity,cesg:eq:quartic,cesg:eq:scale-conditions}, we have
\[
 \bt^T\nabla F(\bt)\geq\frac{c_0}{4}r^4
      -K(h^3r+2h^2r^2+3hr^3).
\]
Since $\bt\in H$, we have $\bt^T\one=0$ and can replace $\bgg(\bp)$ by $\bgg(\bp)-\one$ in the radial derivative.
Since the market clearing error is at most $\varepsilon$ and $\nabla_{\bt}F(\bt)=\bgg(\bp)/m$, we obtain
\begin{equation*}
 |\bt^T\nabla F(\bt)|\leq\frac{\varepsilon\|\bt\|_1}{m}\leq\varepsilon r.
\end{equation*}
For $r\geq h$, we have
$K(h^3r+2h^2r^2+3hr^3)+\varepsilon r\leq7Khr^3$.
Then, we have $r\leq28Kh/c_0\leq R_*h$.
For $r<h$, the inequality $r\leq R_*h$ follows from $R_*\geq1$.

Therefore, we have $r\leq hR_*\leq1/2$, and we can use the derivative bounds
through order five for the functions $\phi_{\bw}$.
The remainder in $\nabla_HF_*(\bt)$ after degree three has Euclidean norm at most $C_{\sigma}r^4$.
The remainder in $\nabla_H(F-F_*)(\bt)$ after degree two has Euclidean norm at most $C_{\sigma}\Delta r^3$.
We also have $\|\nabla_HF(\bt)\|_2\leq\varepsilon$, since $\nabla_{\bt}F(\bt)=\bgg(\bp)/m$.
By dividing these bounds by $h^3$ and using the homogeneity of $Q_4,P_1,P_2,P_3$,
we obtain~\cref{cesg:eq:decoding-error}.
\end{proof}

\paragraph{Rational coefficients and completion of the realization.}
We now choose the required precision and round the disutility coefficients to complete the proof of the main realization proposition.

\begin{proof}[Proof of \cref{cesg:prop:embedding}]
The positive rational coefficients $A,B$ in~\cref{cesg:eq:quartic} depend only on $\sigma,m$.
For the prescribed terms $P_1,P_2,P_3$, take $K$ to be one plus the sum of the absolute values of their coefficients and set $R_*:=\max\{1,32K/c_0\}$.
Thus, $A,B$ are independent of the prescribed terms, and $R_*$ is independent of the target tolerance $\xi$.

Choose $0<h\leq1$ with a power-of-two denominator satisfying~\cref{cesg:eq:scale-conditions} and
\[
 C_{\sigma}h(R_*^4+M_JR_*^3)\leq\frac{\xi}{2},
 \qquad 0<\varepsilon_0\leq\min\left\{h^3,\frac{\xi h^3}{2}\right\}.
\]
By~\cref{cesg:lem:unrounded}, every normalized positive $\varepsilon_0$-approximate equilibrium price vector of this market yields $\|\by\|_2\leq R_*$ and $\|\nabla_HG(\by)\|_2\leq\xi$.

The coefficient rows $\bw$ are rational, but the disutility coefficients $a_{ij}=w_{ij}^{-1/(\sigma-1)}$ may be irrational.
Replace each $a_{ij}$ by a positive rational $\widetilde a_{ij}$ with relative error at most
\[
 \delta:=\min\left\{\frac12,\frac{\varepsilon_0}{32(\sigma-1)}\right\}.
\]
Since $\sigma$ is fixed and rational, we can compute these fixed-degree roots to the required relative accuracy by bisection in polynomial time.
By~\cref{eq:ces-demand}, for every positive price vector,
\[
 \left(\frac{1-\delta}{1+\delta}\right)^{\sigma-1}
 \leq\frac{\widetilde x_{ij}}{x_{ij}}
 \leq\left(\frac{1+\delta}{1-\delta}\right)^{\sigma-1}.
\]
The same bounds hold for aggregate demands and with the two markets interchanged.
With the chosen $\delta$ and $\varepsilon_0\leq1$, these ratios differ from one by at most $\varepsilon_0/4$.
Consequently, every $\varepsilon_0/2$-approximate equilibrium price vector of the rounded market satisfies
\[
 \|\bgg(\bp)-\one\|_\infty
 \leq\frac{\varepsilon_0}{2}
      +\frac{\varepsilon_0}{4}\left(1+\frac{\varepsilon_0}{2}\right)
 <\varepsilon_0
\]
for the unrounded market, and hence has the required localization and gradient bounds.
We return the rounded market with tolerance $\varepsilon:=\varepsilon_0/2$.

Since the bounds in these precision choices have polynomial encoding length, we can compute $h,\varepsilon_0,\delta$ and the rounded market data with polynomial encoding length in polynomial time.
\end{proof}

\section{Exact CES equilibrium computation}\label{sec:ces-exact}\label{app:ces-exact}

An exact equilibrium need not have rational prices, even when $\rho=2$.
For this example, we leave earning requirements unnormalized and consider two agents with $B_1=B_2=1$, two unit-supply chores, and disutilities
\[
D_1(\bx):=\sqrt{x_1^2+x_2^2},\quad
D_2(\bx):=\sqrt{x_1^2+2x_2^2}.
\]
We write $r:=p_1/p_2$ and use $p_1+p_2=2$.
By~\cref{eq:ces-demand}, the market-clearing condition is equivalent to
\begin{equation*}\label{eq:ces-irrational}
4r^3-3r^2+3r-2=0.
\end{equation*}
The polynomial $4r^3-3r^2+3r-2$ changes sign between zero and one and has positive derivative $12(r-1/4)^2+9/4$.
By the rational-root test, this polynomial has no rational root.
Therefore, the unique positive equilibrium ratio $r$ and the prices $p_1=2r/(1+r)$, $p_2=2/(1+r)$ are irrational.
Dividing both earning requirements and prices by two gives a market with total earnings one whose equilibrium prices remain irrational.
For exact CES price computation, we therefore use a representation that permits irrational outputs.
We represent prices and allocations using fixed points of continuous maps from a unit cube to itself, specified by arithmetic circuits with rational constants and gates $+,-,\times,/,\min,\max$.
A polynomial-size fixed-point representation of this kind establishes membership in $\mathsf{FIXP}$~\cite{etessami2010complexity}.

\begin{proposition}\label{prop:ces-fixp}
For all binary-encoded rational input exponents $\rho>1$, exact normalized prices and allocations belong to $\mathsf{FIXP}$ in its standard algebraic-circuit representation.
\end{proposition}
\begin{proof}
We construct such a map by introducing variables for prices and for the rational powers in the demand formula.
We write $\sigma-1=1/(\rho-1)=s/t$, where $s,t$ are positive binary integers.
For $\bz\in[0,1]^m$, we set $\bp:=\operatorname{proj}_{\Delta_1}(\bz)$ and $v_{ij}:=(a_{\min}/a_{ij})p_j$.
Then, we have $v_{ij}\in[D^2/(4m),1]$ for all $i,j$.
We introduce auxiliary variables $\by:=(y_{ij})_{i,j}\in[0,1]^{nm}$ to represent these powers and define
\begin{equation}\label{eq:ces-root-update}
y'_{ij}:=\min\{1,\max\{0,y_{ij}+v_{ij}^{s}-y_{ij}^{t}\}\}.
\end{equation}
Suppose that $y'_{ij}=y_{ij}$ in~\cref{eq:ces-root-update}.
Since $v_{ij}>0$, we cannot have $y_{ij}=0$.
If $y_{ij}=1$, we have $v_{ij}=1$.
If $0<y_{ij}<1$, we have $y_{ij}+v_{ij}^s-y_{ij}^t=y_{ij}$.
Thus, we have $y_{ij}^t=v_{ij}^s$ in each case.
We compute the powers in~\cref{eq:ces-root-update} by repeated squaring, using $O(\log s+\log t)$ gates.
We also compute $\kappa:=(D^2/(4m))^s>0$ by repeated squaring and set $\bar y_{ij}:=\max\{\kappa,y_{ij}\}$.
The value $\kappa$ is the output of a subcircuit of polynomial size, even if its expanded rational encoding is long.
We substitute these auxiliary variables into the demand formula to form $\bcalG(\bp,\by)$, and update the price coordinates as follows:
\[
\mathcal G_j(\bp,\by):=\sum_{i=1}^n\frac{B_i\bar y_{ij}}{\sum_{k=1}^mp_k\bar y_{ik}},
\quad \bz':=\operatorname{proj}_{\Delta_1}\bigl(\bp-\bcalG(\bp,\by)\bigr).
\]
For all cube inputs, we have $\sum_{k=1}^mp_k\bar y_{ik}\ge\kappa$ for all agents $i$.
Therefore, the denominators in $\bcalG$ are positive throughout the cube.

At a fixed point $(\bz,\by)$, we have $\bz=\bp\in \Delta_1$ and $y_{ij}=v_{ij}^{s/t}\ge v_{ij}^s\ge\kappa$.
The common factor $a_{\min}^{\sigma-1}$ cancels from the numerator and denominator of each agent's demand.
Then, we have $\bcalG(\bp,\by)=\bgg(\bp)$.
At a fixed point, the price update therefore satisfies
\begin{equation*}
\bp=\operatorname{proj}_{\Delta_1}(\bp-\bgg(\bp)).
\end{equation*}
The boundary argument in~\cref{sec:ces-membership}, with step size one and zero displacement, rules out $p_j=D/(4m)$.
The projection optimality conditions then imply that every $g_j(\bp)$ equals the same multiplier.
Since $\bp^\top\bgg(\bp)=1$ and $\sum_{j=1}^mp_j=1$, this multiplier is one. Therefore, $\bgg(\bp)=\one$.
Conversely, for any equilibrium price vector $\bp$, the variables $\bz=\bp$ and $y_{ij}=v_{ij}^{s/t}$ form a fixed point.
We recover the prices by taking the $\bz$ coordinates.

To recover allocations, we add variables $u_{ij}\in[0,1]$ with updates
\[
u'_{ij}:=\frac{D B_i\bar y_{ij}}{4m\sum_{k=1}^mp_k\bar y_{ik}}.
\]
Since $p_j\ge D/(4m)$ and $B_i\le1$, we have $0\le u'_{ij}\le1$ for all cube inputs.
At a fixed point, we recover $x_{ij}=4m u_{ij}/D$ by a separable linear map, as permitted in the definition of $\mathsf{FIXP}$.
\end{proof}

\end{document}